\documentclass[a4paper,fleqn,usenatbib]{mnras}
\usepackage{amsmath,graphicx,amssymb,aas_macros}
\usepackage{newtxtext,newtxmath}
\usepackage{nicefrac}
\usepackage{longtable}

\usepackage{multirow}
\renewcommand{\[}{\begin{equation}}
\renewcommand{\]}{\end{equation}}

{\newif\ifnotend
\notendtrue
\def\veclist{ABCDEFGHIJKLMNOPQRSTUVWXYZabcdefghijklmnopqrstuvwxyz.}
\def\top#1#2.{#1}
\def\tail#1#2.{#2.}
\loop\expandafter\xdef\csname v\expandafter\top\veclist\endcsname%
{{\noexpand\bf\expandafter\top\veclist}}
\edef\veclist{\expandafter\tail\veclist}
\if\veclist.\notendfalse\fi\ifnotend\repeat}

\def\Gyr{\,\mathrm{Gyr}}

\def\kpc{\,\mathrm{kpc}}

\def\kms{\,\mathrm{km\,s}^{-1}}
\def\masyr{\,\mathrm{mas\,yr}^{-1}}
\def\mas{\,\mathrm{mas}}
\def\muas{\,\mu\mathrm{as}}
\def\dex{\,\mathrm{dex}}
\def\Kelvin{\,\mathrm{K}}

\def\magn{\,\mathrm{mag}}

\def\msun{\,{\rm M}_\odot}

\def\mh{[\mathrm{M}/\mathrm{H}]}
\def\teff{T_\mathrm{eff}}

\def\pc{\,\mathrm{pc}}

\def\Gaia{{\it Gaia}}
\def\Hipparcos{{\it Hipparcos}}

\def\redTGAS{\alpha}
\def\zTGAS{\beta}
\def\redRAVE{\gamma}
\def\mine{$^1$}
\def\used{$^2$}
\def\DR5{$^3$}
\def\referee{ }
\def\breferee{ }
\def\creferee{ }

\def\teffIRFM{T_\mathrm{eff,IRFM}}

\title[Distances and ages for RAVE-TGAS stars]
{Improved distances and ages for stars common to TGAS and RAVE}

\author[P.~J.~McMillan et al.]{Paul J. McMillan$^{1}$\thanks{E-mail: paul@astro.lu.se},
Georges Kordopatis$^{2}$,
Andrea Kunder$^{3}$,
James Binney$^{4}$, \newauthor
Jennifer Wojno$^{3,5}$,
Toma\v{z} Zwitter$^{6}$, 
Matthias Steinmetz$^{3}$, 
Joss Bland-Hawthorn$^{7}$,    \newauthor
Brad K. Gibson$^{8}$, 
Gerard Gilmore$^{9}$,  
Eva~K.~Grebel$^{10}$,  
Amina Helmi$^{11}$,  \newauthor
Ulisse Munari$^{12}$,   
Julio F. Navarro$^{13}$, 
Quentin A. Parker$^{14,15}$, 
George Seabroke$^{16}$,  \newauthor
Fred Watson$^{17}$,
Rosemary~F.~G.~Wyse$^{5}$  \vspace{1mm}\\ 
	$^1${Lund Observatory, Lund University, Department of Astronomy and Theoretical Physics, Box 43, SE-22100, Lund, Sweden;} \\
	$^2${Universit\'e C\^ote d'Azur, Observatoire de la C\^ote d'Azur, CNRS, Laboratoire Lagrange, France} \\
	$^{3}${Leibniz-Institut f\"ur Astrophysik Potsdam, An der Sternwarte 16, D-14482 Potsdam, Germany}\\
$^{4}${Rudolf Peierls Centre for Theoretical Physics, Keble Road, Oxford, OX1 3NP, UK}\\
$^{5}${Department of Physics and Astronomy, Johns Hopkins University, 3400 N. Charles St, Baltimore, MD 21218, USA} \\
$^{6}${Faculty of Mathematics and Physics, University of Ljubljana, Jadranska 19, 1000 Ljubljana, Slovenia}\\
$^{7}${Sydney Institute for Astronomy, School of Physics A28, University of Sydney, NSW 2006, Australia}\\
$^{8}${E.A. Milne Centre for Astrophysics, University of Hull, Hull, HU6 7RX, U.K.}\\
$^{9}${Institute of Astronomy, University of Cambridge, Madingley Road, Cambridge, CB3 0HA,UK}\\
$^{10}${Astronomisches Rechen-Institut, Zentrum f\"ur Astronomie der Universit\"at Heidelberg, M\"onchhofstr.\ 12--14, 69120 Heidelberg, Germany} \\
$^{11}${Kapteyn Astronomical Institute, University of Groningen, PO Box 800, NL-9700 AV Groningen, the Netherlands}\\
$^{12}${INAF Astronomical Observatory of Padova, 36012 Asiago (VI), Italy} \\
$^{13}${Senior CIfAR Fellow, Department of Physics and Astronomy, University of Victoria, Victoria BC, Canada V8P 5C2} \\
$^{14}${The University of Hong Kong, Department of Physics, Hong Kong SAR, China}\\ 
$^{15}${The University of Hong Kong, Laboratory for Space Research, Hong Kong SAR, China}\\
$^{16}${Mullard Space Science Laboratory, University College London, Holmbury St. Mary, Dorking, Surrey, RH5 6NT, UK}\\
$^{17}${Australian Astronomical Observatory,PO Box 915, North Ryde, NSW 1670, Australia}\\}

\date{}

\pubyear{2017}

\begin{document}
\label{firstpage}
\pagerange{\pageref{firstpage}--\pageref{lastpage}}
\maketitle

\begin{abstract}
We combine parallaxes from the first \Gaia\ data release with the spectrophotometric distance estimation framework for stars in the fifth RAVE survey data release. The combined distance estimates are more accurate than either determination in isolation -- uncertainties are on average two times smaller than for RAVE-only distances (three times smaller for dwarfs), and 1.4 times smaller than TGAS parallax uncertainties (two times smaller for giants). We are also able to compare the estimates from spectrophotometry to those from \Gaia, and use this to assess the reliability of both catalogues and improve our distance estimates. We find that the distances to the lowest $\log g$ stars are, on average, overestimated and caution that they may not be reliable. We also find that it is likely that the \Gaia\ random uncertainties are smaller than the reported values. As a byproduct we derive ages for the RAVE stars, many with relative uncertainties less than 20 percent. These results for $219\,566$ RAVE sources have been made publicly available, and we encourage their use for studies that combine the radial velocities provided by RAVE with the proper motions provided by \Gaia. A sample that we believe to be reliable 
can be found by taking only the stars with the flag notification `flag\_any=0'.
\end{abstract}

\section{Introduction}\label{sec:intro}
ESA's \Gaia\ mission \citep{GaiaMission} is an enormous project that is revolutionising Milky Way astronomy. \Gaia\ will provide a wide range of data about the stars of the Milky Way, including photometry and spectroscopy. However it is the astrometry -- and in particular the parallaxes -- from \Gaia\ that are the cause of the most excitement. It is very difficult to determine the distances to stars, and not knowing the distance to a star means that one knows neither where it is nor how fast it is moving, even if the proper motion of the star is known.

The RAVE survey \citep[Radial Velocity Experiment:][]{RAVE1} is a spectroscopic survey that took spectra for $\sim$$500\,000$ stars. From these one could determine for each star its line-of-sight velocity and the structural parameters, such as its effective temperature ($\teff$), surface gravity ($\log g$) and metallicity ($\mh$). These can be used to derive the distances to stars, and since RAVE's fourth data release \citep{RAVEDR4} these have been provided by the Bayesian method that was introduced by \cite{BuJJB10}, and extended by \cite{JJBea14}. {\referee Bayesian methods had previously been used for distance estimation in astrophysics for small numbers of stars of specific classes \citep{Th03,Baea03}, and the \citeauthor{BuJJB10} method is similar to an approach that } had previously been used to determine the ages of stars \citep{PoEy04,JoLi05}. Closely related approaches have since been used by numerous studies \citep[e.g.,][]{Seea13,ScBe14,Waea16,Saea16,Scea17,MiHe17,Quea17}. The method produces a probability density function (pdf) for the distance, and these pdfs were tested by, amongst other things, comparison of some of the corresponding parallax estimates to the parallaxes found by \Gaia's predecessor \Hipparcos\ \citep{HipparcosCatalogue,vL07}. RAVE's most recent data release was the fifth in the series (henceforth DR5), and included distance estimates found using this method \citep{RAVEDR5}. The RAVE sample appears to be kinematically and chemically unbiased \citep{Woea17}.

\Gaia's first data release \citep[\Gaia\ DR1, ][]{GaiaDR1, GaiaDR1:TGAS} includes parallaxes and proper motions for $\sim$$2\,000\,000$ sources. These were available earlier than full astrometry for the other $\sim$$1$ billion sources observed by \Gaia, because the sources were observed more than twenty years earlier by the \Hipparcos\ mission, and their positions at that epoch (and proper motions) appear in either the \Hipparcos\ catalogue or the, less precise, Tycho-2 catalogue \citep{Tycho2}, which used data from the \Hipparcos\ satellite's star mapper. This means that the proper motions of the stars can be derived using this very long time baseline, which breaks degeneracies between proper motion and parallax that made the determination of these parameters for the other sources impossible. The resulting catalogue is known as the Tycho-\Gaia\ Astrometric solution \citep*[TGAS: ][]{TGAS15}.

Since RAVE and TGAS use fundamentally different methods for deriving the distances to stars, it is inevitable that these have different precisions for different types of stars. The \cite{BuJJB10} method relies, fundamentally, on comparing the observed magnitude to the expected luminosity. The uncertainty in distance modulus, which is roughly equivalent to a relative distance uncertainty, is therefore approximately independent of the distance to the star. The parallax uncertainty from TGAS, on the other hand, is independent of the parallax value, so the relative precision declines with distance -- large distances correspond to small parallaxes, and therefore large relative uncertainties.

\begin{figure}
  \centerline{
    \resizebox{\hsize}{!}{\includegraphics{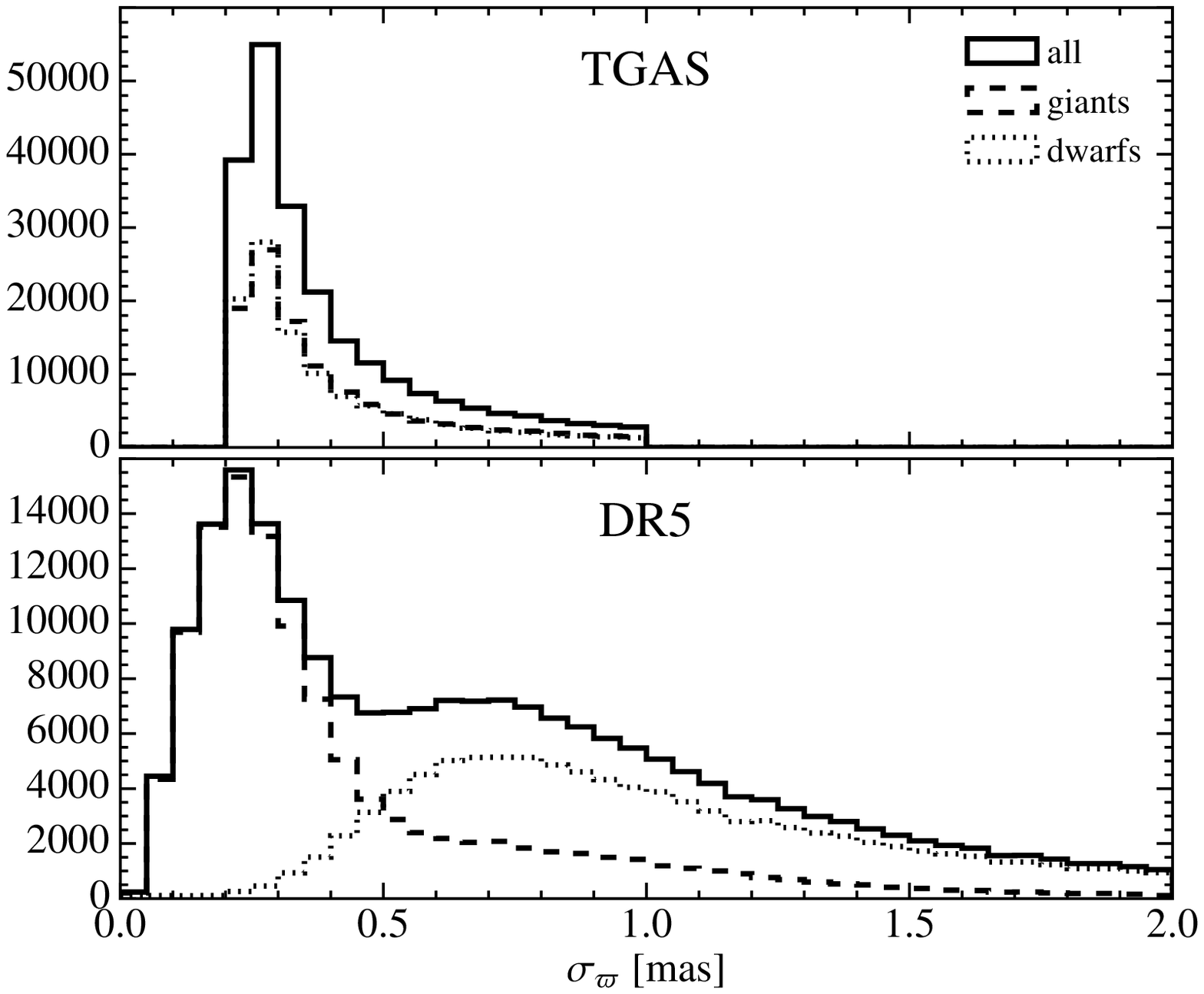}}}
  \caption{
  	Histograms of the quoted random parallax uncertainties ($\sigma_{\varpi}$) from TGAS and those from RAVE DR5 for stars common to the two catalogues. We show histograms of the uncertainties for all stars (solid), and separately for giants ($\log g_{\rm DR5}<3.5$) and dwarfs ($\log g_{\rm DR5}\geq 3.5$). The $y$-axis gives the number of stars per bin, and there are 40 bins in total in both cases. The cut-off at $1\mas$ for the TGAS parallaxes is due to a filter applied by the \Gaia\ consortium to their DR1. For RAVE sources we make the standard cuts to the catalogue described in \protect\cite{RAVEDR5}. TGAS parallaxes are more precise than RAVE's for dwarfs, but not necessarily for giants.
  \label{fig:DR5uncerts}
}
\end{figure}

In Figure~\ref{fig:DR5uncerts} we show the quoted parallax uncertainty from both TGAS and DR5 for the sources common to both catalogues. In the case of TGAS we use the quoted statistical uncertainties (see Section~\ref{sec:TGAS} for further discussion). We also divide this into the uncertainty for giant stars (DR5 $\log g<3.5$) and dwarfs (DR5 $\log g\geq 3.5$). We see that for TGAS this distinction is immaterial, while it makes an enormous difference for DR5. The DR5 parallax estimates tend to be less precise than the TGAS ones for dwarfs (which tend to be nearby because the survey is magnitude limited), but as precise, or more, for the more luminous giants, especially the more distant ones.

It is worth noting that TGAS provides only parallax measurements, not \emph{distance} estimates and, {\referee as discussed by numerous authors at various points over the last century}, the relationship between one and the other is non-trivial when one takes the uncertainties into account \citep[e.g.][]{St27,LuKe73,LuAr97,CBJ15}. \cite{AsCBJ16} looked at how the distances derived from TGAS parallaxes depend on the prior probability distribution used for the density of stars, but did not use any information about a star other than its parallax.

For this reason, and because TGAS parallaxes have large relative errors for distant stars, when studying the dynamics of the Milky Way using stars common to RAVE and TGAS, it has been seen as advantageous to use distances from DR5 rather than those from TGAS parallaxes \citep[e.g.,][]{Heea17,Huea16}. It is therefore important to improve these distance estimates and to check whether there are any systematic errors associated with the DR5 distance estimates.

\cite{RAVEDR5} discusses the new efforts in RAVE DR5 to reconsider the parameters of the observed stars. They provided new $\teff$ values {\referee derived from the Infrared Flux Method \citep[IRFM:][]{Blea79} using an updated version of the implementation described by \cite{Caea10:IRFM}. } Also provided in a separate data-table were new values of $\log g$ following a re-calibration for red giants from the \cite{Vaea17} study of 72 stars with $\log g$ values derived from asteroseismology of stars by the K2 mission \citep{K2}. These were not used to derive distances in the main DR5 catalogue, and we now explore how using these new data products can improve our distance estimates.

In this study, we compare parallax estimates from TGAS and RAVE to learn about the flaws in both catalogues. We then include the TGAS parallaxes in the RAVE distance estimation, to derive more precise distance estimates than are possible with either set of data in isolation.

It is also possible to derive ages for stars from the same efforts, indeed the use of Bayesian methods to derive distances was preceded by studies using them to determine ages \citep{PoEy04,JoLi05}. RAVE DR4 included the age estimates derived alongside the distances, but these were recognised as only being indicative \citep{RAVEDR4}. In this study we show the substantial improvement that is possible using TGAS parallaxes and a more relaxed prior.


In Section~\ref{sec:Bayes} we describe the method used to derive distances. In Section 
\ref{sec:DR5} we compare results from DR5 to those from TGAS, which motivates us to look at improving our parallax estimates using other RAVE data products in Section~\ref{sec:improved}.  In Section~\ref{sec:altprior} we explore the effect of varying our prior. In Section~\ref{sec:TGAS} we look at what we can learn about TGAS by comparison with these new parallax estimates. Finally, Sections \ref{sec:Combined}, \ref{sec:ages} and  \ref{sec:reverse} demonstrate the improvements made possible by using the TGAS parallaxes as input to the Bayesian scheme.
 
\section{Bayesian estimation}\label{sec:Bayes}

Since RAVE DR4, distances to the stars in the RAVE survey have been determined using the Bayesian method developed by \cite{BuJJB10}. This takes as its input the stellar parameters $\teff$, $\log g$ and $\mh$ determined from the RAVE spectra, and $J$, $H$ and $K_{\rm s}$ magnitudes from 2MASS \citep{2MASS}. This method was extended by \cite{JJBea14} to include dust extinction in the modelling, and introduce an improvement in the description of the distance to the stars by providing multi-Gaussian fits to the full probability density function (pdf) in distance modulus.\footnote{While the distance estimates always use 2MASS (and, in this study, AllWISE) photometry, we will refer to them as `RAVE-only' at various points in this paper, to distinguish them from those found using TGAS parallaxes as input too.} 

In this paper we extend this method, principally by including the parallaxes found by TGAS as input, but also by adding AllWISE W1 and W2 mid-infrared photometry \citep{AllWISE}. We will explore improvements made possible by using IRFM $\teff$ values given in RAVE DR5, rather than $\teff$ derived from the spectra. We expect that the IRFM values can be more precise than those from the RAVE spectra, which only span a narrow range in wavelength (8410-8795\AA)



Because the original intention of this pipeline was to estimate distances, we often refer to it as the `distance pipeline'. In practice we are now often as interested in its other outputs as we are in the distance estimates. The pipeline applies the simple Bayesian statement
\[ \label{eq:bayes} 
P(\hbox{model}|\hbox{data})=\frac{P(\hbox{data}|\hbox{model})P(\hbox{model})}{P(\hbox{data})},
\]
where in our case ``data'' refers to the inputs described above (and shown in Table~\ref{tab:data}) for a single star, and ``model'' comprises a star of specified initial mass
${\cal M}$, age $\tau$, metallicity $\mh$, and location relative to the Sun (where Galactic coordinates $l$ and $b$ are treated as known and distance $s$ is unknown), observed through a specified line-of-sight extinction, which we parametrise by extinction in the $V$-band, $A_V$. The likelihood $P(\hbox{data}|\hbox{model})$ is determined assuming uncorrelated Gaussian uncertainties on all inputs, and using isochrones to find the values of the stellar parameters and absolute magnitudes of the model star. The isochrones that we use are from the PARSEC v1.1 set \citep{Brea12}, and the metallicities of the isochrones used are given in Table~\ref{tab:isochrones}. 
$P(\hbox{model})$ is our prior which we discuss below, and $P(\hbox{data})$ is a normalisation constant which we can ignore. The assumption of uncorrelated Gaussian errors on the stellar parameters is one which is imperfect \citep[see e.g.][]{ScBe14,Scea17}, but it is the best approximation that we have available for RAVE.

Putting this in a more mathematical form and defining the notation for a single Gaussian distribution
\begin{equation}
G(x,\mu,\sigma) = \frac{1}{\sqrt{2\pi\sigma^2}} \exp\left(\frac{(x-\mu)^2}{2\sigma^2}\right),
\end{equation}
 we have
\begin{multline}\label{eq:maths}
P({\cal M},\tau,\mh,s,A_V|\,\hbox{data}) \propto \; P({\cal M},\tau,\mh,s,A_V | l,b) \\
	\times \prod_i G(O^T_i ({\cal M},\tau,\mh,s,A_V),O_i, \sigma_i)
\end{multline}
where the prior $ P({\cal M},\tau,\mh,s,A_V | l,b)$ is described in Section~\ref{sec:prior}, and the inputs $O_i$, $\sigma_i$ are those given in Table~\ref{tab:data} (the cases where any of these inputs are unavailable or not used can be treated as the case where $\sigma_i\rightarrow\infty$). The theoretical values of these quantities -- $O^T_i ({\cal M},\tau,\mh,s,A_V)$ -- are found using the isochrones and the relations between extinctions in different bands given in Section~\ref{sec:prior}.

Once we have calculated the probability density functions $P(\hbox{model}|\hbox{data})$ for the stars we can characterise them however we wish. In practice, we characterise them by the expectation values and standard deviation (i.e., estimates and their uncertainties) for all parameters, found by marginalising over all other parameters.

For distance we find several characterisations of the pdf: expectation values and standard deviation for the distance itself ($s$), for distance modulus ($\mu$) and for parallax $\varpi$.  The characterisation in terms of parallax is vital for comparison with TGAS parallaxes.

In addition we provide multi-Gaussian fits to the pdfs in distance modulus because a number of the pdfs are multi-modal, typically because it is unclear from the data whether a star is a main sequence star or a (sub-)giant. Therefore a single expectation value and standard deviation is a poor description of the pdf.
The multi-Gaussian fits to the pdfs in $\mu$ provide a compact representation of the pdf, and 
{\referee following \cite{JJBea14} we write them as} 
\[\label{eq:defsfk}
P(\mu) = \sum_{k=1}^{N_{\rm Gau}} {f_k} G(\mu,\widehat{\mu_k},\sigma_k),
\]
where the number of components $N_{\rm Gau}$, the means $\widehat{\mu_k}$, weights $f_k$, and dispersions $\sigma_k$ are determined by the pipeline. 


To determine whether a distance pdf is well represented by a given multi-Gaussian representation in $\mu$ we take bins in distance modulus of width $w_i = 0.2\magn$, which contain a fraction $p_i$ of the total probability taken from the computed pdf and a fraction $P_i$ from the Gaussian representation, and compute the goodness-of-fit statistic 
\[\label{eq:defsF}
F = \sum_i \left(\frac{p_i}{w_i}-\frac{P_i}{w_i}\right)^2\tilde{\sigma} w_i
\]
where the weighted dispersion
\[
\tilde{\sigma}^2 \equiv \sum_{k=1,N_{\rm Gau}} f_k \sigma_k^2
\]
is a measure of the overall width of the pdf. Our strategy is to represent the pdf with as few Gaussian components as possible, but if the value of $F$ is greater than a threshold value ($F_t=0.04$), or the standard deviation associated with the model differs by more than 20 percent from that of the complete pdf, then we conclude that the representation is not adequate, and add another Gaussian component to the representation (to a maximum of 3 components, which we have found is almost always enough). {\referee We fit the multi-Gaussian representation to the histogram using the Levenberg-Marquandt algorithm \citep[e.g.][]{NumRec},  which we apply multiple times with different starting points estimated from the modes of the distribution. In this way we can take the best result and therefore avoid getting caught in local minima. The relatively broad bins mean that we only use more than one Gaussian component if the pdf is significantly multi-modal, though this comes at the cost of reducing the accuracy of the fit when a peak is narrow.}

These multi-Gaussian fits were particularly important in previous RAVE data releases. In DR5 we found that a single Gaussian component proved adequate for only 45 percent of the stars, while around 51 percent are fit with two Gaussians, and only 4 percent require a third component. In Section~\ref{sec:Combined} we show that the addition of TGAS parallaxes substantially reduces the number of stars for which more than one Gaussian is required.

The value of $F$ is provided in the database as FitQuality\_Gauss, and we also include a flag (denoted Fit\_Flag\_Gauss) which is non-zero if the standard deviation of the final fitted model differs by more than 20
percent from that of the computed pdf. Typically the problems flagged are rather minor \citep[as shown in fig.~3 of][]{JJBea14}.

\begin{table*}
  \begin{center}
    \caption{Data used to derive the distances to our stars, and their source.\label{tab:data}}
    \begin{tabular}{ccc}
      \hline
      Data & Symbol & Notes \\
      \hline
      Effective temperature & $T_{\rm eff}$ & RAVE DR5 -- either from spectrum (DR5) or IRFM  \\ 
      Surface gravity & $\log{g}$ & RAVE DR5 \\ 
      Metallicity & $\mh$ & RAVE DR5   \\ 
      $J$-band magnitude & $J$ & 2MASS  \\
      $H$-band magnitude & $H$ & 2MASS  \\
      $K_s$-band magnitude & $K_s$ & 2MASS  \\
	$W_1$-band  magnitude & $W_1$ & AllWISE -- not used for DR5 distances  \\
	$W_2$-band  magnitude & $W_2$ & AllWISE  -- not used for DR5 distances\\
	Parallax & $\varpi_{\rm TGAS}$ & \Gaia\ DR1 -- not used for DR5 distances or in comparisons \\

         \hline
    \end{tabular}
  \end{center}
\end{table*}

The uncertainties of the RAVE stellar parameters are assumed to be the quadratic sum of the quoted internal uncertainties and the external uncertainties  (Table 4 of DR5). The external uncertainties are those calculated from stars with SNR$>40$, except in the case of the IRFM temperatures for which a single uncertainty serves for stars of every SNR since the IRFM temperatures are not extracted from the spectra. We discard all observations with a signal-to-noise ratio less than 10, or where the RAVE spectral pipeline returns a quality flag (AlgoConv) of `1', because the quoted parameters for these observations are regarded as unreliable.

For the 2MASS and AllWISE photometry we use the quoted uncertainties. We discard the AllWISE magnitudes if they are brighter than the expected saturation limit in each band, which we take to be $W_{1,{\rm sat}}=8.1\magn$,  $W_{2,{\rm sat}}=6.7\magn$ \citep[following][]{WISETech}.

When using the TGAS parallaxes, we consider only the quoted statistical uncertainties. We will show that these appear to be, if anything, slight overestimates of the uncertainty. 

{\referee The posterior pdf (eq.~\ref{eq:maths}) is calculated on an grid of isochrones at metallicities as given in Table~\ref{tab:isochrones} and ages spaced by $\delta\log_{10}(\tau/{\mathrm  yr})=0.04$ for $\tau<1 \Gyr$ and $\delta\log_{10}(\tau/{\mathrm  yr})=0.01$ for  $\tau>1 \Gyr$. For each of these isochrones we take grid points in initial mass ${\cal M}$ such that there is no band in which any magnitude changes by more than 0.005 mag. We then evaluate the posterior on an informed grid in $\log A_V$ and distance, which is centred on the expected $\log A_V$ from the prior at an estimated distance (given the observed and model $J$-band magnitude) and then the estimated distance (given each $\log A_V$ value evaluated).  }

Where stars have been observed more than once by RAVE, we provide distance estimates for the quoted values from each spectrum. We provide a flag `flag\_dup' which is 0 if the spectrum is the best (or only) one for a given star, as measured by the signal-to-noise ratio, and 1 otherwise. Where one wishes to avoid double counting stars one should only use rows where this flag is 0.\footnote{We have based this on the RAVEID number for each source. It is worth noting that the cross-matching of stars is not perfect, and so despite our best attempts to clean duplicate entries, there may be a few percent of stars that are in fact listed twice.}

\subsection{Standard prior} \label{sec:prior}
For our standard results, we use the prior that was used for DR4 and DR5. We do this for consistency, and because we find that this provides good results. The prior reflects some elements of our existing understanding of the Galaxy, at the cost of possibly biasing us against some results that run counter to our expectations (for example, metal rich or young stars far from the plane). In Section~\ref{sec:altprior} we consider alternative priors. Although the prior is described in \cite{JJBea14}, we describe it here for completeness, and to enable comparisons with alternative priors considered.

The prior considers all properties in our model, and can be written as
\[ \label{eq:priorAll}
\begin{split}
P(\hbox{model})  & = P({\cal M},\tau,\mh,s,A_V | l,b) \\
	& = P({\cal M}) \times P(A_V |\, s, l ,b) \times P(s,\mh,\tau |\, l,b)
\end{split}
\]
 with the prior on initial mass being a \cite{Kr01} initial mass function (IMF), as modified by \cite{AuJJB09}
 \begin{equation} \label{eq:priorMass}
  P({\cal M}) \propto 
  \begin{cases} 0 & {\rm if}\; {\cal M}<0.1\,M_\odot \\
  {\;\cal M}^{-1.3} & {\rm if}\; 0.1\,M_\odot \le {\cal M}<0.5\,M_\odot, \\
    \;0.536 \,  {\cal M}^{-2.2} & {\rm if}\; 0.5\,M_\odot \le {\cal M}<1\, M_\odot,\\
    \;0.536 \,  {\cal M}^{-2.519} & {\rm otherwise}. \\
    \end{cases}
\end{equation}

We describe extinction in terms of the value $A_V$ for the Johnson $V$ band, and, since extinction is necessarily non-negative, we take our prior to be Gaussian in $\ln A_V$ around an expected value which varies with the model star's position in the Galaxy, $\ln A_V^{\rm pr}(s,l,b)$.

To find the expected value $A_V^{\rm pr}(s,l,b)$ we start from an expected value at infinity, $A_V^{\rm pr}(\infty,l,b)$, which we take from the \cite*{ScFiDa98} values of $E(B-V)$, with a correction for high extinction sightlines following \cite{ArGo99} and \cite{Shea11}, leaving us with 
\[ \label{eq:priorExtinction}
\begin{split}
 A_V^{\rm pr}(\infty,l,b) =  & \;3.1 \times E(B-V)_{\rm SFD}\; \times \\ 
 &	\left\{0.6+0.2\left[1-\tanh\left(\frac{E(B-V)_{\rm SFD}-0.15}{0.3}\right)\right]\right\},
 \end{split}
\]
We then determine the expected extinction at a given distance $s$ in the direction $l,b$, which is some fraction of the total extinction along that line of sight. We take this to be the fraction of the total extinguishing material along that line of sight that lies closer than $s$ in a 3D dust model of the Milky Way taken from \cite{Shea11}. For details of the model see \cite{JJBea14}.

As in \cite{JJBea14} we take the uncertainty in $\ln A_V$ to be $\sqrt{2}$. We can then write the prior on $A_V$ to be
\begin{equation}
P(A_V | s,l,b) = G(\ln A_V,\; \ln(A_V^{\rm pr}(s,l,b)),\;\sqrt{2}).
\end{equation}

Extinction varies between different photometric bands. For a given extinction value $A_V$, from \cite{RiLe85} we take the extinctions to be
 \begin{equation}
 \begin{aligned}
A_J&=0.282A_V\\
A_H&=0.175A_V\\
A_{K_s} &= 0.112A_V,
\end{aligned}
\end{equation}
and, following from this, and using the results of \cite*{YuLiXi13}, we have extinction in the WISE photometric bands of 
 \begin{equation}
 \begin{aligned}
A_{W1}&=0.0695A_V \\
A_{W2}&=0.0549A_V.
\end{aligned}
\end{equation}

The other term in the prior is related to the probability of there being a star of a given $\tau$, $\mh$ and position. It also contains a factor of $s^2$, to reflect the conical shape of the surveyed volume.\footnote{This factor was stated by \cite{BuJJB10}, but not directly noted by either \cite{Buea11} or \cite{JJBea14}, who simply stated the density profile associated with the prior on position. This oversight meant that \cite{Saea16} noted the absence of this factor as a difference between the \cite{JJBea14} values and their own, closely related, results. The factor of $s^2$ was, however, used in all of these studies.} 

The prior on distance, $\mh$ and age can then be written as:
 \begin{equation}\label{eq:priorofx}
  P(s,\mh,\tau |\, l,b) \propto s^2 \sum_{i=1}^3 N_i P_i(\mh) \, P_i(\tau) \, P_i(\mathbf{r}),
\end{equation}
 where $i=1,2,3$ correspond to a thin disc, thick disc and stellar halo,
respectively and where $\mathbf{r}$ is the Galactocentric position of the star. We then have

\paragraph*{Thin disc ($i=1$):}
\begin{eqnarray} \label{eq:thindisc}
  P_1(\mh) &=& G(\mh, 0, 0.2), \nonumber \\
  P_1(\tau)  &\propto& \exp(0.119 \,\tau/\mbox{Gyr}) \quad \mbox{for $\tau \le 10$\,Gyr,}  \\
  P_1(\mathbf{r}) &\propto& \exp\left(-\frac{R}{R_d^{\rm{thin}}} - \frac{|z|}{z_d^{\rm{thin}}}  \right);  \nonumber
\end{eqnarray}

\paragraph*{Thick disc ($i=2$):}
\begin{eqnarray}\label{eq:thickdisc}
  P_2(\mh) &=& G(\mh, -0.6, 0.5), \nonumber \\
  P_2(\tau)  &\propto& \mbox{uniform in range $8 \le \tau \le 12$\,Gyr,} \\
  P_2(\mathbf{r}) &\propto& \exp\left(-\frac{R}{R_d^{\rm{thick}}} - \frac{|z|}{z_d^{\rm{thick}}}  \right); \nonumber
\end{eqnarray}

\paragraph*{Halo ($i=3$):}
\begin{eqnarray} \label{eq:halo}
  P_3(\mh) &=& G(\mh, -1.6, 0.5), \nonumber \\
  P_3(\tau)  &\propto& \mbox{uniform in range $10 \le \tau \le 13.7$\,Gyr,} \\ 
  P_3(\mathbf{r}) &\propto& r^{-3.39}; \nonumber
\end{eqnarray}
 where $R$ signifies Galactocentric cylindrical radius, $z$ cylindrical
height and $r$ spherical radius. We take $R_d^{\rm thin}=2\,600\pc$, $z_d^{\rm thin}=300\pc$, $R_d^{\rm thick}=3\,600\pc$, $z_d^{\rm thin}=900\pc$. These values are taken from the analysis of SDSS data
in \cite{Juea08}. The metallicity and age distributions for the thin disc
come from \mbox{\cite{Ha01}} and \cite{AuJJB09}, while the radial density of the
halo comes from the `inner halo' detected in \cite{Caea10}. The metallicity
and age distributions of the thick disc and halo are influenced by
\cite{Re10} and \cite{Caea10}. {\referee The halo component tends towards infinite density as $r\rightarrow 0$, so we apply an arbitrary cut-off for $r<1\kpc$ -- a region which the RAVE sample does not, in any case, probe.}

The normalizations $N_i$ were then adjusted so that at the Solar position, taken as
$R_0=$~8.33\,kpc (\citealt{Giea09}), $z_0=$~15\,pc
\citep*{JJBGeSp97}, we have number
density ratios $n_2 /n_1 = 0.15$ \citep{Juea08}, $n_3 /n_1 = 0.005$
\citep{Caea10}.

\begin{table}
  \begin{center}
    \caption{Metallicities of isochrones used, taking $Z_\odot = 0.0152$ and applying scaled solar composition, with  $Y=0.2485+1.78Z$. Note that the minimum metallicity is $\mh =-2.2$, significantly lower than for the \protect\cite{JJBea14} distance estimates where the minimum metallicity used was $-0.9$, which caused a distance underestimation for the more metal poor stars \protect\citep{Anea15}. \label{tab:isochrones}}
    \begin{tabular}{rrr}
      \hline
      $Z$ & $Y$ & $\mh$ \\
      \hline
      0.00010 & 0.249 & -2.207 \\ 
0.00020 & 0.249 & -1.906 \\ 
0.00040 & 0.249 & -1.604 \\ 
0.00071 & 0.250 & -1.355 \\ 
0.00112 & 0.250 & -1.156 \\ 
0.00200 & 0.252 & -0.903 \\ 
0.00320 & 0.254 & -0.697 \\ 
0.00400 & 0.256 & -0.598 \\ 
0.00562 & 0.259 & -0.448 \\ 
0.00800 & 0.263 & -0.291 \\ 
0.01000 & 0.266 & -0.191 \\ 
0.01120 & 0.268 & -0.139 \\ 
0.01300 & 0.272 & -0.072 \\ 
0.01600 & 0.277 & 0.024 \\ 
0.02000 & 0.284 & 0.127 \\ 
0.02500 & 0.293 & 0.233 \\ 
0.03550 & 0.312 & 0.404 \\ 
0.04000 & 0.320 & 0.465 \\ 
0.04470 & 0.328 & 0.522 \\ 
0.05000 & 0.338 & 0.581 \\ 
0.06000 & 0.355 & 0.680 \\ 
      \hline
    \end{tabular} 
  \end{center}
\end{table}

\section{Comparison of DR5 and TGAS parallaxes} \label{sec:DR5}

For RAVE DR5 the distance estimation used the 2MASS $J$, $H$, $K_s$ values, and the $\teff$, $\log g$ and $\mh$ values calculated from RAVE spectra. The parallaxes computed were compared with the parallaxes obtained by the {\it Hipparcos} mission \citep{HipparcosCatalogue}, specifically those found by the new reduction of \cite{vL07} for the $\sim$$5000$ stars common to both catalogues. The parallaxes were compared by looking at the statistic
\[ \label{eq:Delta}
\Delta = \frac{\left\langle \varpi_{\rm sp} \right\rangle - \varpi_{\rm ref}} { \sqrt{\sigma_{\varpi,{\rm sp}}^2+\sigma_{\varpi,{\rm ref}}^2} },
\] 
where $\varpi_{\rm sp}$ and $\sigma_{\varpi,{\rm sp}}$ are the spectrophotometric parallax estimates and their uncertainties. In \cite{RAVEDR5} the reference parallax $\varpi_{\rm ref}$ and its uncertainty $\sigma_{\varpi,{\rm ref}}$ were from {\it Hipparcos}, but henceforth in this paper they will be from TGAS. A negative value of $\Delta$, therefore, corresponds to an overestimate of distance from RAVE (compared to the reference parallaxes), and a positive value corresponds to an underestimate of distance. We would hope that the mean value of $\Delta$ is zero and the standard deviation is unity (consistent with the uncertainties being correctly estimated).

\begin{figure*}
  \centerline{
    \resizebox{0.33\hsize}{!}{\includegraphics{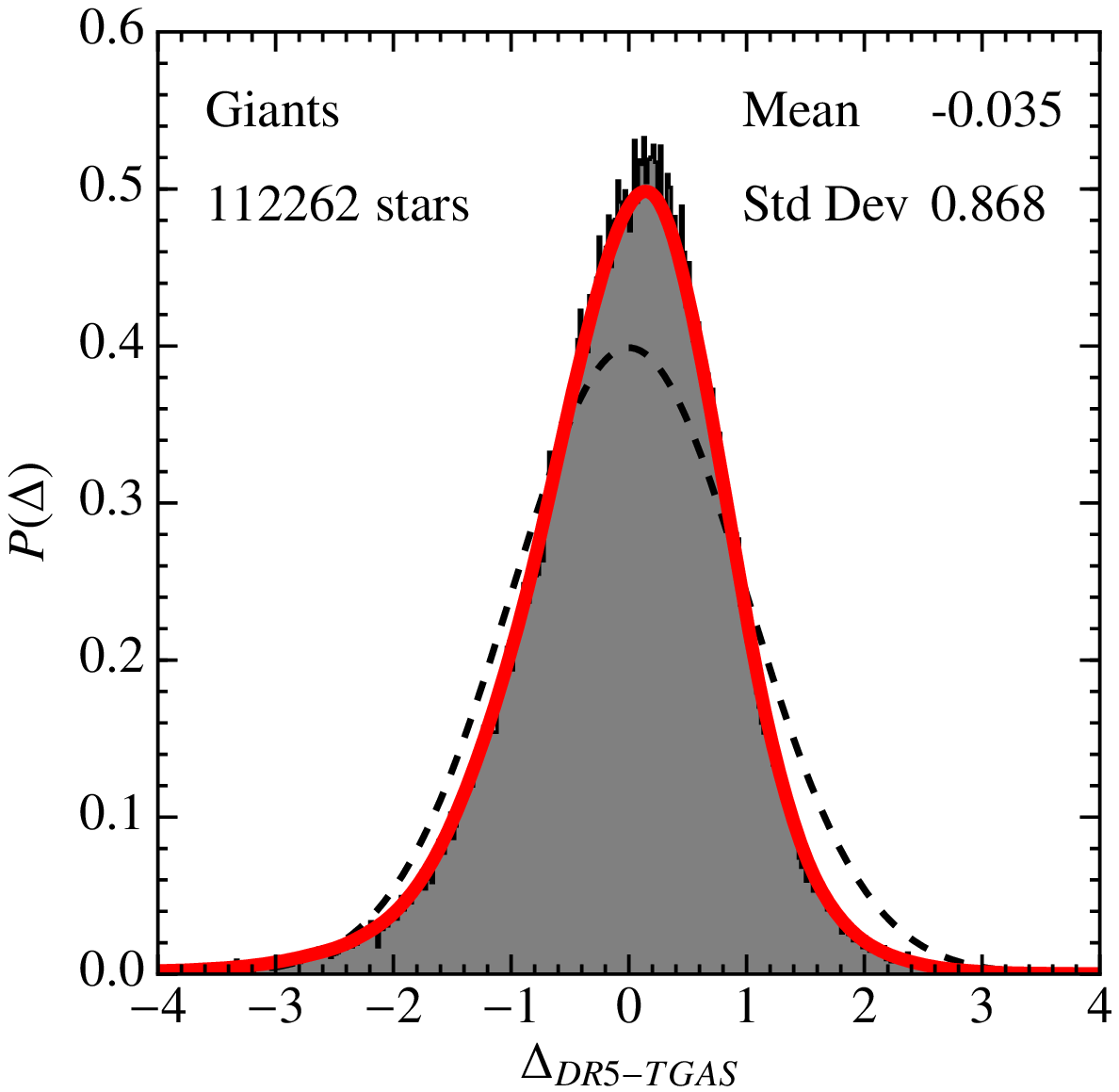}}
    \resizebox{0.33\hsize}{!}{\includegraphics{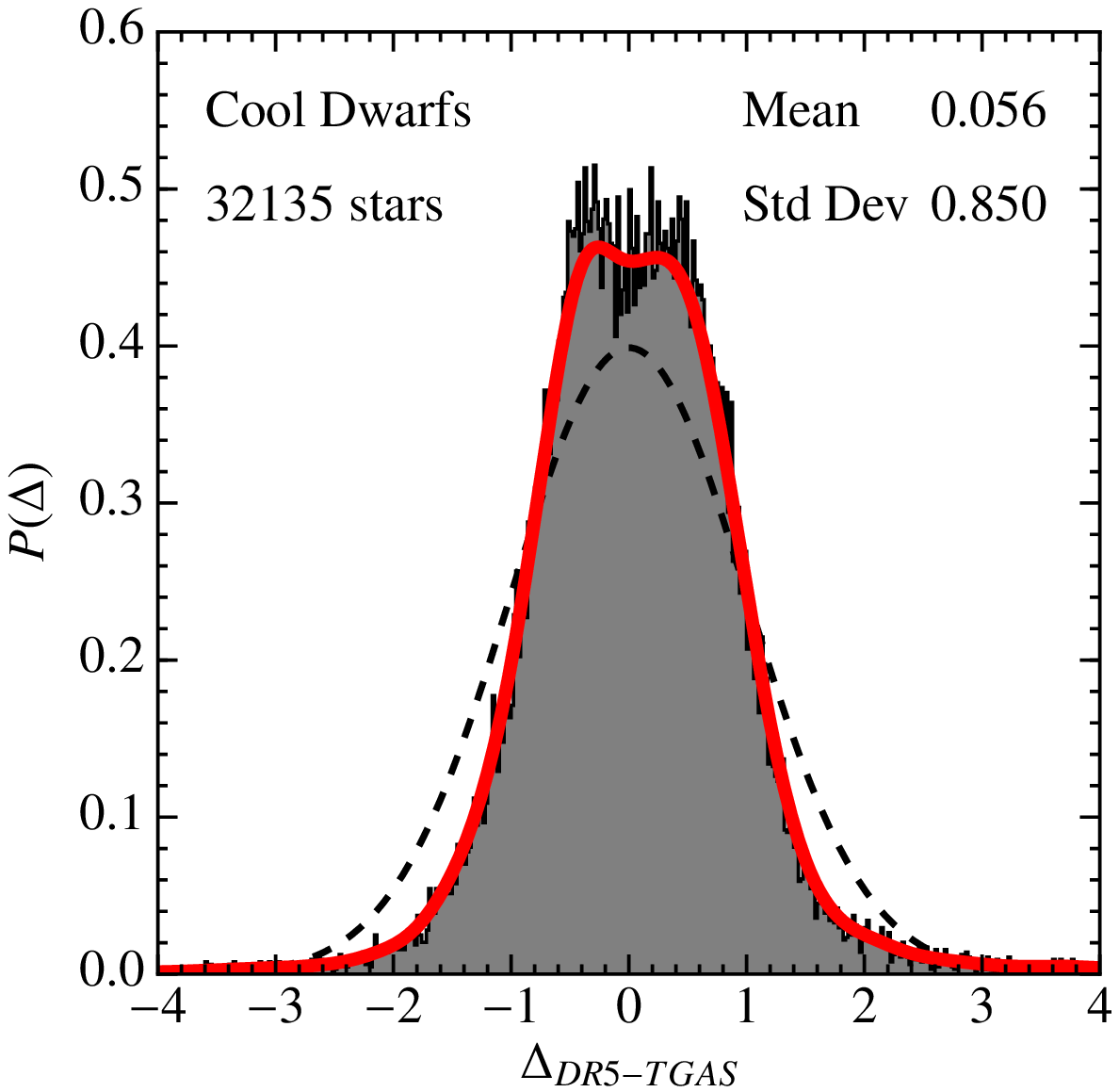}}
    \resizebox{0.33\hsize}{!}{\includegraphics{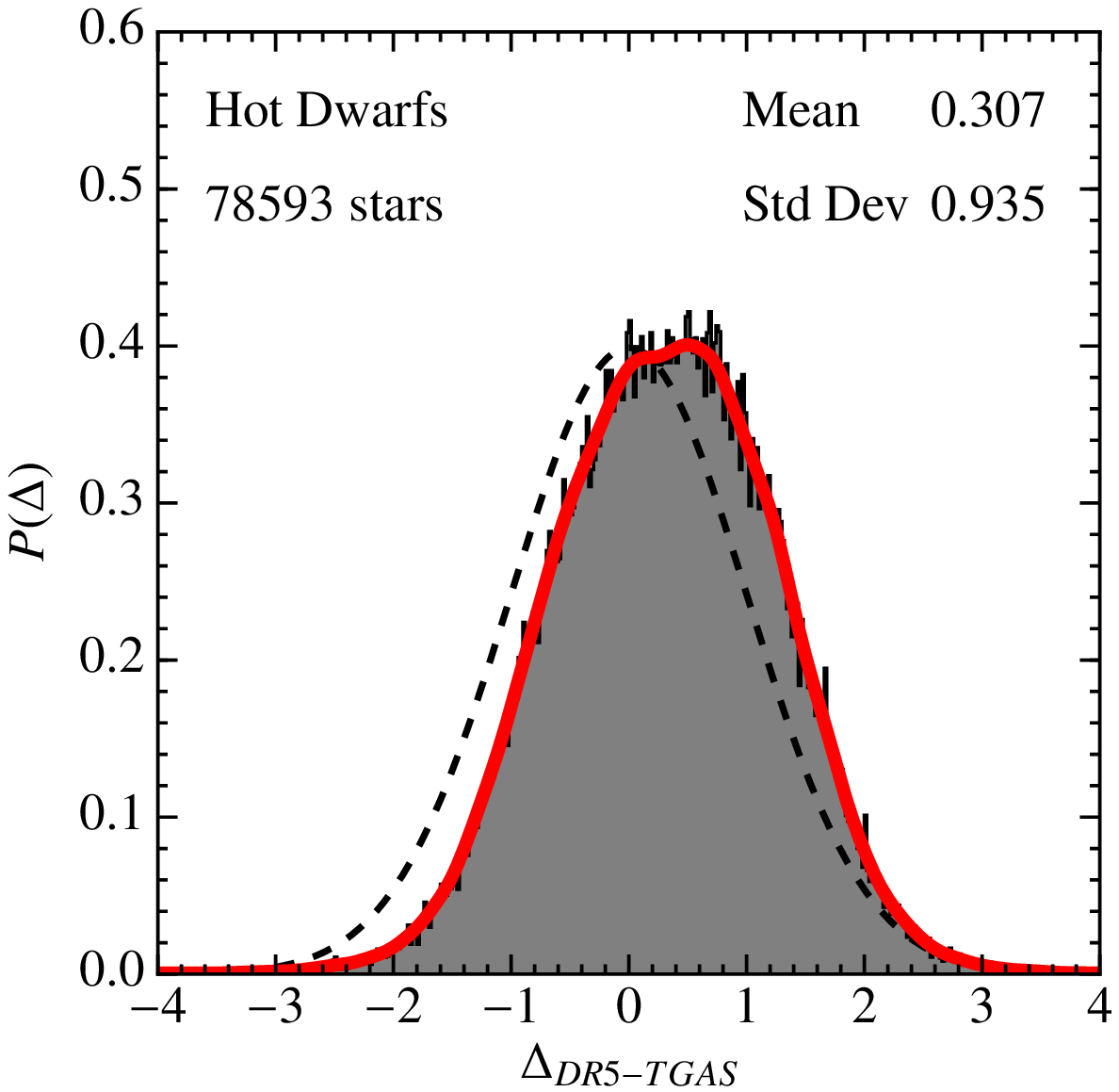}}}
  \caption{
  	Comparison of parallax estimates from RAVE DR5 and those from TGAS. We divide the stars into giants ($\log g<3.5$), cool dwarfs ($\log g\ge3.5$ and $\teff\le5500$$\,\mathrm{K}$) and hot dwarfs ($\log g\ge3.5$ and $\teff>5500$$\,\mathrm{K}$) and provide pdfs of $\Delta$ (i.e. difference between spectrophotometric parallax and TGAS parallax, normalised by the combined uncertainty, see Eq.~\ref{eq:Delta}) in each case. The \emph{red} lines show the kernel density estimate of this pdf in each case, with the finely-binned grey histogram shown to give a indication of the variation around this smooth estimate. The \emph{black dashed line} is a Gaussian with mean 0 and standard deviation of unity. The means and standard deviations shown in the top right are for stars with $-4<\Delta_{\rm DR5}<4$, to avoid high weight being given to outliers. Positive values of $\Delta$ correspond to parallax overestimates (i.e. distance or luminosity underestimates).
  \label{fig:DR5}
}
\end{figure*}

\begin{figure*}
  \centerline{
    \resizebox{0.5\hsize}{!}{\includegraphics{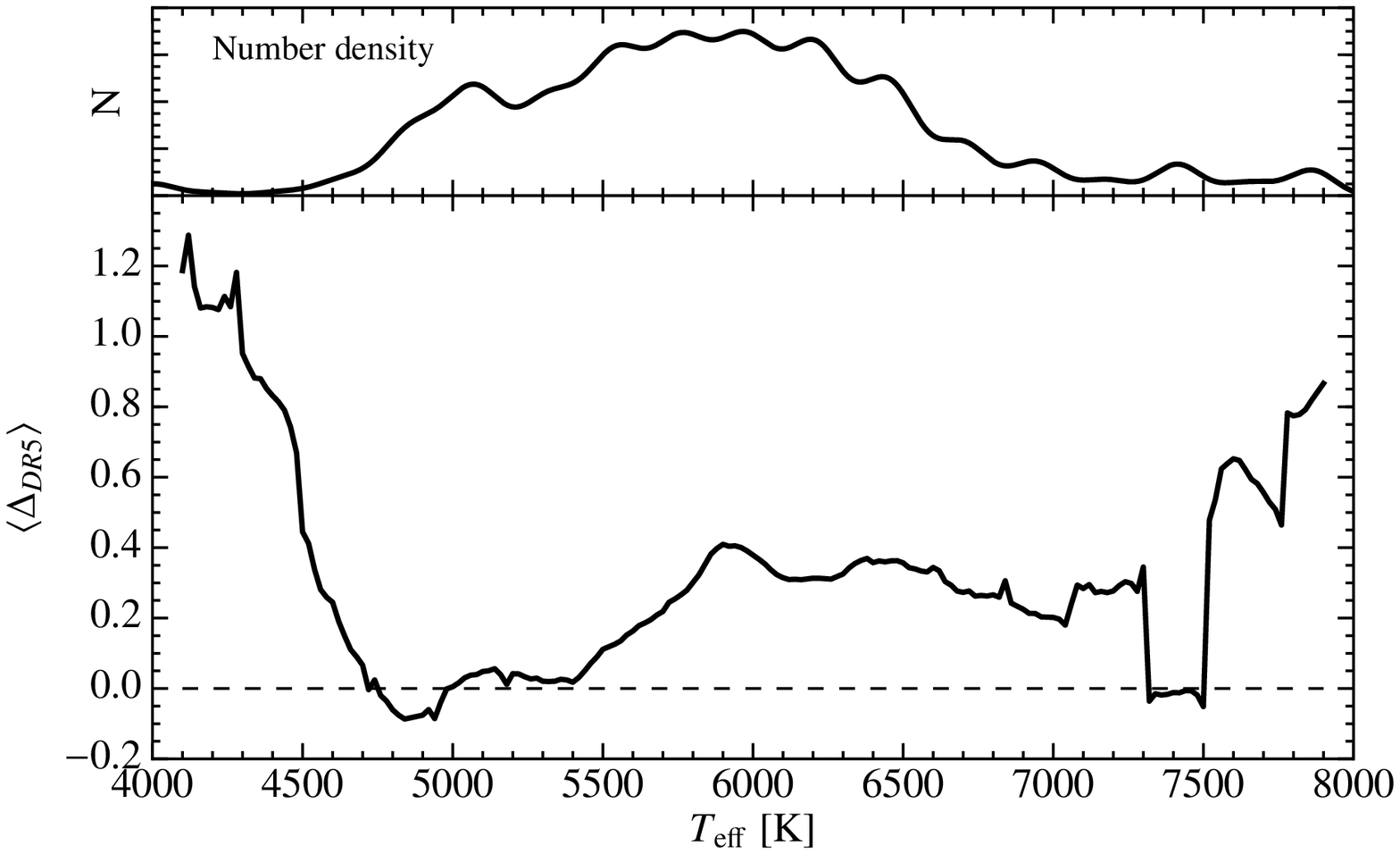}}
    \resizebox{0.5\hsize}{!}{\includegraphics{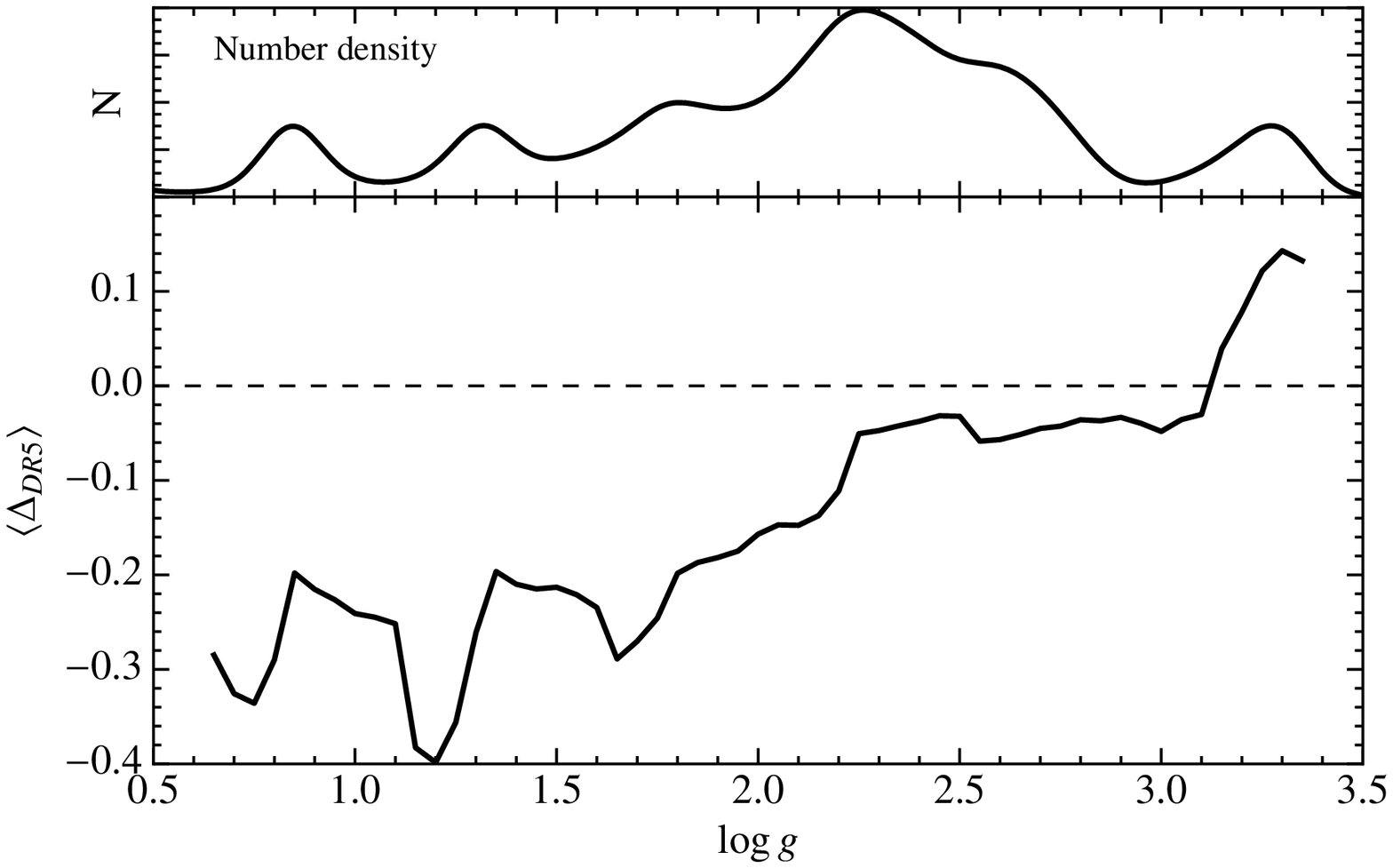}}}
  \caption{
  	Running average of $\Delta$ (i.e. difference between spectrophotometric parallax and TGAS parallax, normalised by the combined uncertainty, see Eq.~\ref{eq:Delta}) as a function of $\teff$ for dwarfs (left lower) and $\log g$ for giants (right lower), comparing DR5 values to those from TGAS. The running averages are computed for widths of $200$$\,\mathrm{K}$ and 0.3 dex respectively. The plot also shows the number density as a function of $\teff$ and $\log g$ respectively for reference. Means are only calculated for stars with $-4<\Delta_{\rm DR5-TGAS}<4$. Note that positive values of $\Delta$ correspond to parallax overestimates (i.e. distance or luminosity underestimates).
  	   \label{fig:DR5fTg}
}
\end{figure*}

Here, as in \cite{RAVEDR5} we divide the stars into dwarfs ($\log g\ge3.5$) and giants ($\log g < 3.5$), and further subdivide dwarfs into hot ($\teff >5500$$\,\mathrm{K}$) and cool ($\teff \le 5500$$\,\mathrm{K}$). It is worth noting that this means that main-sequence turn-off stars are likely to be put in the 'dwarf' category. In Figure~\ref{fig:DR5} we show a comparison between the DR5 parallaxes and the TGAS parallaxes described by this statistic (which we call $\Delta_{\rm DR5-TGAS}$ in this case). The figures show kernel density estimates \citep[KDEs:][]{KDE}, which provide an estimate of the pdf of $\Delta_{\rm DR5}$ for each group, along with finely binned histograms (which are used to give a sense of the variation around the smooth KDE). These are generally encouraging for both cool dwarfs and giants, with a mean value that is close to zero (meaning that any parallax, and therefore distance, bias is a small fraction of the uncertainty), and a dispersion that is slightly smaller than unity (implying that the uncertainties of one or both measurements are overestimated).

For hot dwarfs there is a clear difference between the DR5 parallaxes and the TGAS parallaxes. The mean value of $\Delta$ is $0.301$, meaning that the systematic error in parallax is a significant fraction of the uncertainty, with the DR5 parallaxes being systematically larger than the TGAS parallaxes (corresponding to systematically smaller distance estimates from DR5).

The typical combined quoted uncertainty on the parallaxes for hot dwarfs is $\sim1\mas$, so this systematic difference is $\sim0.3\mas$, which is comparable to the size of the colour-dependent and spatially correlated uncertainties identified by \cite{GaiaDR1:TGAS}. It was therefore not immediately obvious whether the difference seen here is due to a systematic error with the DR5 parallaxes, or with the TGAS parallaxes.

However, we have indications from \cite{RAVEDR5} that the effective temperatures found by by the RAVE pipeline tend to be underestimates for $\teff\gtrsim5300$$\,\mathrm{K}$. The effective temperatures determined using the IRFM are systematically \emph{higher} than those found from the RAVE pipeline \citep[fig. 26,][]{RAVEDR5}. If the effective temperature used in the distance estimation is systematically lower than the true value, then this will cause us to systematically underestimate the luminosity of the star, and thus underestimate its distance (overestimate its parallax). Therefore a systematic underestimate of $\teff$ by the RAVE pipeline can explain the difference with the IRFM $\teff$ values \emph{and} the systematic difference with the TGAS parallaxes. This motivates us to investigate the IRFM temperatures in Section~\ref{sec:improved} for an improved estimate of $\teff$, and thus more accurate distance estimates.

We can investigate this more closely by looking at how an average value of $\Delta_{\rm DR5}$ (which we write as $\langle\Delta_{\rm DR5}\rangle$) varies with $\teff$ for dwarfs or with $\log g$ for giants. In Figure~\ref{fig:DR5fTg} we show the running average of this quantity in windows of width $200$$\,\mathrm{K}$ in $\teff$ for dwarfs and 0.3 dex in $\log g$ for giants.  For reference we also include the number density as a function of these parameters in each case.

The left panel of Figure~\ref{fig:DR5fTg} shows the value of $\langle\Delta_{\rm DR5-TGAS}\rangle(\teff)$ for dwarfs. As we expect, we see that for $\teff\gtrsim5500$$\,\mathrm{K}$ we have a parallax offset of $\sim$0.3 times the combined uncertainty, which has a small dip around $7400$$\,\mathrm{K}$ \footnote{The sharp edges are due to the fact that a relatively large number of sources are assigned temperatures very near to $7410$$\,\mathrm{K}$, due to the pixelisation produced by the fitting algorithm -- see \cite{Koea11}}. The vast majority of what we termed `cool dwarfs' are in the temperature range $4600\lesssim\teff<5500$$\,\mathrm{K}$, where TGAS and RAVE clearly agree nicely. 

Below $\sim4600$$\,\mathrm{K}$ the value of $\langle\Delta\rangle(\teff)$ goes to very large values, corresponding to a substantial underestimate of distance by RAVE DR5. This was not clearly seen in Figure~\ref{fig:DR5} because there are very few dwarfs in this temperature range. It is not clear what causes this, though it could occur if 1) there is a tendency to underestimate the $\teff$ for these stars, which is not something which has been noted before; 2) stars with quoted $\log g$ values between the dwarf and giant branches have been given too high a probability of being dwarfs by the pipeline, and/or 3) the pipeline assigns too low a luminosity to stars near this part of the main sequence -- possibly because many of them are still young and perhaps still settling onto the main-sequence \citep[see][]{Zeea17}. 

The right panel of Figure~\ref{fig:DR5fTg} shows the value of $\langle\Delta_{\rm DR5}\rangle(\log g)$ for giants. In the range $2.2\lesssim\log g\lesssim3.0$ (which is a region with a high number of stars) we can see that the DR5 parallaxes more-or-less agree with those from TGAS. However, at high $\log g$ RAVE parallaxes are on average larger than those from TGAS (corresponding to an underestimate of the luminosity), whereas at low $\log g$ RAVE parallaxes are on average smaller than those from TGAS (i.e. the luminosity is overestimated). We will discuss this difference in Section \ref{sec:Giants}.

It is worth emphasising that the effects we see here for low $\teff$ or low $\log g$ are not ones that we would simply expect to be caused by the statistical uncertainties in the RAVE parameters (e.g., the stars with the lowest quoted $\log g$ values being only the ones scattered there by measurement error). The Bayesian framework compensates for exactly this effect, so the problem we are seeing is real.

\section{Using other RAVE data products for distance estimation} \label{sec:improved}

\begin{figure*}
  \centerline{
    \resizebox{0.33\hsize}{!}{\includegraphics{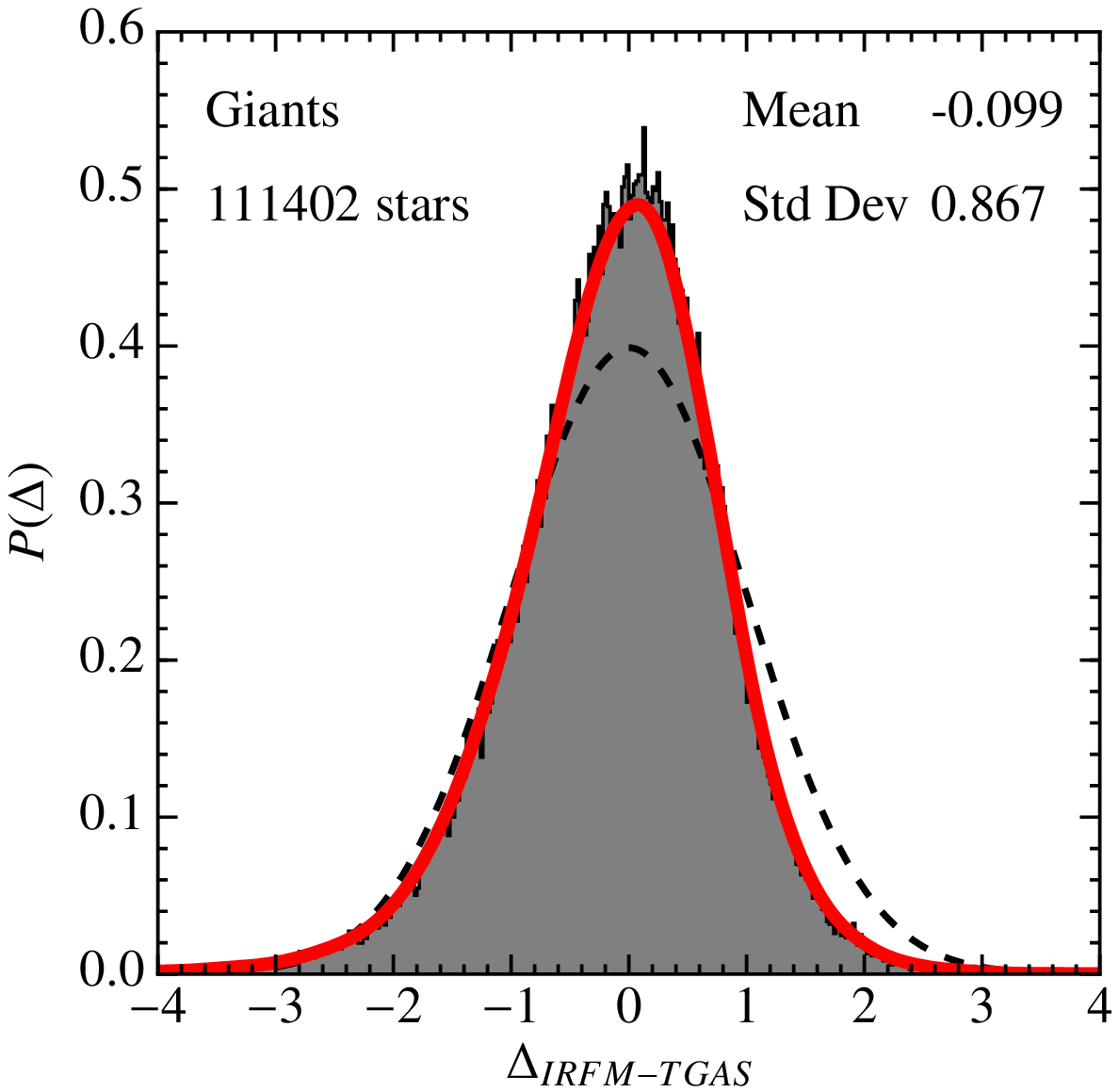}}
    \resizebox{0.33\hsize}{!}{\includegraphics{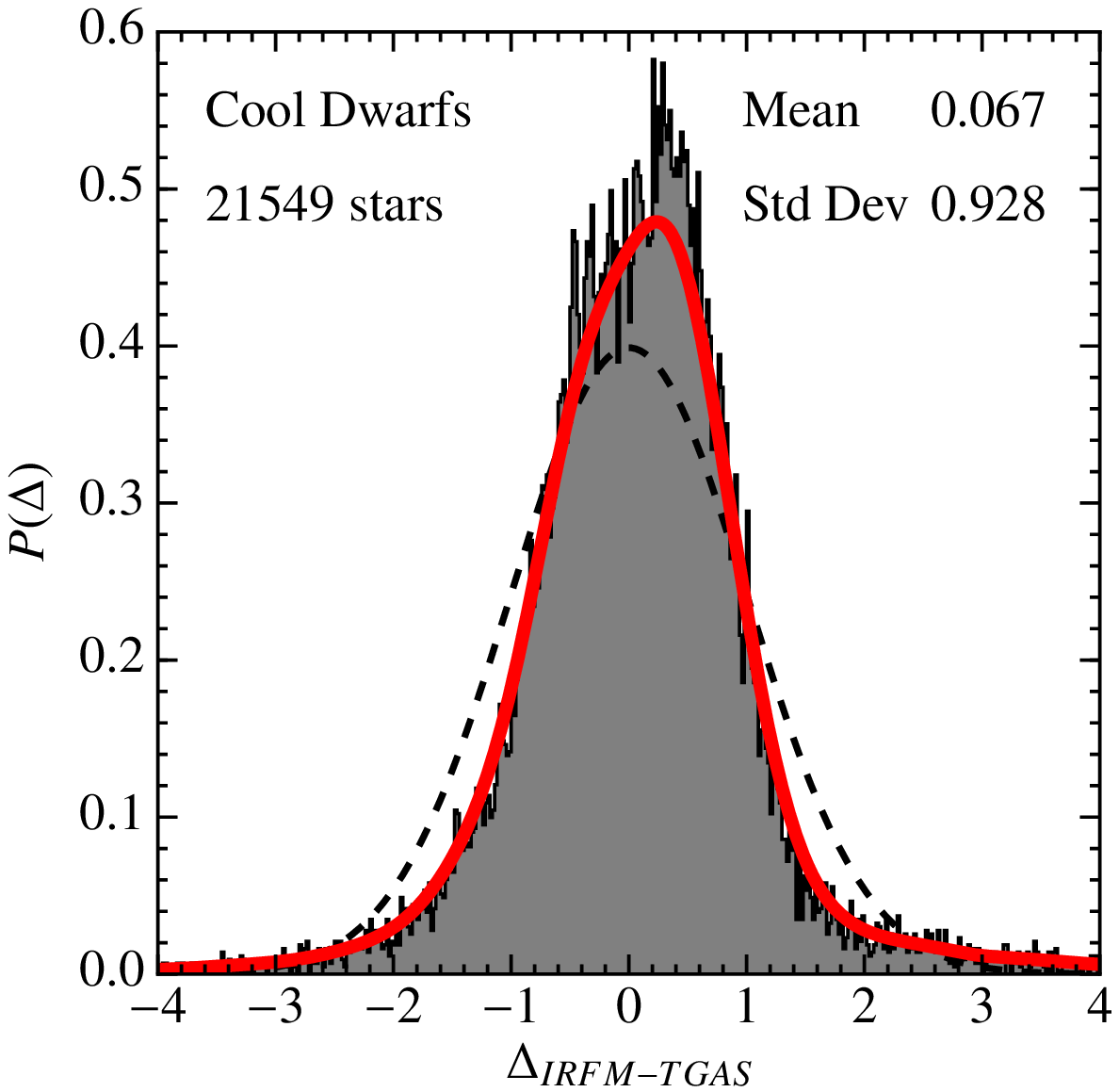}}
    \resizebox{0.33\hsize}{!}{\includegraphics{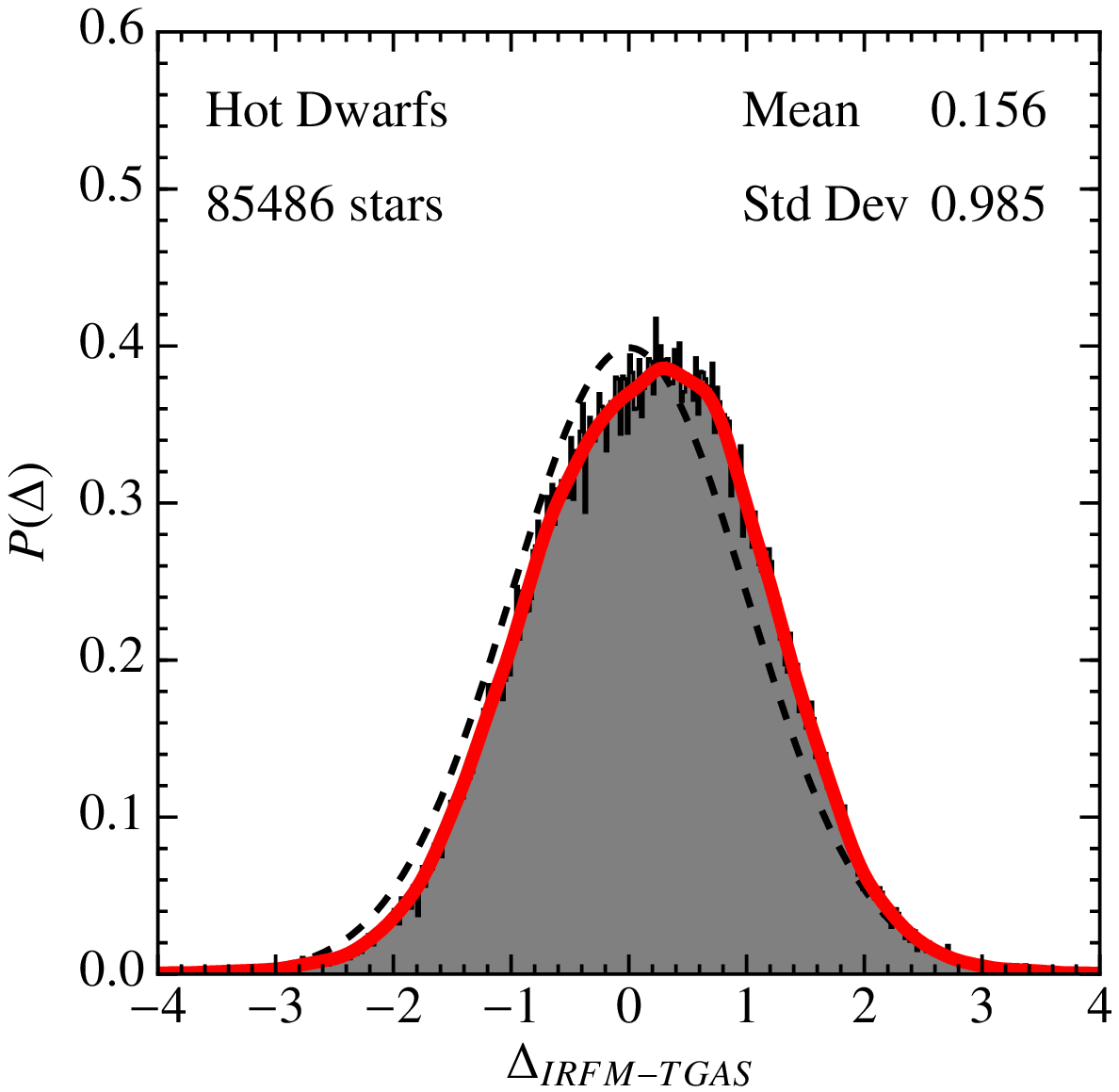}}}
  \caption{
  	Comparison of parallax estimates from RAVE with temperatures taken from the IRFM and parallax measurements from TGAS. This plot shows the same statistics as in Figure~\ref{fig:DR5}, and again we divide the stars into giants ($\log g<3.5$), cool dwarfs ($\log g\ge3.5$ and $\teffIRFM\le5500$$\,\mathrm{K}$) and hot dwarfs ($\log g\ge3.5$ and $\teffIRFM>5500$$\,\mathrm{K}$) and provide pdfs of $\Delta$ (Eq.~\ref{eq:Delta}) in each case -- positive values of $\Delta$ correspond to parallax overestimates (i.e. distance or luminosity underestimates). The main difference we can see is that the parallax estimates for hot dwarfs are substantially improved.
  \label{fig:IRFM}
}
\end{figure*}

\begin{figure}
  \centerline{
    \resizebox{\hsize}{!}{\includegraphics{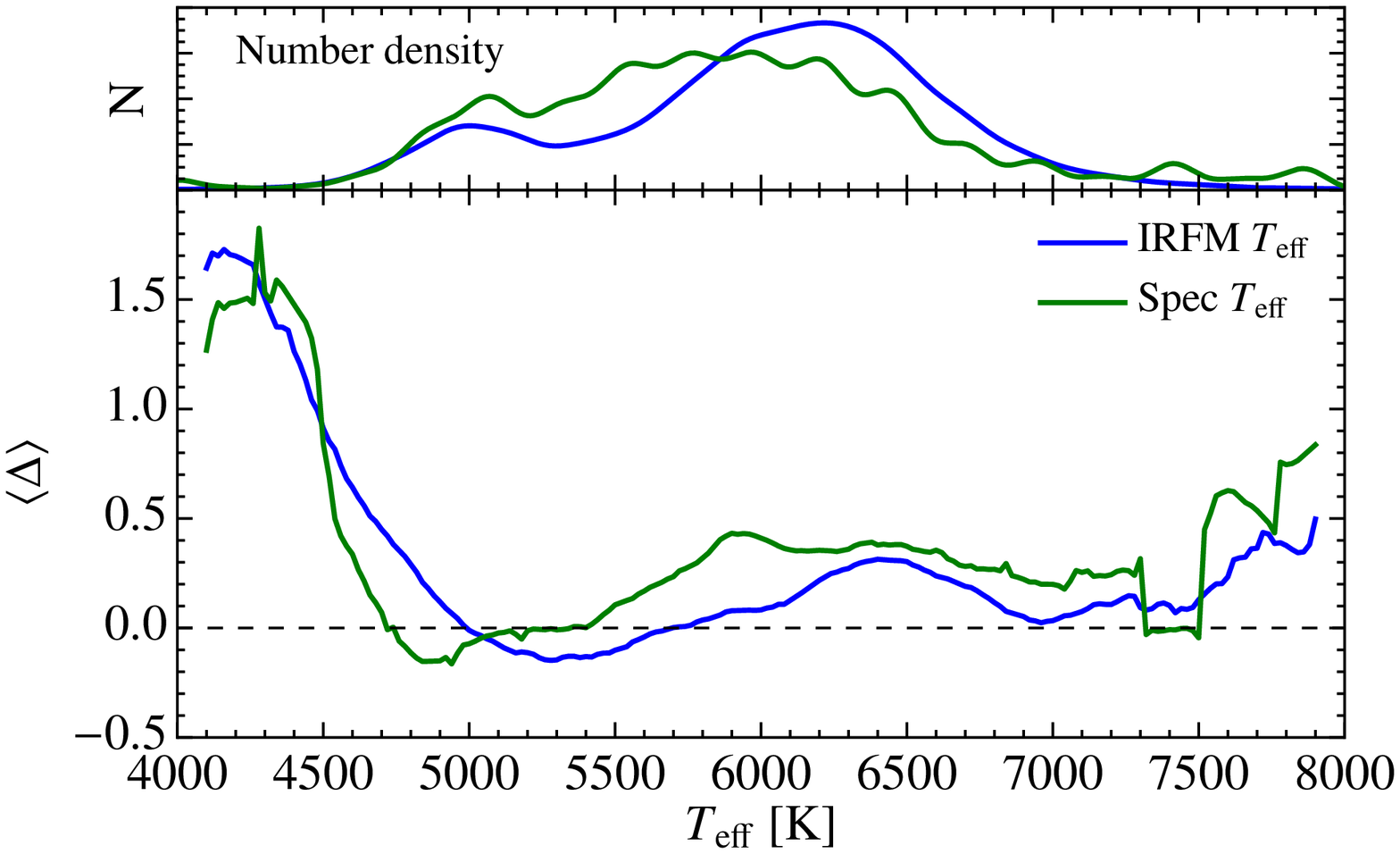}}}
  \caption{
  	As Figure~\ref{fig:DR5fTg} (left panel), this is a running average of $\Delta$ as a function of $\teff$ for dwarfs ($\log g\ge3.5$), but here we are using $\teff$ values determined by the IRFM (blue) or from the RAVE spectra (green). Again we plot also show the number density of dwarfs as a function of $\teff$ for reference. Use of the IRFM temperatures reduces the bias seen for hot dwarfs.
  	   \label{fig:NewfT}
}
\end{figure}

We now look at how the difference between parallaxes derived from RAVE and those from TGAS compare if we use $\teff$ values derived from the IRFM, rather than those derived from the spectrum directly. We also include WISE photometry in the W1 and W2 bands in both cases (as discussed in Section~\ref{sec:Bayes}).

Figure~\ref{fig:IRFM} again shows the difference between the parallaxes we derive and those found by TGAS, divided into the same three categories. We can see that the disagreement for hot dwarfs is significantly reduced from that found for DR5, with a systematic offset that is half that seen when using the spectroscopic $\teff$ values. However we can also see that the agreement between the two values is now slightly less good than before for cool dwarfs and for giants.

We can explore this in more detail by, again, looking at how the average value of $\Delta$ varies as we look at different $\teff$ for all dwarfs. In Figure~\ref{fig:NewfT} we show how a running average, $\langle\Delta\rangle(\teff)$, varies for dwarfs when we use the IRFM or the spectroscopic $\teff$ values.\footnote{Note that the $\langle\Delta\rangle$ values using the spectroscopic $\teff$ values are now not those given in DR5, but new ones, found when we include the WISE photometry. These prove to be very similar to those found by DR5.} It is clear that whatever we choose as a $\teff$ value, our parallax estimates differ dramatically from those from TGAS for dwarfs with $\teff\lesssim4600$$\,\mathrm{K}$, but there are very few dwarfs with these temperatures. For $4600$$\,\mathrm{K}$$\lesssim\teff\lesssim5500$$\,\mathrm{K}$ the values found using the spectroscopically determined $\teff$ values are better than those found using the IRFM values, while for $\teff\gtrsim5500$$\,\mathrm{K}$ the IRFM values are better. Even using the IRFM temperatures, the parallaxes found at $\teff\sim6400$$\,\mathrm{K}$ are still somewhat larger than those found by TGAS. 

\subsection{Giants} \label{sec:Giants}

\begin{figure}
  \centerline{
    \resizebox{\hsize}{!}{\includegraphics{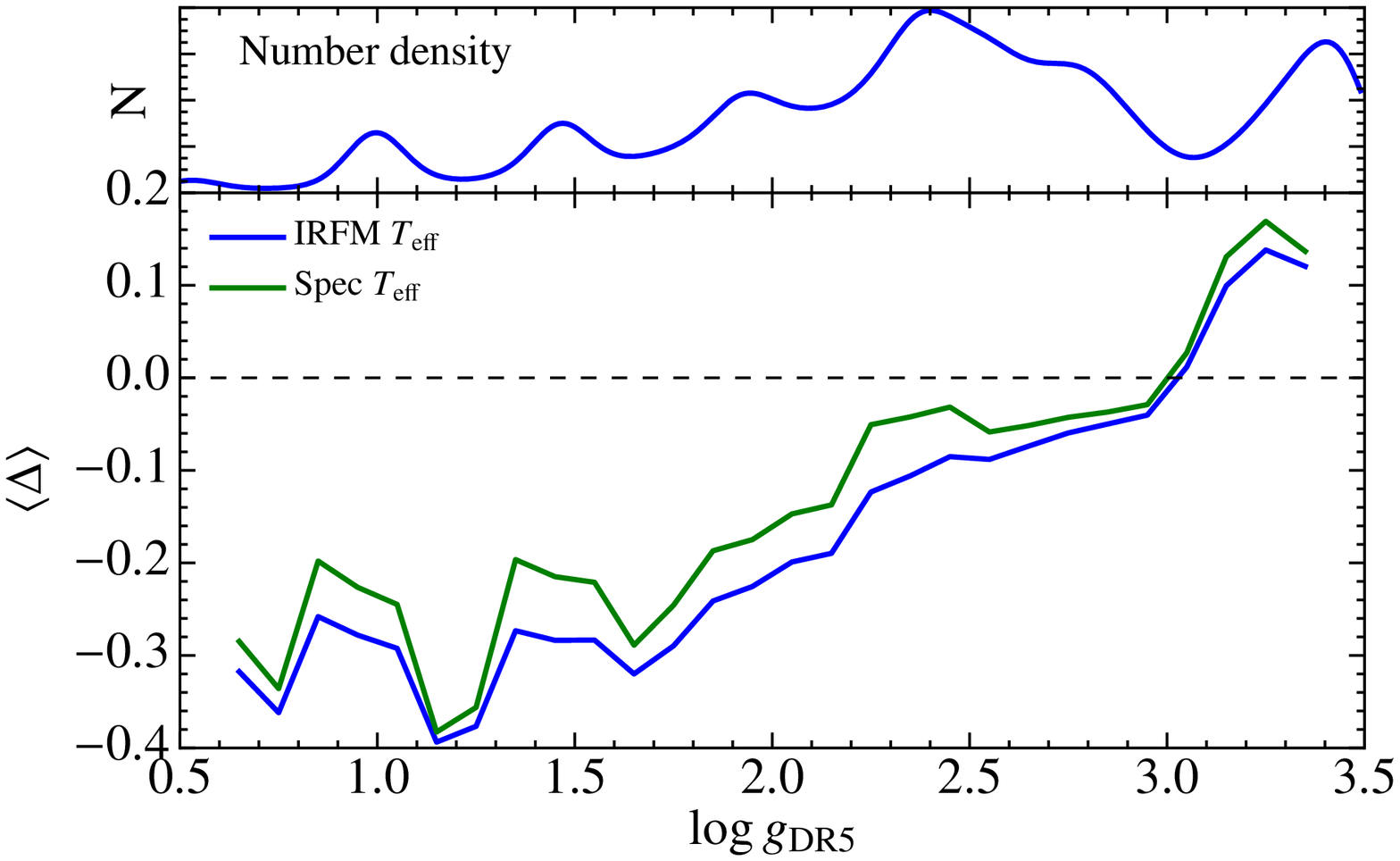}}}
  \caption{
  	As Figure~\ref{fig:DR5fTg} (right panel), this is a running average of $\Delta$ as a function of $\log g$ for giants ($\log g<3.5$), but here we are using $\teff$ values determined by the IRFM (blue) or from the RAVE spectra (green). Again, the plot also shows the number density as a function of $\log g$ respectively for reference. Means are calculated for stars with $-4<\Delta<4$.
  	   \label{fig:Newfg}
}
\end{figure}

\begin{figure}
  \centerline{
    \resizebox{\hsize}{!}{\includegraphics{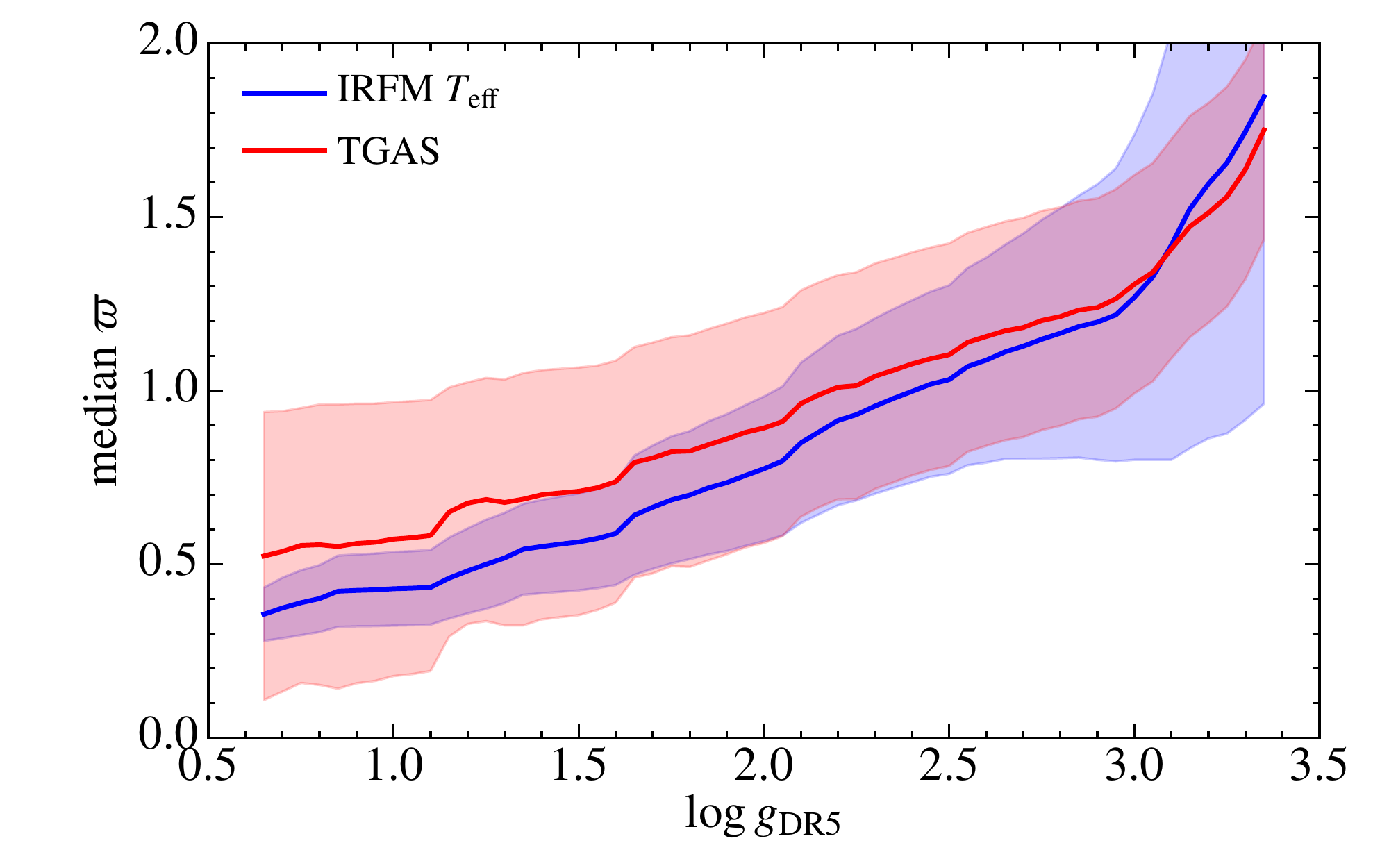}}}
  \caption{
  	Median parallax (solid line) and median parallax uncertainty (shaded region) for the RAVE pipeline using IRFM $\teff$ values (blue) and TGAS (red) as a function of $\log g$. The quoted parallax uncertainty from RAVE becomes much smaller than that from TGAS as $\log g$ becomes small. This means that when we use the TGAS parallaxes to improve the distance estimates, they will have little influence at the low $\log g$ end.
  	   \label{fig:Respective}
}
\end{figure}

\begin{figure}
  \centerline{
    \resizebox{\hsize}{!}{\includegraphics{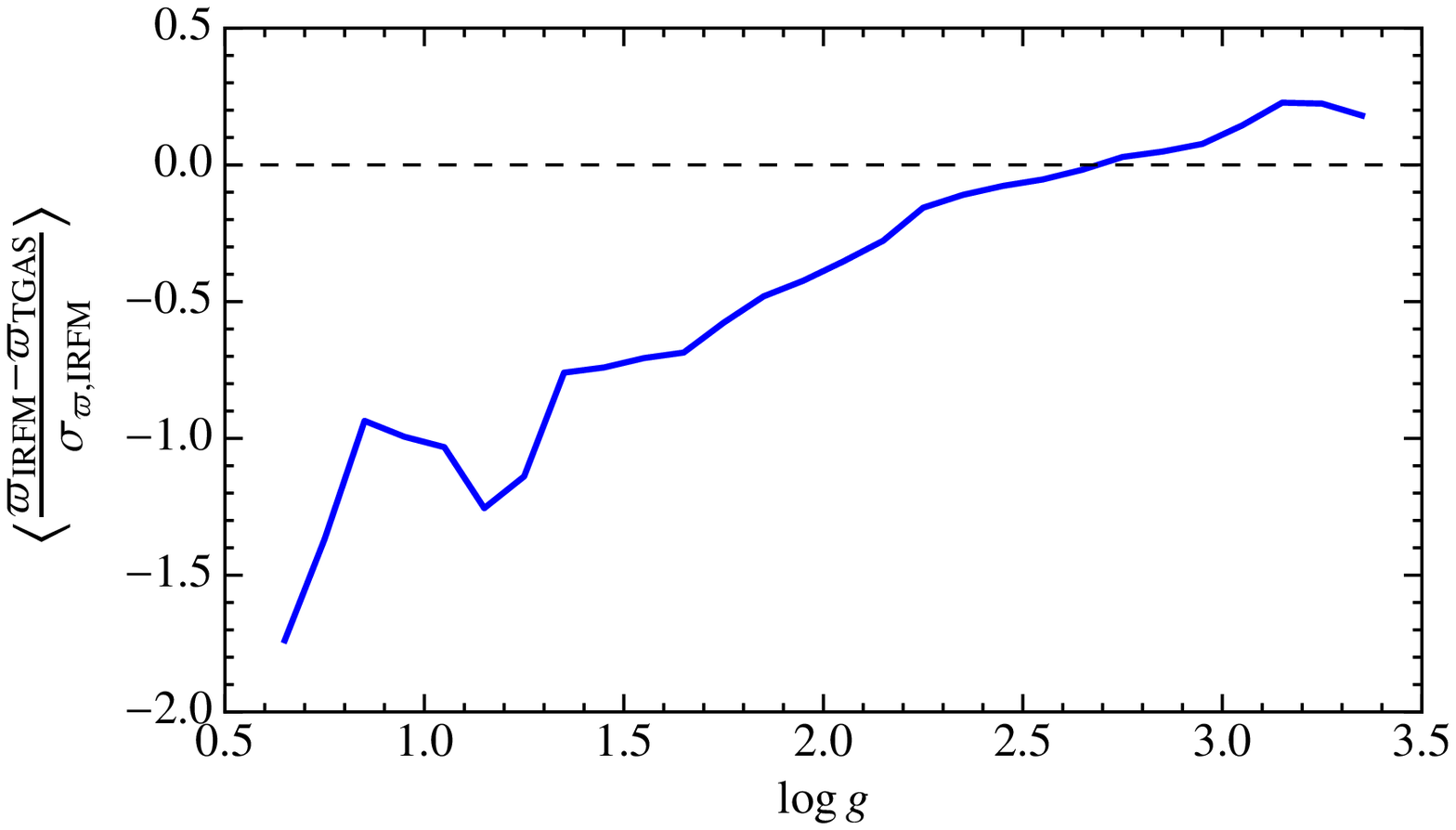}}}
  \caption{
  Running average of $(\varpi_{\rm IRFM}-\varpi_{\rm TGAS})/\sigma_{\varpi, \rm IRFM}$ as a function of $\log g$ for giants ($\log g<3.5$) -- this statistic is similar to $\Delta$ used elsewhere, but does not include the TGAS uncertainty. It therefore shows the typical systematic offset of the RAVE parallax estimates as a function of the quoted uncertainty. 
  For the lowest $\log g$, the two values are comparable.
  	   \label{fig:offset}
}
\end{figure}

We can now turn our attention to the giant stars. When we simply divide the stars into dwarfs and giants -- as was done with \Hipparcos\ parallaxes by \cite{JJBea14} and \cite{RAVEDR5}, and with TGAS parallaxes in Figures \ref{fig:DR5} and \ref{fig:IRFM} of this study -- any biasses appear small. However, when we study the trend with $\log g$, as in Figures \ref{fig:DR5fTg} and \ref{fig:Newfg}, we  see that while the stars with $\log g\gtrsim2.2$ have RAVE parallaxes that are very similar to those from TGAS (with a moderate overestimate for $\log g<3$), the stars with lower $\log g$ values have RAVE parallaxes which seem to be systematically underestimated (corresponding to distance overestimates).

We can understand how this may have come about if we look at the comparison of the RAVE $\log g$ values with those found by GALAH \citep{GALAH17} or APOGEE \citep{APOGEE} for the same stars  -- as presented in figs. 17 \& 19 of \cite{RAVEDR5}. In both cases there appears to be a trend that the other surveys find larger $\log g$ values for stars assigned RAVE $\log g \lesssim 2$. A systematic underestimate of the $\log g$ values of these stars would lead to exactly this effect. In Section~\ref{sec:ASCal} we will look at the asteroseismic re-calibration of RAVE $\log g$ found by \cite{Vaea17}, which also suggests that these $\log g$ values may be underestimated.

It is important to note that these low $\log g$ stars are intrinsically luminous, and therefore those observed by RAVE tend to be distant. This means they have relatively small parallaxes, and so the quoted TGAS uncertainties are a large fraction of true parallax, while those from RAVE are relatively small. Figure~\ref{fig:Respective} illustrates this point by showing the median parallax and uncertainty for each method as a function of $\log g$. 

{\referee
A consequence of this is that the combined parallax uncertainty used to calculate $\Delta$ is dominated by that from TGAS.} 
{\referee We illustrate this in Figure~\ref{fig:offset}, which shows the median value of the alternative statistic $(\varpi_{\rm IRFM}-\varpi_{\rm TGAS})/\sigma_{\varpi, \rm IRFM}$, where $\varpi_{\rm IRFM}$ is the parallax estimate using the IRFM $\teff$ value, and $\sigma_{\varpi, \rm IRFM}$ is the corresponding uncertainty.\footnote{Because the TGAS uncertainty is far smaller than the RAVE uncertainty for dwarfs, the equivalent plot for them is very similar to that in Figure~\ref{fig:NewfT}.} This shows that the systematic error for the lowest $\log g$ stars is comparable to the quoted statistical uncertainty. }

This also means that when we include the TGAS parallaxes in the distance pipeline for these objects, it will typically have a rather limited effect, and so the bias that we see here will persist.

\subsubsection{Asteroseismic calibration} \label{sec:ASCal}
The $\log g$ values given in the main table of RAVE DR5 have a global calibration applied, which uses both the asteroseismic $\log g$ values of 72 giants from \cite{Vaea17} and those of the \Gaia\ benchmark dwarfs and giants \citep{Heea15}. 
This leads to an adjustment to the raw pipeline values (which were used in RAVE DR4, so we will refer to them as $\log g_{\rm DR4}$) such that
\[ \label{eq:DR5Cal}
\log g_{DR5} = \log g_{\rm DR4}+0.515-0.026\times\log g_{\rm DR4}-0.023\times\log g_{\rm DR4}^2. 
\]
A separate analysis by \cite{Vaea17} which focussed only on the 72 giants with asteroseismic $\log g$ values, which are only used to recalibrate stars  with dereddened colours $0.50 <(J-K_s)_0<0.85\magn$, and found that for these stars a much more drastic recalibration was preferred, with the recalibrated $\log g$ value being 
\[ \label{eq:ASCal}
\begin{split}
\log g_{\rm AS} & = \log g_{\rm DR4} - 0.78 \log g_{\rm DR4} + 2.04 \\
		& \approx 2.61 + 0.22\times ( \log g_{\rm DR4} - 2.61)
\end{split} 
\]
This has the effect of increasing the $\log g$ values for stars in the red clump and at lower $\log g$ -- thus decreasing their expected luminosity and distance, and increasing their expected parallax. It has the opposite effect on stars at higher $\log g$. It is clear, therefore, that this recalibration is in a direction required to eliminate the trend in $\Delta$ with $\log g$ for giants seen in the right panel of Figure~\ref{fig:DR5fTg}. It is also worth noting that \cite{RAVEDR5} compared $\log g_{\rm AS}$ to literature values and found a clear trend in the sense that $\log g_{\rm AS}$ was an overestimate for stars with literature $\log g < 2.3$ , and an underestimate for literature $\log g > 2.8$.

\begin{figure}
  \centerline{
    \resizebox{\hsize}{!}{\includegraphics{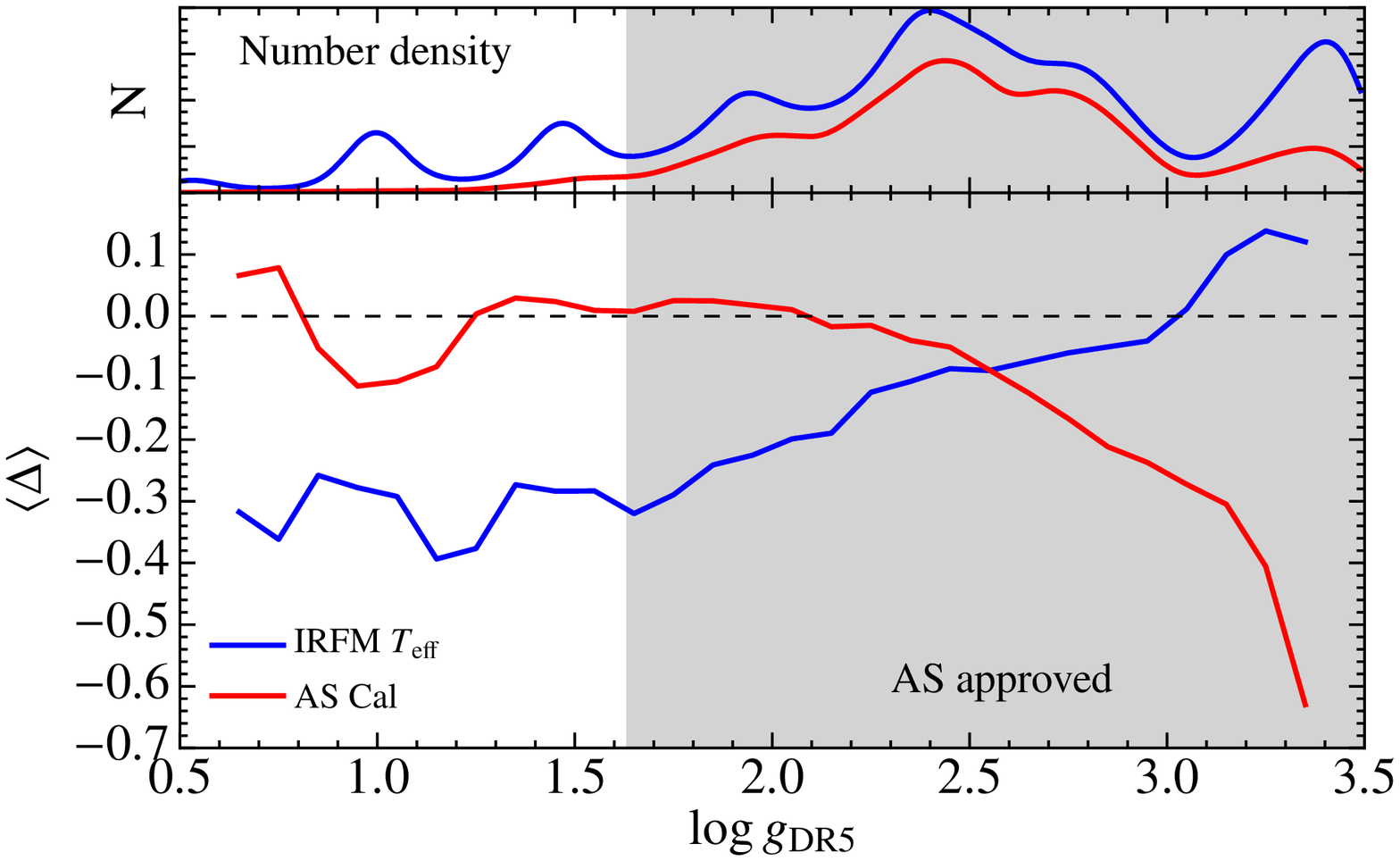}}}
  \caption{
  	As Figure~\ref{fig:Newfg} this is a running average of $\Delta$ as a function of $\log g_{\rm DR5}$ for giants where the $\log g$ values 
	used come from the main DR5 calibration (blue; eq~\ref{eq:DR5Cal}) or the asteroseismic calibration (red, Eq.~\ref{eq:ASCal}). Note that the $x$-axis gives the DR5 $\log g$ value in each case - this is to enable a side-by-side comparison. 
	In both cases we have used $\teff$ values determined by the IRFM. The grey region indicates the range in $\log g$ over which the asteroseismic calibration appears to work reasonably well for the reference stars considered by \protect\cite{RAVEDR5}.
	The running averages are computed for over a width of 0.3 in $\log g$. The plot also shows the number density as a function of $\log g$ respectively for reference. Means are calculated for stars with $-4<\Delta<4$. Using the asteroseismically calibrated $\log g$ values for stars clearly improves the distance estimates for $\log g_{\rm DR5}\lesssim2.5$, which is the point where the two values are equal, but makes them worse for $\log g_{\rm DR5}\gtrsim2.5$.
  	   \label{fig:NewfgAS}
}
\end{figure}

In Figure~\ref{fig:NewfgAS} we show $\Delta$ as a function of $\log g_{DR5}$ for stars using the recalibrated $\log g_{\rm AS}$ values given by \cite{Vaea17} (along with those when using the DR5 $\log g$ values for reference). We use the DR5 $\log g$ value on the x-axis to provide a like-for-like comparison, and the grey region in Figure~\ref{fig:NewfgAS} is equivalent to the range $2.3<\log g_{\rm AS}<2.8$. It is clear that the asteroseismically calibrated $\log g$ values improve the distance estimation for stars with low $\log g$ values -- even beyond the range of $\log g$ values where these $\log g$ values disagree with other external catalogues \citep[as found by][]{RAVEDR5} -- though it should be noted that these stars (with $0.50 <(J-K_s)_0<0.85\magn$) represent a small fraction of the stars with these low $\log g$ values.

However, for gravities greater than $\log g_{\rm DR5}\simeq2.5$ (which is the point where $\log g_{\rm AS}=\log g_{\rm DR5}$), the asteroseismic calibration makes the $\log g$ values significantly worse in the sense that the spectrophotometric parallaxes are underestimates (i.e. the distances are typically overestimated). Inspection of the comparison of RAVE DR5 $\log g$ values to those from GALAH or APOGEE in \cite{RAVEDR5} appears to indicate that those with $\log g_{\rm DR5}\approx3$ are split into two groups (one with higher $\log g$ found by the other surveys, one with lower) -- i.e. these are a mixture of misidentified dwarfs/sub-giants and giants. The asteroseismic calibration is blind to this difference, and it seems likely that it does a reasonable job of correcting the $\log g$ values for the giants, at the cost of dramatically underestimating the $\log g$ values for the dwarfs/sub-giants at the same $\log g_{\rm DR5}$. 

The \cite{Vaea17} catalogue comes with an entry `flag\_050' which is true if the difference between $\log g_{\rm DR5}$ and $\log g_{\rm AS}$ is less than 0.5, and it is recommended that only stars with this flag are used. This sets a upper limit of $\log g_{\rm DR5}\simeq3.5$ for sources where the asteroseismic calibration can be applied. Our work here implies that the asteroseismic calibration should not be used for sources with $\log g_{\rm DR5}\gtrsim2.7$. 

\subsection{Outliers} \label{sec:Outliers}
\begin{figure}
  \centerline{
    \resizebox{\hsize}{!}{\includegraphics{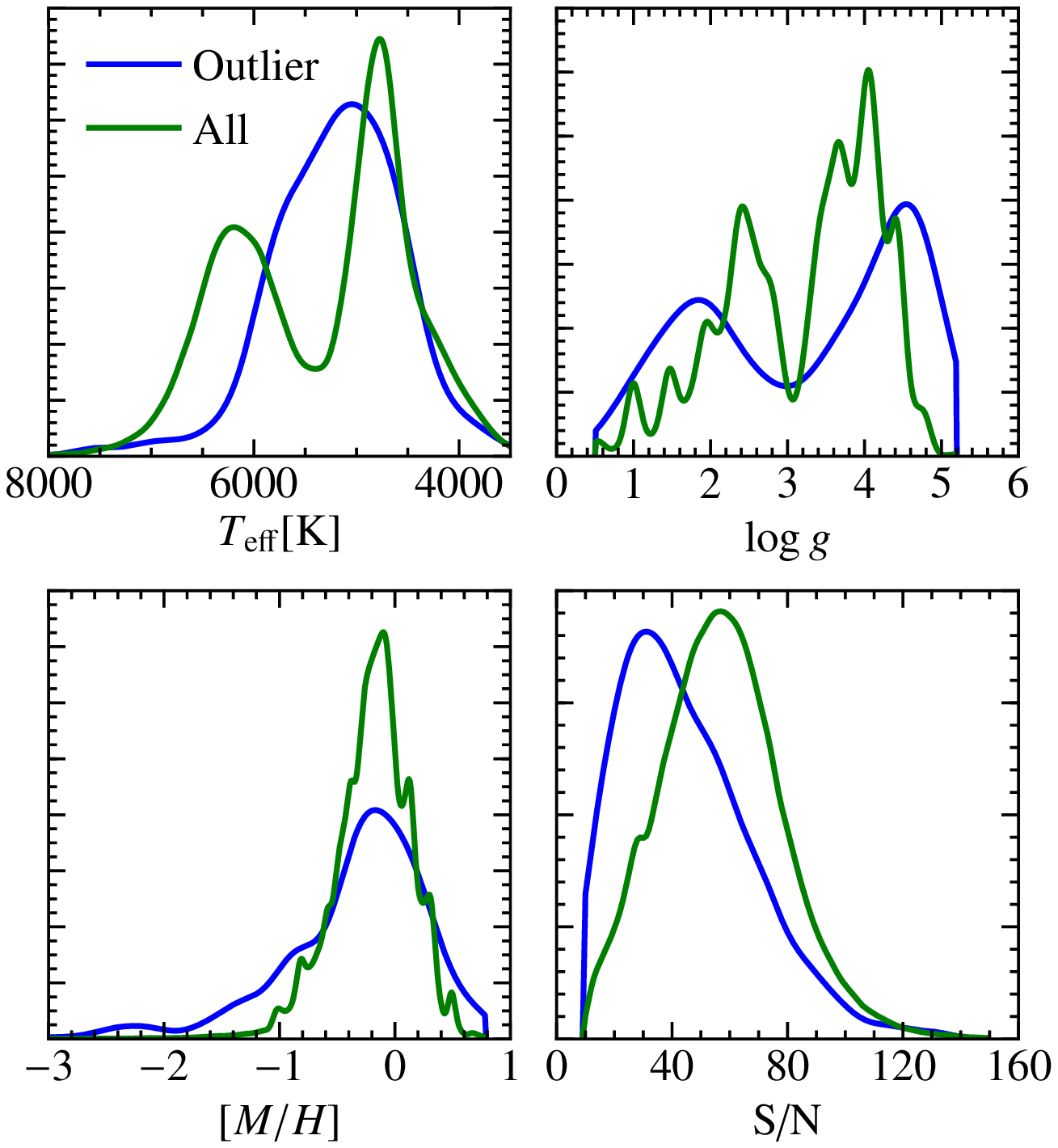}}}
  \caption{
  	Distributions of the quoted parameters of the $\sim$$1000$ stars that outliers in the sense that they have $|\Delta|>4$ (\emph{blue} lines), and of all stars in the study, for reference,  (\emph{green} lines). The plots are pdfs (so the area is normalised to 1 in all cases) produced using a kernel density estimate. The distributions shown are in $\teffIRFM$ (top left), $\log g_{\rm DR5}$ (top right), $\mh$ (bottom left) and $S/N$ (bottom right). The outliers cover a wide range of these parameter spaces, and do not come from any clearly distinct population.
  	   \label{fig:outliers}
}
\end{figure}

We have $\sim$$1000$ stars for which the quoted parallaxes from RAVE and TGAS differ by more than 4$\sigma$. We will refer to these as `outliers'. We would only expect $\sim$$12$ such objects if the errors were Gaussian with the quoted uncertainties. In Figure~\ref{fig:outliers} we show pdfs indicating how these stars are distributed in quoted $\teffIRFM$,  $\log g$,  and $\mh$. They cover a wide range of these parameters, and no clear problematic area is evident. They do tend to have relatively low $\teff$ values, and constitute a relatively large fraction of stars with quoted $\mh$ values towards either end of the full range.

We also show the distribution of these stars in terms of $S/N$, and we can see that while they tend to have relatively low $S/N$ values, they are certainly not limited to such stars. We have also looked at the values of the AlgoConv quality flag, which is provided with RAVE parameters, and find that the outliers are indicated as unreliable around the same rate as the rest of the sources.  Around $26$ percent of the outliers have flags $2$ or $3$, which indicate that the stellar parameters should be used with caution, as compared to $\sim23$ percent of all other sources, which suggests that this is not the problem. There is also no indication that they are particularly clustered on the sky.

There is some indication that the outliers tend to be problematic sources as labelled by the flags from \cite{Maea12}, which are provided with DR5. These flags are based on a morphological classification of the spectra, and can indicate that stars are peculiar (e.g., have chromospheric emission or are carbon stars) or that the spectra have systematic errors (e.g., poor continuum normalisation). $\sim$$20$ percent of the outliers are flagged as binary stars, and $\sim$$35$ percent are flagged as having chromospheric emission (compared to $\sim$$2$ percent and $\sim$$6$ percent of all sources, respectively). Similarly, $\sim$$40$ percent of the outliers are in the catalogue of stars with chromospheric emission from \cite{Zeea13,Zeea17}. The chromospheric emission can only have affected the RAVE distance estimates. However binarity can affect either the RAVE distance (by affecting the parameter estimates and/or observed magnitudes) or the TGAS parallaxes (by altering the star's path across the sky, thus changing the apparent parallax).

\subsection{Metallicity} \label{sec:HRetc}
\begin{figure*}
  \centerline{
    \resizebox{0.30769\hsize}{!}{\includegraphics{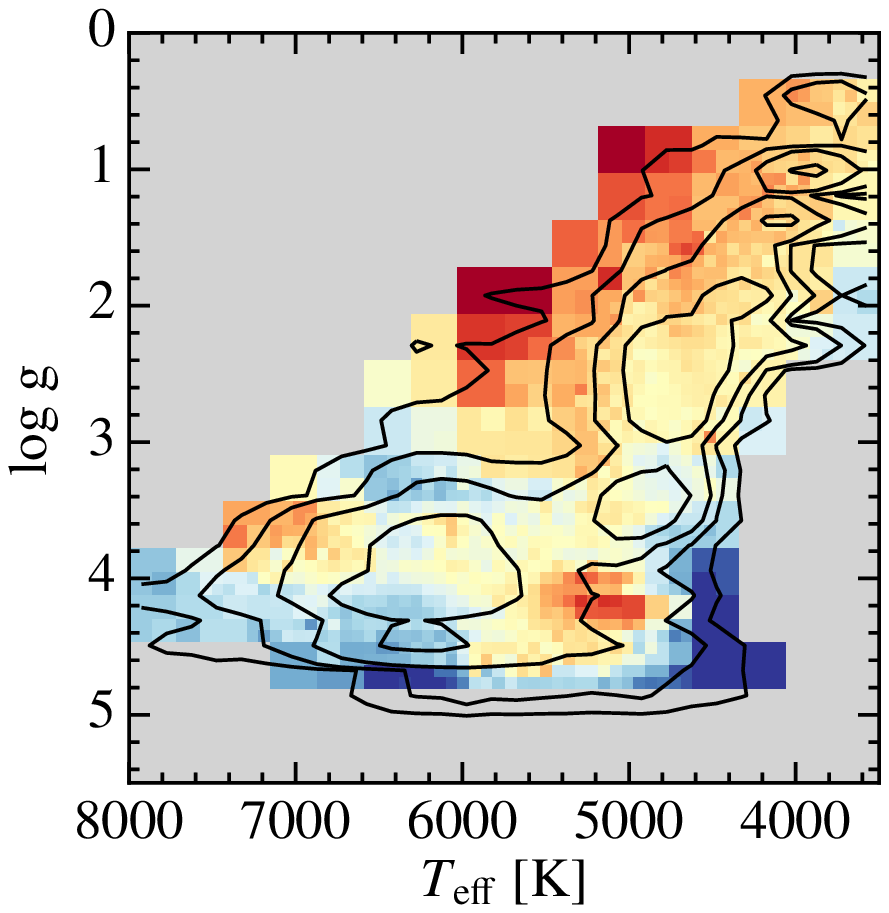}}
    \resizebox{0.30769\hsize}{!}{\includegraphics{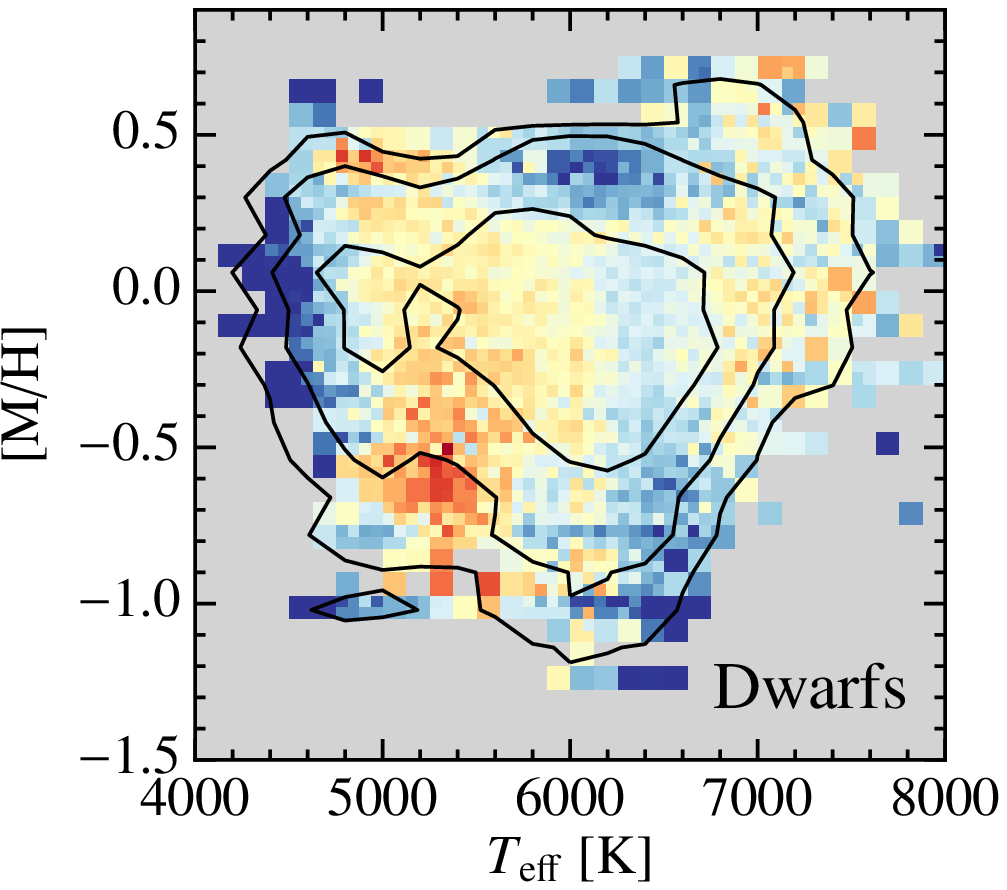}}
    \resizebox{0.30769\hsize}{!}{\includegraphics{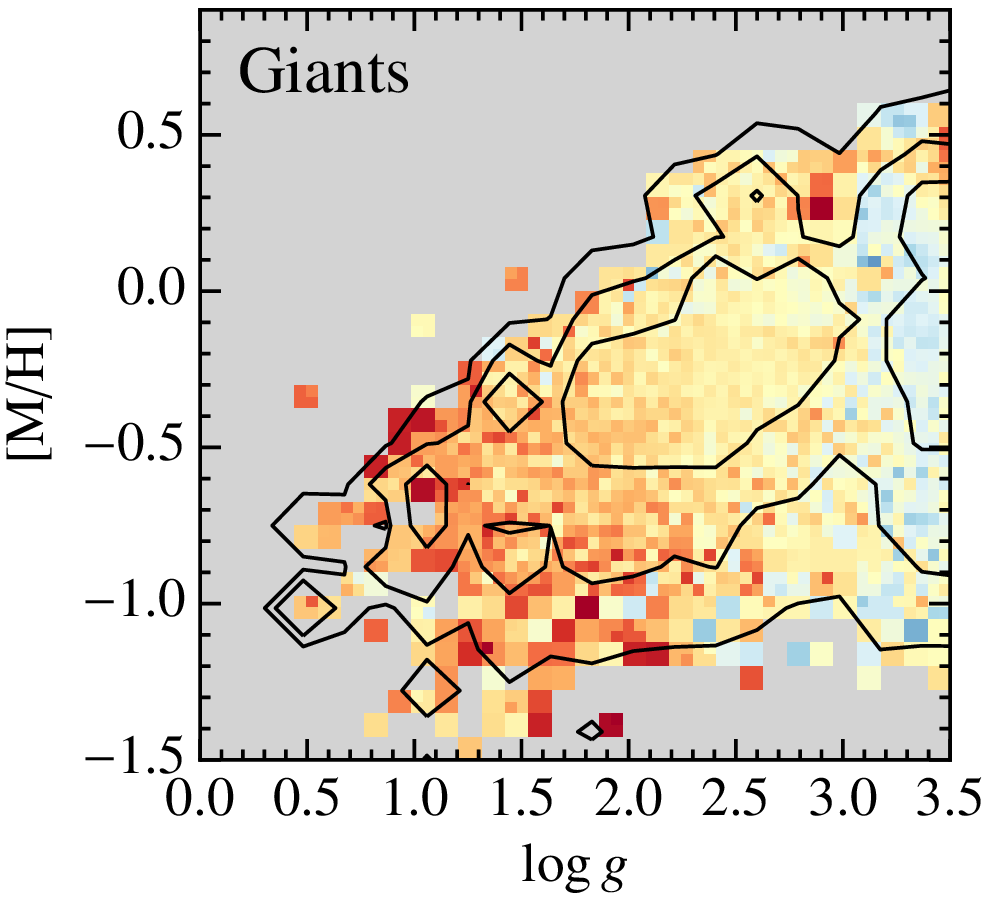}}
    \resizebox{0.07692\hsize}{!}{\includegraphics{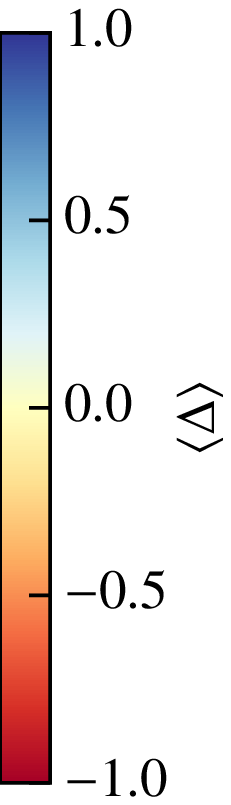}}
    }
  \caption{
  	{\referee Median values of $\Delta$, using IRFM temperatures, as a function of the stellar parameters $\teff$, $\log g$ and $\mh$. Pixel sizes are adapted such that there is never fewer than ten stars in a pixel for which we show the median. For the variation with metallicity we have, as before, divided the stars into dwarfs and giants, to show the more relevant parameter in each case. The grey areas contain very few stars. Density contours are shown as a guide to the location of the majority of the sources in these plots (this shows signs of the pixelisation of these parameters produced by the fitting algorithm used in the RAVE spectroscopic pipeline).}
  	   \label{fig:2d}
}
\end{figure*}

{\referee Finally we can look at the variation of $\Delta$ with more than one stellar parameter. In Figure~\ref{fig:2d} we show the variation of $\Delta$ in the Herzsprung-Russell (HR) diagram ($\teff$ against $\log g$) for all stars. We also show the variation of $\Delta$ in the $\mh$-$\teff$ plane for dwarfs and the $\mh$-$\log g$ plane for giants. In all cases we just show the statistics when we use the IRFM temperatures.}

{\referee
The HR diagram shows some areas where RAVE parallaxes appear to be particularly discrepant. We had already seen that low temperature dwarfs ($\teffIRFM\lesssim4500\,$$\,\mathrm{K}$ are have overestimed parallaxes. The sources with $\teffIRFM\sim5000\,$$\,\mathrm{K}$ and $\log g_{\rm DR5}\sim4.2$ have underestimated parallaxes. These sources are between the dwarf and subgiant branches, and it appears that they are typically assigned too high a probability of belong to the subgiant branch. These will be greatly improved when we include the TGAS parallax in our estimates. Sources at the upper edge of the giant branch (high quoted $\teff$ for their quoted $\log g$) also have very small RAVE parallaxes compared to those from TGAS, but these are a small fraction of giant stars.
}

There are no clear trends with metallicity for giants. For the dwarfs it is perhaps notable that there are significant parallax underestimates for metal poor stars at $\teff\sim5200$$\,\mathrm{K}$ and parallax overestimates for both unusually metal poor and metal rich stars at  $\teff\sim6200$$\,\mathrm{K}$. Again these do not comprise a particularly large fraction of all sources, and will be corrected when we include the TGAS parallax in our estimates. It is worth noting that selection effects mean that the more metal-poor stars (which tend to be further from the Sun in the RAVE sample) are likely to be higher temperature dwarfs, and (particularly) lower $\log g$ giants, and this affects any attempts to look at variation of $\Delta$ with metallicity independent of the other stellar parameters. 

{\creferee Since the most metal-poor stars tend to be cool giants which, as we have noted, are assigned distances in our output that are systematically too large, a sample of our stars which focusses on the metal-poor ones will suffer from particularly serious distance overestimates. Any prior which (like our standard one) assumes that metal-poor stars are the oldest will have a similar overestimate for the stars that are assigned the oldest ages in the sample. Note, however, that the age estimates we provide are found using a prior which assumes no such age-metallicity relation (see Section~\ref{sec:choice}), so the most metal-poor stars are not necessarily assigned the oldest ages in our catalogue.}

\subsection{Which to use?} \label{sec:which}

It is clear that adopting the IRFM temperature estimates improves the distance estimates for stars that have $T_{\rm eff,Spec}>5500$$\,\mathrm{K}$. Use of the IRFM temperatures does make the problems at low $\log g$ somewhat worse than they already were, but this is a smaller effect. We feel that switching from one temperature estimate to another at different points in the HR diagram would be a mistake, so we use the IRFM temperature in all cases. For $\sim 5000$ sources there is no IRFM $\teff$ available, so we do not provide distance estimates.

For sources recognised as outliers ($|\Delta |>4$) we assume that the RAVE parameters are unreliable, in the published catalogue these are flagged, and we provide distances estimated using only TGAS parallaxes and the 2MASS and WISE photometry. Similarly, we recognise that there is a systematic problem with dwarfs at $\teff<4600\,\mathrm{K}$, so for these stars we exclude the RAVE $\teff$ and $\log g$ from the distance estimation, and add an (arbitrary) $0.5$$\dex$ uncertainty in quadrature with the quoted RAVE uncertainty on metallicity.

We have seen that sources with $\log g_{\rm DR5}<2.0$ show a systematic difference between our parallax estimates and those found by TGAS. This is probably due to a systematic underestimate of log g for these stars by RAVE. We will determine distances to these stars in the same way as to the others, but they will be flagged as probably unreliable. While the asteroseismic recalibration clearly helps for these stars, it is not helpful at high $\log g$, and is applicable to a dwindling fraction of sources as we go to lower $\log g$. We therefore do not attempt to use this recalibration in our distance estimates, though it certainly indicates the direction we must go to improve the RAVE $\log g$ estimates.


\section{Alternative priors} \label{sec:altprior}

\begin{figure*}
  \centerline{
    \resizebox{0.5\hsize}{!}{\includegraphics{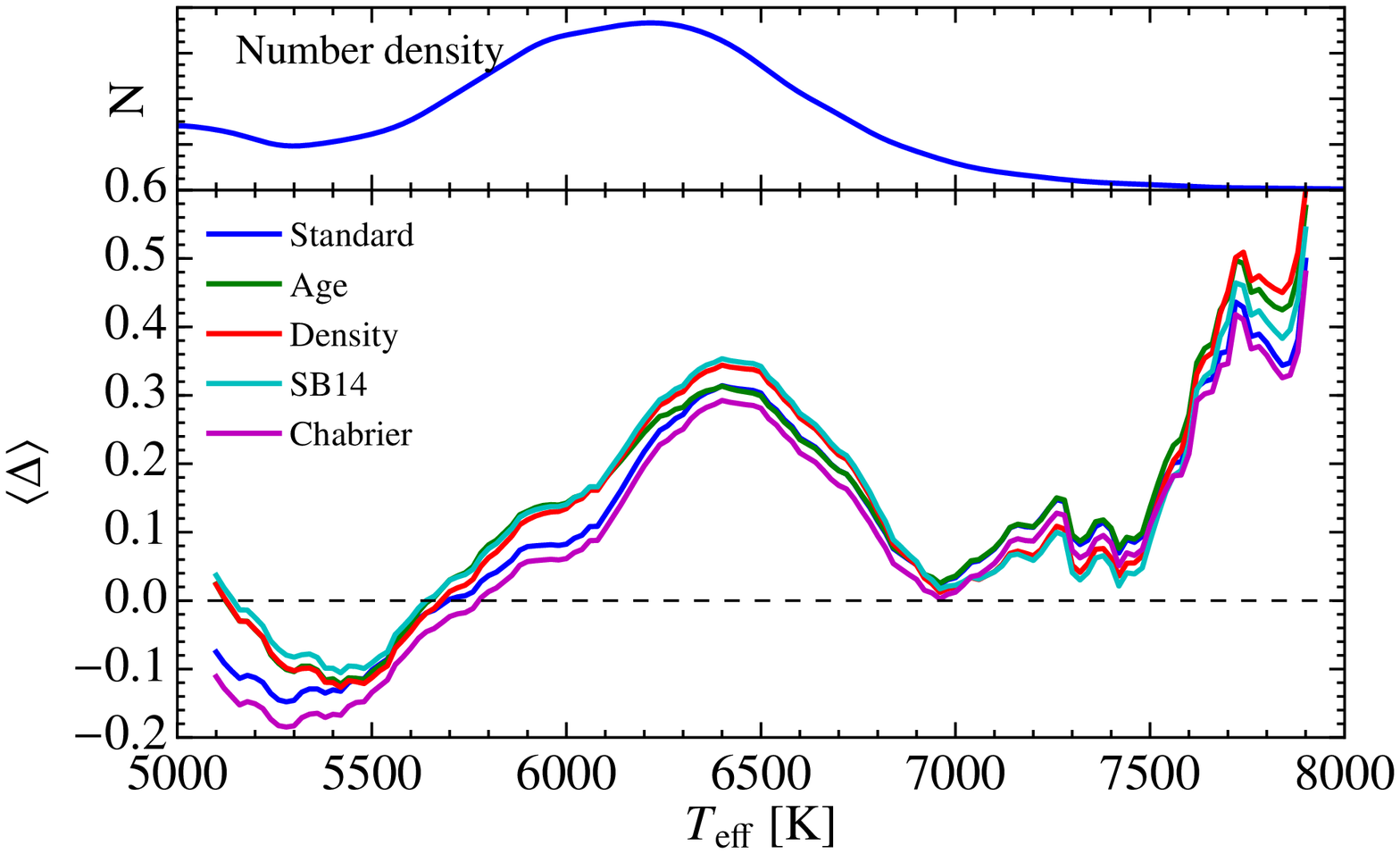}}
    \resizebox{0.5\hsize}{!}{\includegraphics{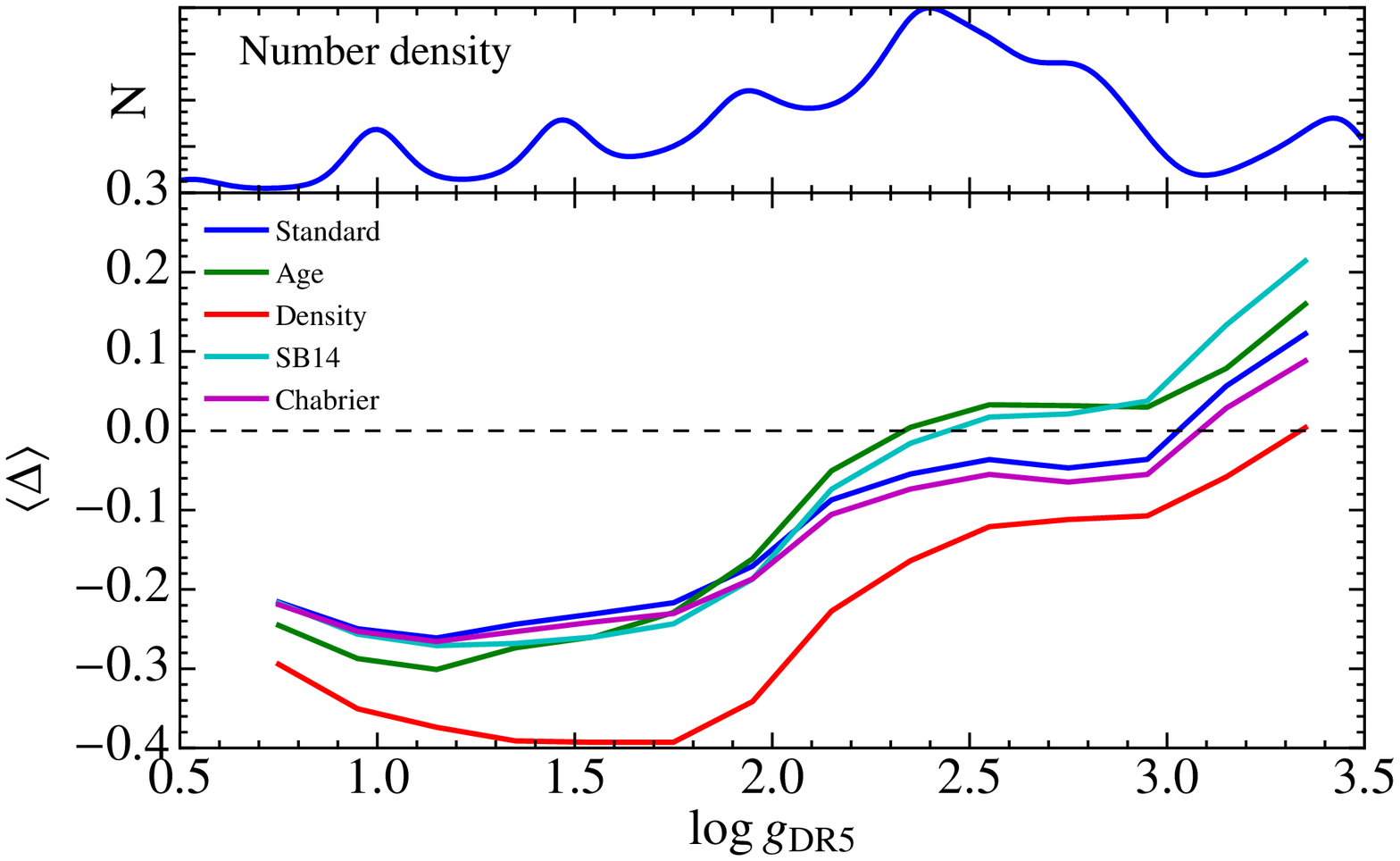}}}
  \caption{
  	As Figure~\ref{fig:DR5fTg}, this is the running average of $\Delta$ as a function of $\teff$ for dwarfs (left lower) and $\log g$ for giants (right lower) when using the alternative priors described in Section~\ref{sec:altprior}. In general, the RAVE distance estimates are reasonably robust to a change of prior.
  	   \label{fig:ComparePriors}
}
\end{figure*}

It would be very troubling if our results were strongly dependant on our choice of prior. We therefore explore the effect of our prior by considering alternative forms. We will call our standard prior `Standard', and describe the differences from this prior.  We consider four main alternative forms:
\begin{enumerate}
\item `Density' prior.  As Standard except that we set the prior on $\mh$ and $\tau$ to be uniform, with a maximum age of $13.8\Gyr$. The minimum and maximum metallicities are effectively set by the isochrone set used (Table~\ref{tab:isochrones}).\footnote{{\referee It is possible to remove this limitation, under the assumption that the stellar models do not change much at lower or higher metallicities, but the effect is limited, and it is not implemented here.}} This leaves the density profile, initial mass function (IMF) and dust model unchanged.
\item `Age' prior. As Standard except that the age prior is the same for all components and simply reflects the assumption that the star formation rate has declined over time, following the same functional form as for the thin disc in the Standard prior i.e.,
\[
 P(\tau)  \propto \exp(0.119 \,\tau/\mbox{Gyr}) \quad \mbox{for $\tau \le 13.8$\,Gyr,}  
\]

\item `SB14' prior. As Standard, except that we set the prior on  $\mh$ and $\tau$ identically for all components, following \cite{ScBe14}. This is uniform in $\mh$ over the metallicity range set by the isochrones, and such that
\[
P(\tau\,|\,\mh) \propto  \begin{cases} 0 &{\rm if}\; \tau>13.8\Gyr \\
1 & {\rm if}\; 11\Gyr \leq \tau \leq 13.8\Gyr \\
\exp\left[\frac{(\tau-11\Gyr)}{\sigma_\tau(\mh)} \right] & {\rm if}\; \tau \leq 11\Gyr, \\
\end{cases}
\]
where 
\[
\sigma_\tau = \begin{cases} 1.5\Gyr & {\rm if}\; \mh < -0.9 \\
\left(1.5 + 7.5 \times \frac{0.9+\mh}{0.4}  \right)\Gyr & {\rm if}\; -0.9 \leq \mh \leq -0.5 \\
9 \Gyr &  {\rm otherwise}. \\
    \end{cases}
\]

\item `Chabrier' prior. As Standard, except that we use a \cite{ChabrierIMF} IMF rather than a \cite{Kr01} IMF, where, following  \cite{Roea05} we take
 \begin{equation} \label{eq:priorMassChab}
  P({\cal M}) \propto 
  \begin{cases} 0 &{\rm if}\; {\cal M}<0.1\,M_\odot \\
  \frac{A_{\rm c}}{\cal M} \exp\left(\frac{\log_{10} {\cal M} - \log_{10} {\cal M}_{\rm c}}{\sigma_{\rm c}}\right)^2 & {\rm if }\;  0.1\,M_\odot \le {\cal M}<M_\odot, \\
    \;B_{\rm c}\,  {\cal M}^{-2.3} &  {\rm otherwise}. \\
    \end{cases}
\end{equation}\end{enumerate}

In Figure~\ref{fig:ComparePriors} we compare the values of $\Delta$ that we derive under all of these priors, in each case using the sets of input parameters described in Section~\ref{sec:which}, and excluding sources where we ignore the RAVE parameters. 

It is clear from the left-hand panel of Figure~\ref{fig:ComparePriors} that the priors make a very limited difference for the dwarfs, except at the low $\teff$ end, where contamination by giants is becoming more important. 

The right-hand panel of Figure~\ref{fig:ComparePriors} shows that for giants, a prior that is uniform in both $\mh$ and stellar age -- i.e., the Density prior -- provides even worse results for the low $\log g$ giants than the Standard prior. The other priors provide very similar results to one another at low $\log g$, but differ somewhat at the higher $\log g$ end -- the two priors where $P(\mh)$ is a function of position (Standard and Chabrier) tend to have lower $\Delta$ values, i.e. greater distances to these stars derived from RAVE.

We have also explored the effect of changing the power-law slope of the halo within our Standard prior (Eq.~\ref{eq:halo}) to either  $P_3(\mathbf{r}) \propto r^{-3.9}$ or $P_3(\mathbf{r}) \propto r^{-2.5}$ (compared to the usual $r^{-3.39}$). The results were essentially indistinguishable from those using the Standard prior, even if we isolate the metal-poor stars. Similarly, a decrease of 50 percent for the thin and thick disc scale heights has almost no effect -- the mean and standard deviations of the $\Delta$ values for a given population of stars (as shown in e.g., Figure~\ref{fig:IRFM}) change by $\sim$$0.001$ at most.

\subsection{Choice of prior} \label{sec:choice}
In the interests of consistency with past studies, we use the Standard prior when producing our distance estimates. However, it is clear that this choice of prior imposes a strong relationship between age and metallicity. Therefore we also provide age estimates (Section~\ref{sec:ages}) using our `Age' prior. The results presented in this section make it clear that results using this prior are roughly as reliable as those from our Standard prior, at least in terms of typical parallax error.

\section{Using RAVE parallaxes to learn about TGAS} \label{sec:TGAS}

\begin{figure}
  \centerline{ 
  \resizebox{\hsize}{!}{\includegraphics{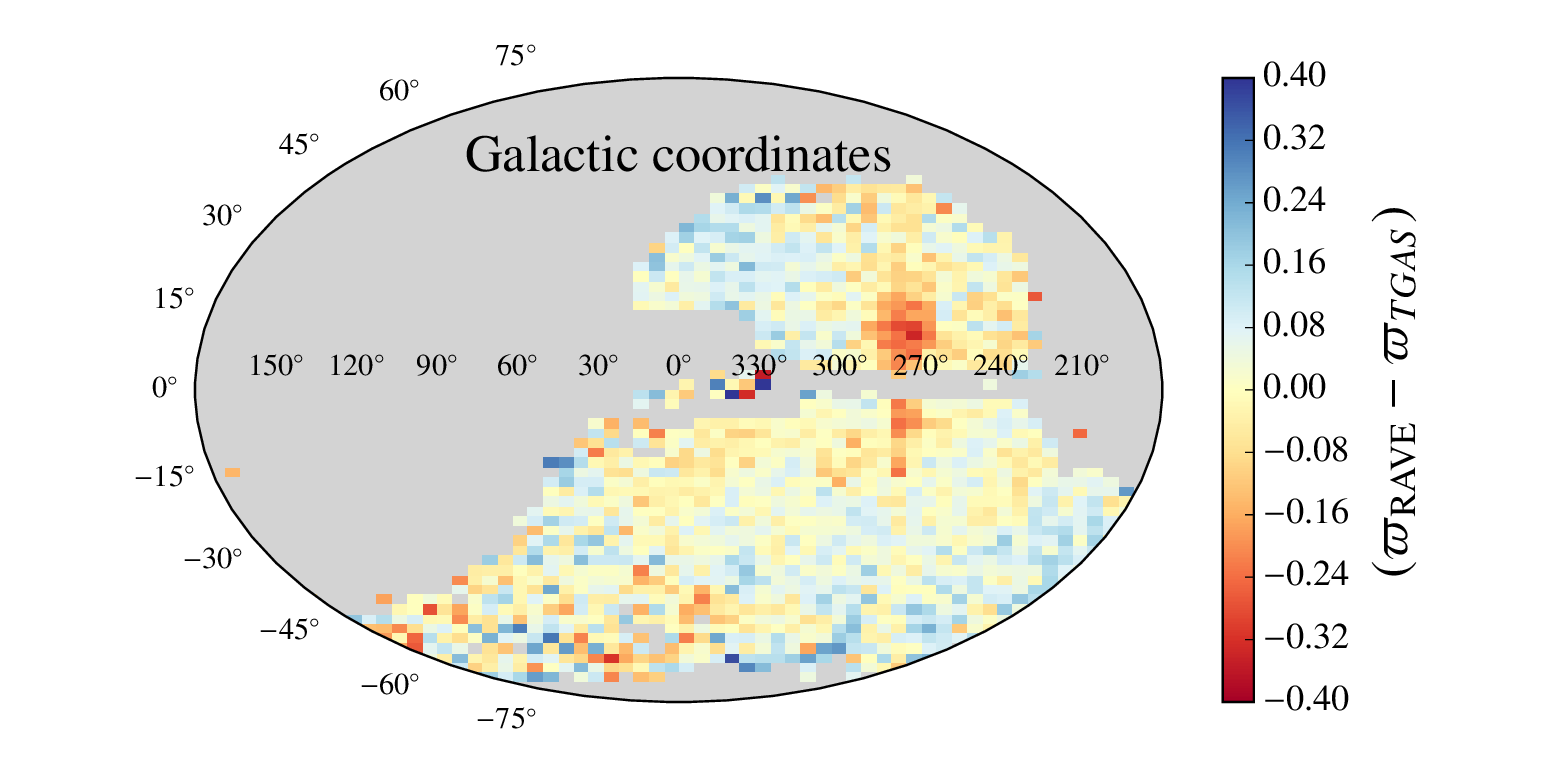}}}
   \centerline{
   \resizebox{\hsize}{!}{\includegraphics{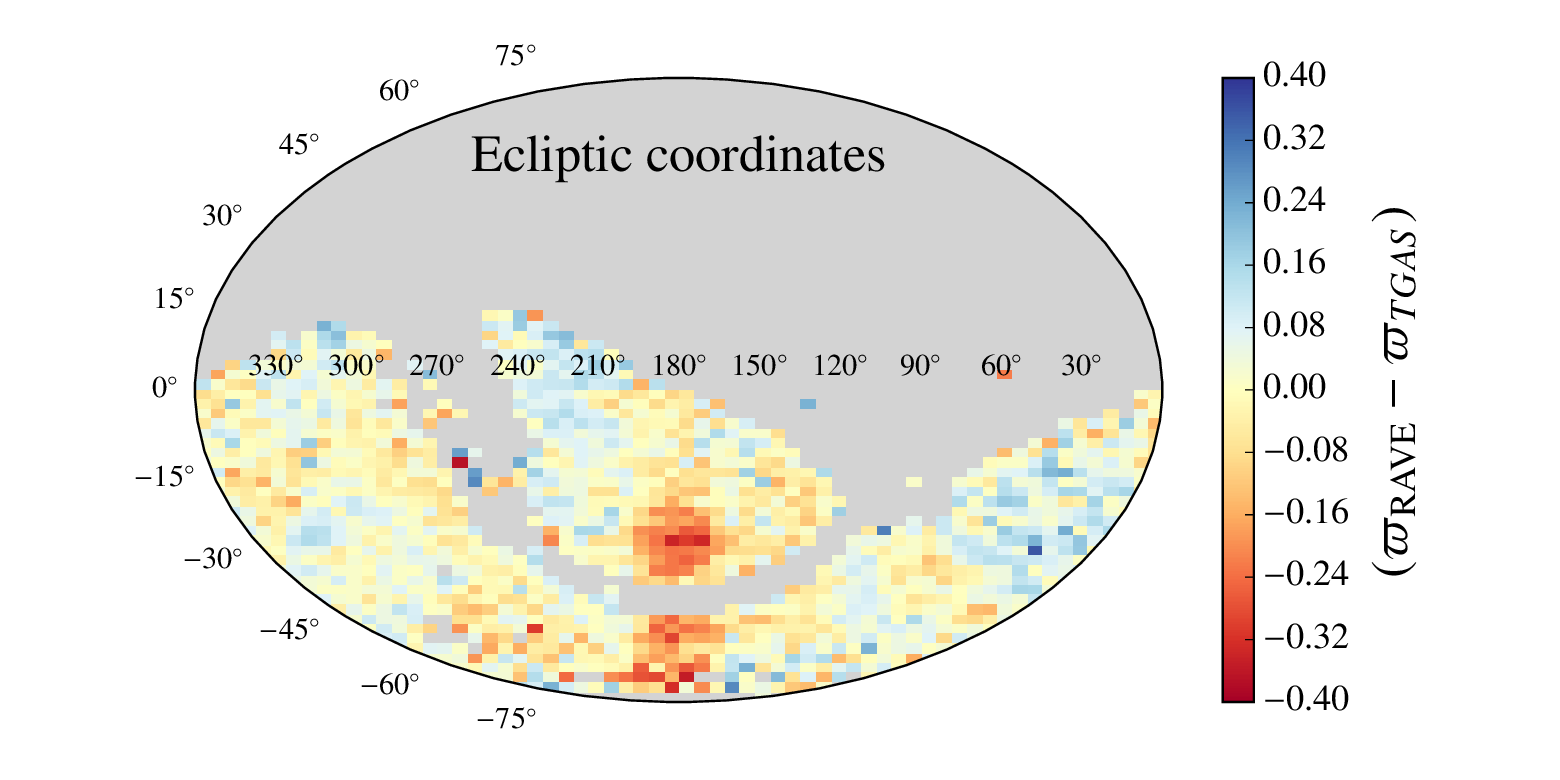}}}
  \caption{
  	Absolute difference between TGAS parallaxes and the new RAVE-only parallax estimates averaged (median) in bins on the sky in an Aitoff projection, shown in Galactic coordinates ($l,b$, upper) and ecliptic coordinates ($\lambda,\beta$, lower -- note that we have placed $\lambda=180^\circ$ at the centre of this plot to clearly show the feature).  In each plot the grey area is where there are few or no stars. The clearest feature is the patch near $l\sim 280^\circ$, $b\sim 0^\circ$ where TGAS parallaxes appear to be systematically larger than those from RAVE. When looked at in ecliptic coordinates this area can be seen to run from ecliptic pole to ecliptic pole, and is therefore likely to be related to \Gaia's scanning law \protect\citep{GaiaDR1:Validation}.
  	   \label{fig:ComboDiffSky}
}
\end{figure}

\begin{figure}
  \centerline{
    \resizebox{0.7\hsize}{!}{\includegraphics{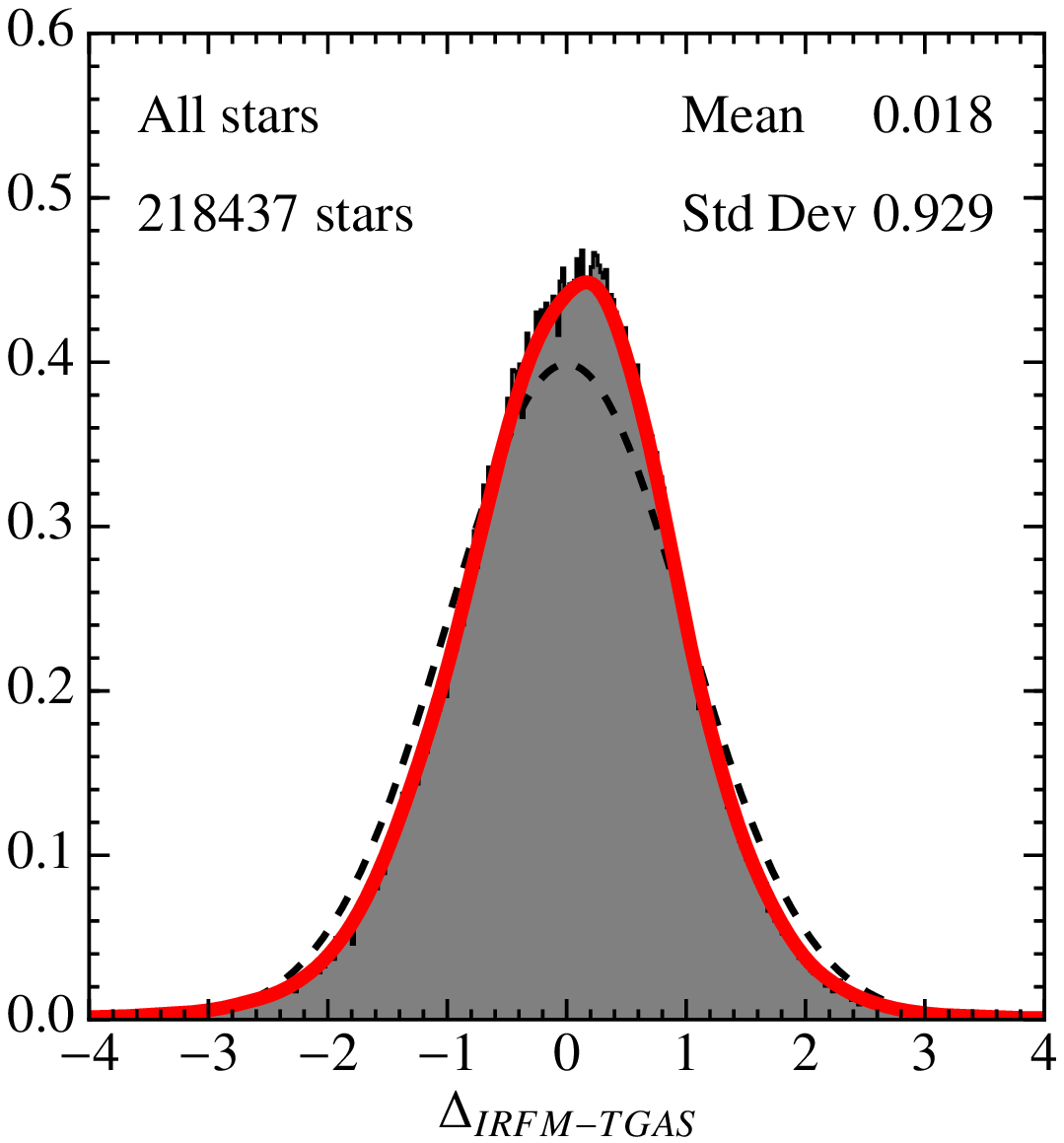}}}
  \caption{
  	Distribution of $\Delta$ for all stars using the new RAVE-only parallax estimates compared to TGAS. The standard deviation is less than unity, implying that the uncertainties of at least one of the parallax estimates have been overestimated.
  	   \label{fig:ComboDiff}
}
\end{figure}

\begin{figure*}
  \centerline{
    \resizebox{0.2\hsize}{!}{\includegraphics{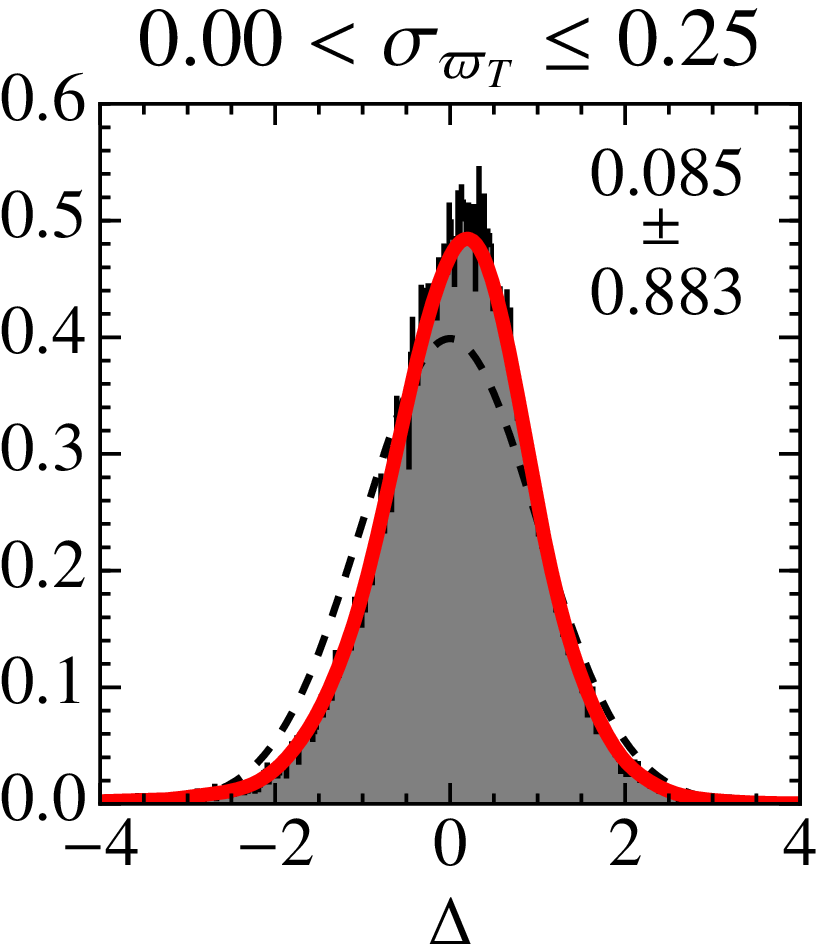}}
    \resizebox{0.2\hsize}{!}{\includegraphics{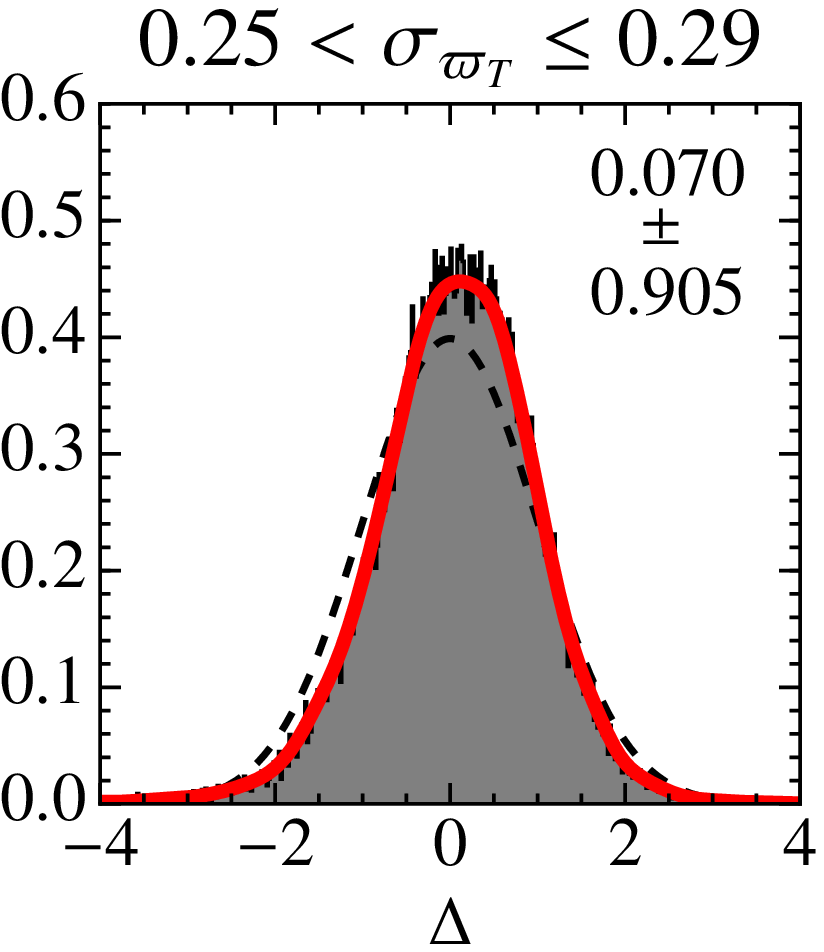}}
    \resizebox{0.2\hsize}{!}{\includegraphics{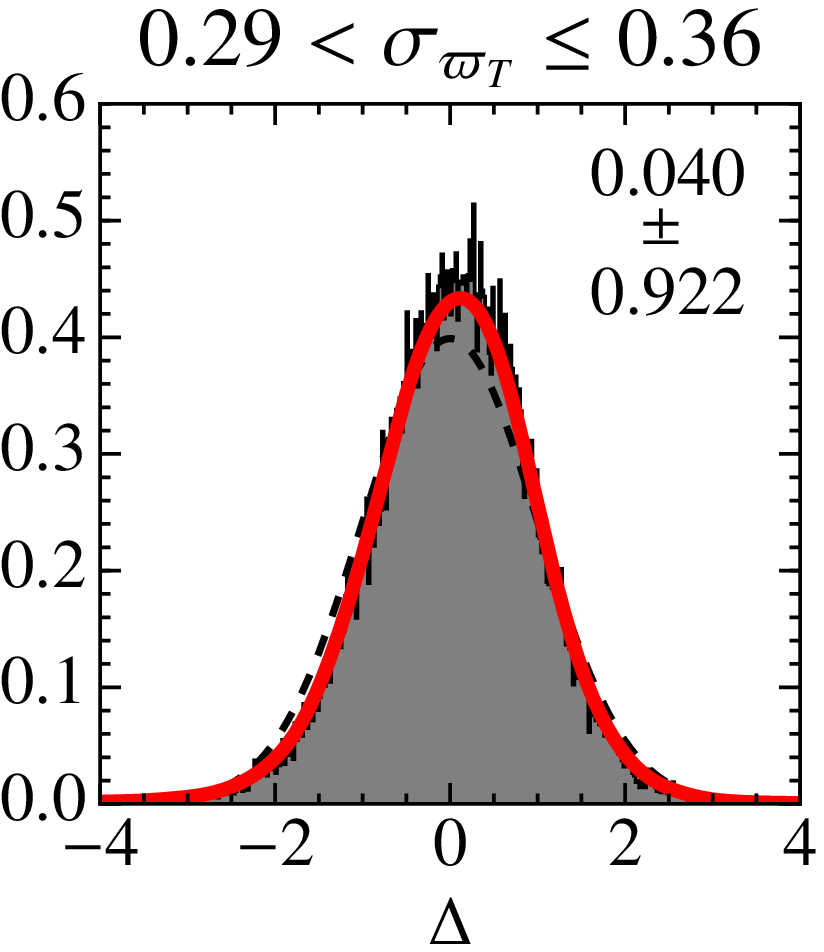}}
    \resizebox{0.2\hsize}{!}{\includegraphics{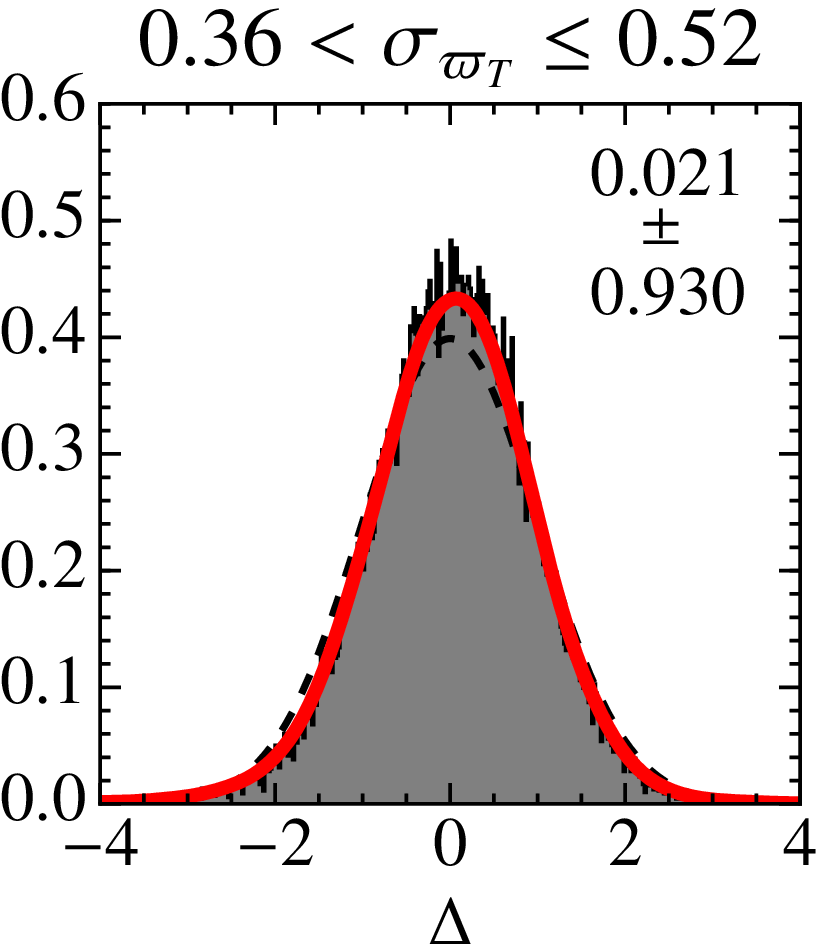}}
    \resizebox{0.2\hsize}{!}{\includegraphics{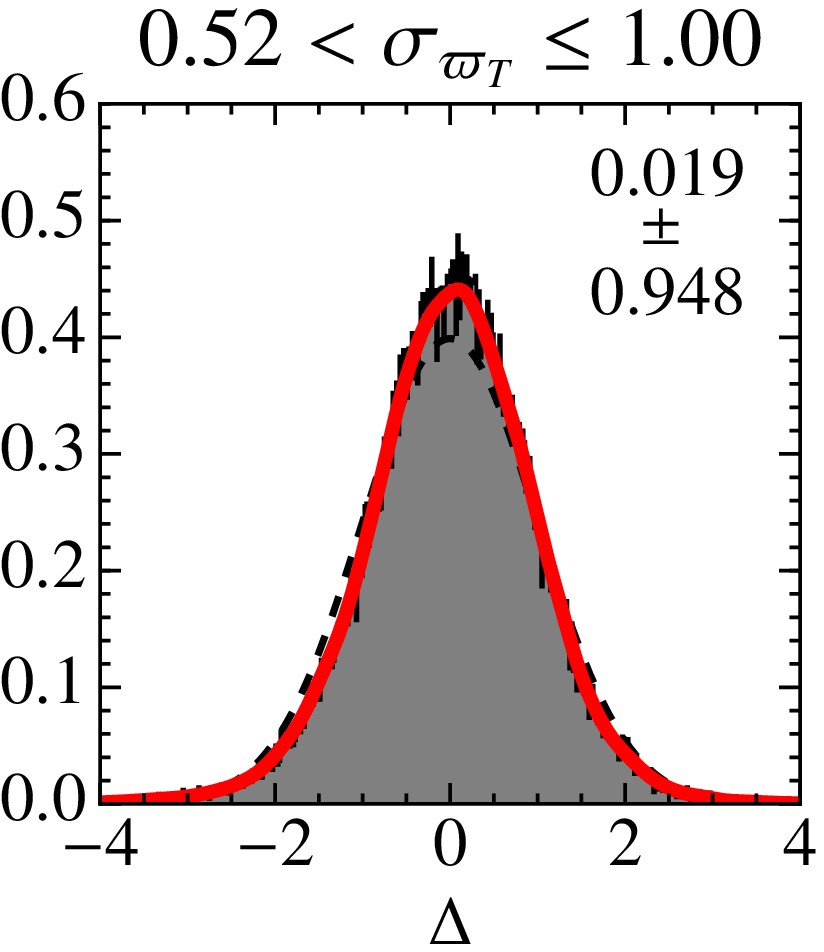}}}
  \caption{
  	Distribution of $\Delta$ using the new RAVE-only parallax estimates, separated into bins by TGAS parallaxes $\varpi_T$. The standard deviation is less than unity in each case, but increases as $\varpi_T$ increases. This could be because the RAVE uncertainties are consistently overestimated, or because the TGAS uncertainties are particularly overestimated for the smallest uncertainties.
		   \label{fig:ComboDiffPar}	
}
\end{figure*}

\begin{figure}
  \centerline{
    \resizebox{\hsize}{!}{\includegraphics{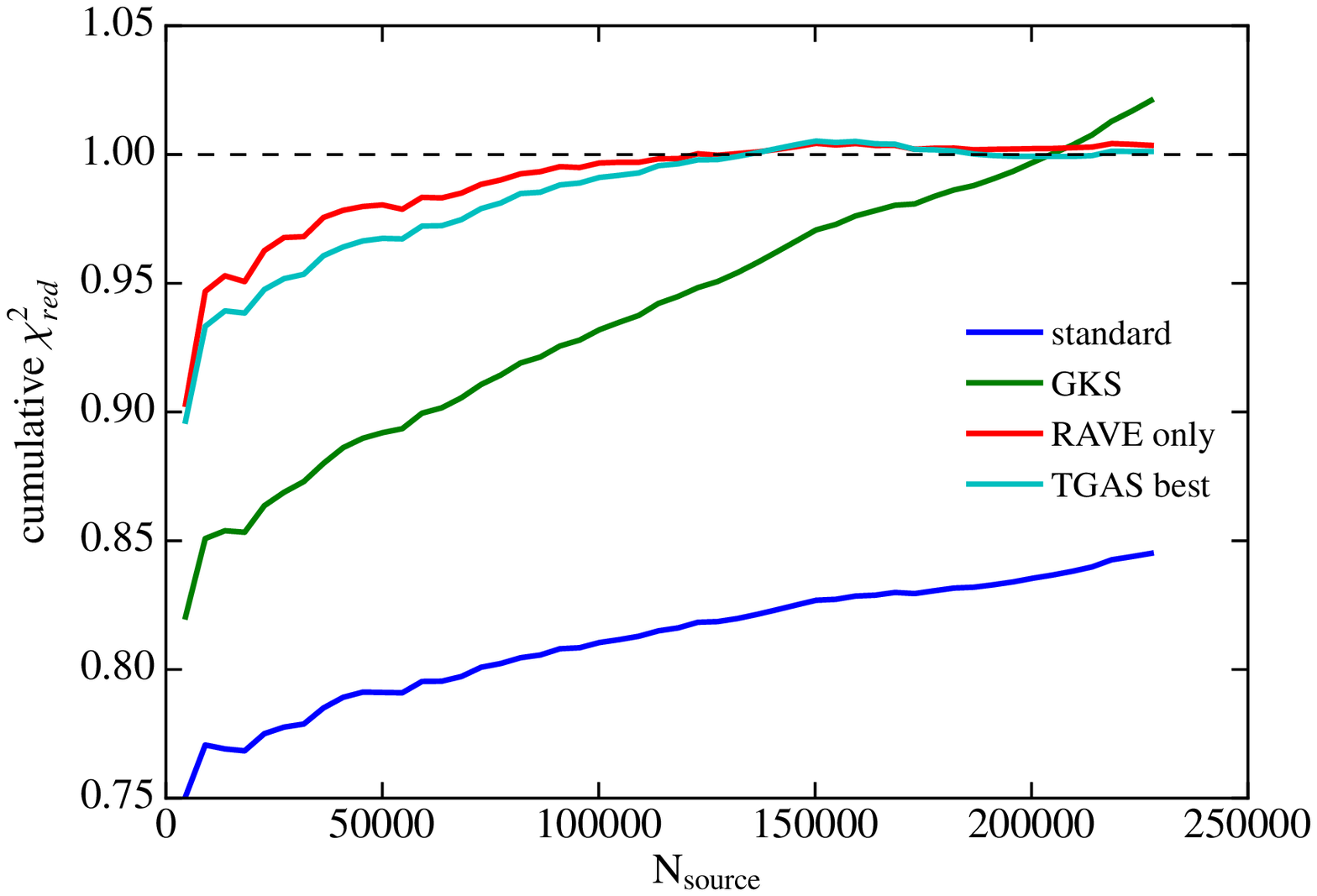}}}
  \caption{
  	The cumulative reduced $\chi^2$ of the TGAS parallax measurements with the new RAVE-only parallax estimates (Eq.~\ref{eq:redchi2}) when the uncertainties of one set or the other have revised downwards. The different lines correspond to the original values (dark blue), the correction from \protect\citet[green; $\redTGAS=0.79$ and $\zTGAS = 0.10$ in Eq.~\ref{eq:deflateTGAS}]{Goea16}, the best correction available when only `deflating' RAVE uncertainties (red, $\redRAVE=0.86$ in Eq.~\ref{eq:deflateRAVE}), and the best available only adjusting TGAS uncertainties (light blue; $\redTGAS=0.95$ and $\zTGAS = 0.20$)
  	   \label{fig:deflate}
}
\end{figure}

In Section~\ref{sec:improved} we used the TGAS parallaxes to investigate the RAVE distance estimation, but we can turn this around and use the RAVE distance estimation to learn about TGAS. TGAS is an early release of \Gaia\ data and is therefore expected to contain strong systematic errors \citep{GaiaDR1,GaiaDR1:TGAS}. Various studies have looked at these systematic errors \citep[including the \Gaia\ consortium itself:][]{GaiaDR1:Validation}, by comparison to distances derived for RR Lyrae stars \citep*{Goea16}, red clump stars \citep{Daea17,GoMo17} or eclipsing binaries \citep{StTo16} or, in the case of \cite{ScAu17}, using a statistical approach based on the correlations between velocity components produced by distance errors. Our approach allows us to study a large area in the southern sky using many sources, spanning a wide range in colour, without any assumptions about kinematics.

In Figure~\ref{fig:ComboDiffSky} we plot the average difference between the TGAS parallax and that from this study, binned on the sky. Zonal differences are unlikely to be produced by any particular issues with the RAVE distance estimation, but may be related to the way in which the sky has been scanned by \Gaia. We can clearly see a stripe showing a substantial difference at  $l\sim280^\circ$, which corresponds to a stripe near the ecliptic pole, as can seen when this diagram is shown in ecliptic coordinates. A similar figure was shown in \citet[fig. 28]{GaiaDR1:Validation}, using the RAVE DR4 parallax estimates, where this feature was attributed to the ``ecliptic scanning law followed early in the mission", and it was noted that a corresponding feature can be found in the median parallaxes of quasar sources. This is also likely to be related to the anomaly reported by \cite{ScAu17}.

We can also look again at the width of the distribution of $\Delta$. As we have seen already, the width of the distribution of $\Delta$, when comparing TGAS and DR5, is less than unity. In Figure~\ref{fig:ComboDiff} we show this width for all stars in our new RAVE-only parallax estimates, and it is again less than unity. This indicates that the uncertainties of one or other measurements have been overestimated. When we divide the distribution by quoted TGAS parallax uncertainty (Figure~\ref{fig:ComboDiffPar}) we can see that the problem is particularly acute for sources with small quoted TGAS uncertainties.


As discussed in \cite{GaiaDR1:TGAS}, uncertainties in the final TGAS catalogue are designed to be conservative,  and have been `inflated' from the formal uncertainties derived internally. This was to take account of uncertainties that are not allowed for in the formal calculation (such as contributions from uncertainties in \Gaia's calibration and attitude). The scheme used was derived from a comparison to the (independent) \Hipparcos\ parallaxes, and the quoted uncertainties were determined from the formal uncertainties using the formula
\[
\sigma_{\varpi, {\rm TGAS}}^2 = a^2 \varsigma_{\varpi, {\rm TGAS}}^2 + b^2
\]
where  $\varsigma_{\varpi, {\rm TGAS}}$ is the formal parallax error derived internally, $a=1.4$ and $b=0.2\mas$.

\cite{Goea16} looked at the reported parallaxes of RR Lyrae stars in TGAS, and used the known period-luminosity relationship for these stars to provide an independent estimate of the uncertainties in parallax. They found that for these sources $a=1.1$, $b=0.12\mas$ provides a better description of the true TGAS uncertainties, and therefore recommended that the TGAS parallax estimates should be reduced to a value $\sigma_{\varpi, {\rm TGAS, sc}}$ given by the formula 
\[ \label{eq:deflateTGAS}
\sigma_{\varpi, {\rm TGAS, sc}}^2 =  \redTGAS^2 \sigma_{\varpi, {\rm TGAS}}^2 - \zTGAS^2
\]
with $\redTGAS=0.79$ and $\zTGAS = 0.10$. They investigated this by looking at how the sum of values of (their equivalent to) $\Delta^2$ varied as they increased the number of values that they summed over (ordered by nominal parallax uncertainty). This was done in the expectation that it should increase linearly with slope unity. For ease of plotting we consider the closely related statistic
\[ \label{eq:redchi2}
\chi^2_{\rm red,n} = \frac{1}{n} \sum_i^n \Delta_i^2
\]
where the sum is over the $n$ sources with the lowest quoted TGAS uncertainty. This should have a constant value of unity as we sum over increasing numbers of sources.

\begin{figure*}
  \centerline{
    \resizebox{0.5\hsize}{!}{\includegraphics{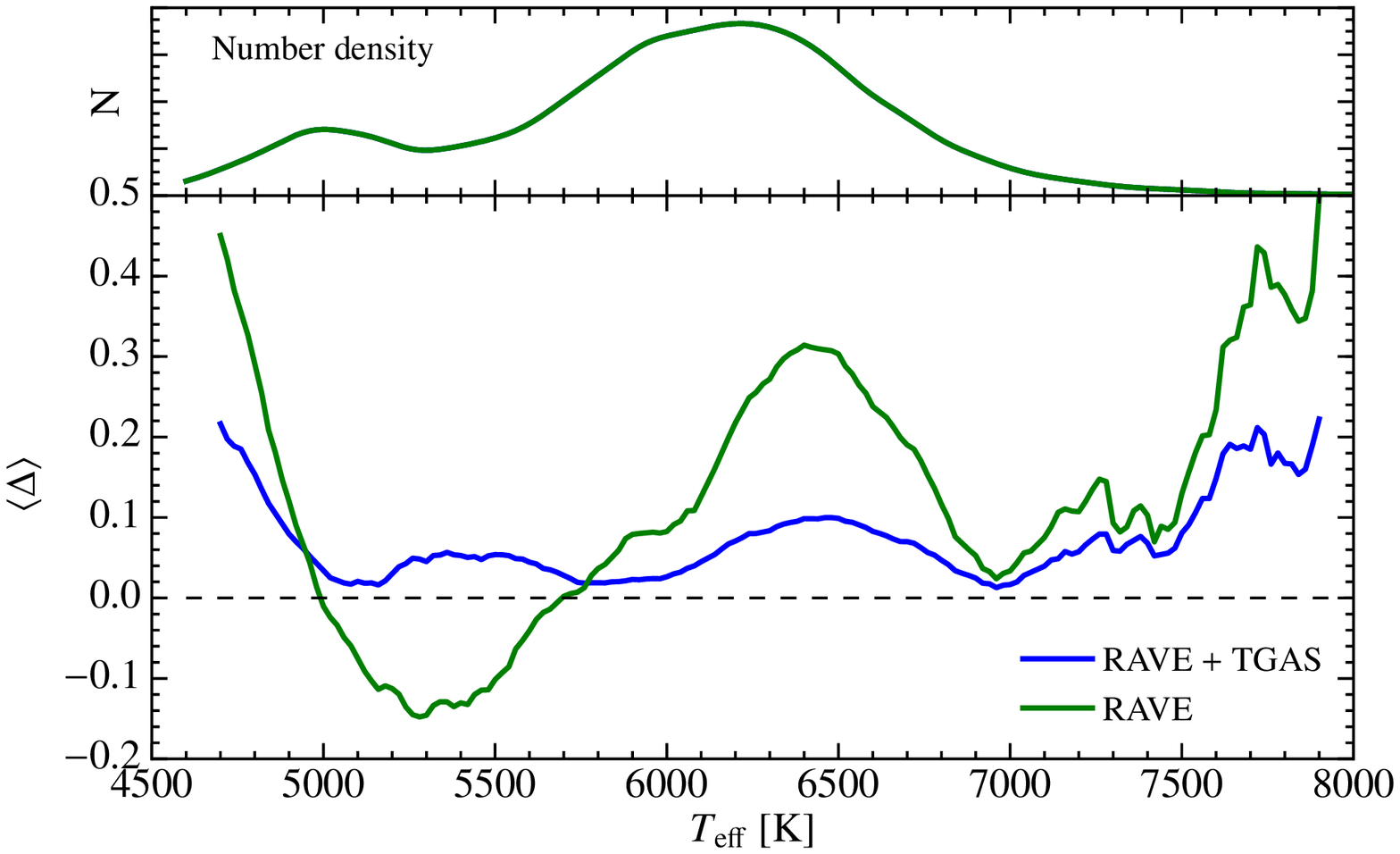}}
    \resizebox{0.5\hsize}{!}{\includegraphics{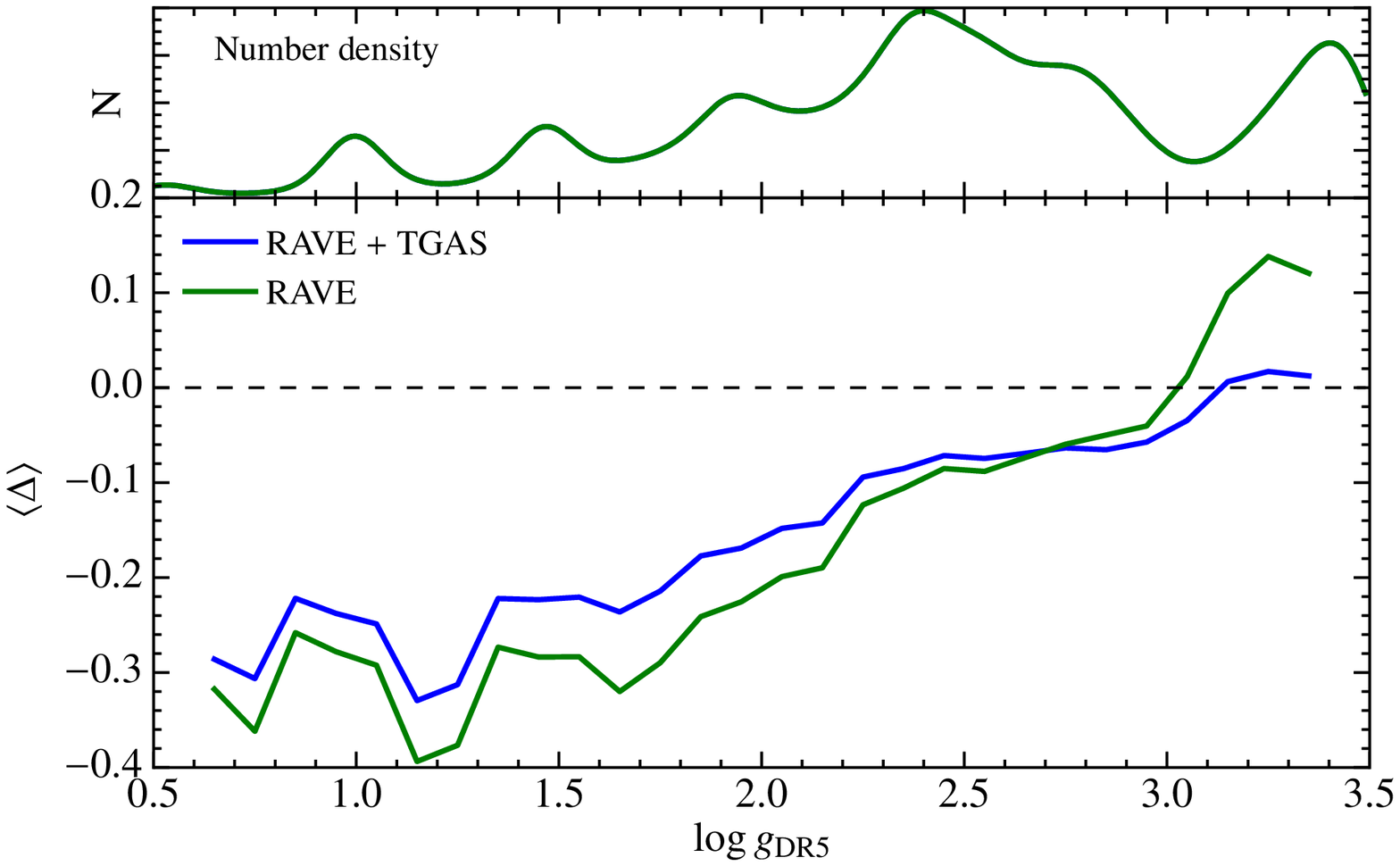}}}
  \caption{
  The variation of the average value of $\Delta$ (Eq.~\ref{eq:Delta}) as a function of $\teff$ for dwarfs ($\log g\ge3.5$, left) and as a function of $\log g$ for giants ($\log g<3.5$, right). The $\teff$ values come from the IRFM. In blue and labelled RAVE+TGAS we show our combined parallax estimates -- we also show the RAVE only estimates (using IRFM $\teff$ values, in green and labelled RAVE, and as shown in Figures~\ref{fig:NewfT}~and~\ref{fig:Newfg}) to guide the eye. For low $\log g$ giants, TGAS parallax uncertainties are too large to have a significant effect on the bias seen in RAVE.
  \label{fig:FinalDelta}
}
\end{figure*}

In Figure~\ref{fig:deflate} we show that, if we use the quoted uncertainties for both RAVE and TGAS, $\chi^2_{\rm red}$ remains smaller than unity for all stars. If we use the prescription from \cite{Goea16} then we come closer to unity when we consider all stars, but  $\chi^2_{\rm red}$ is clearly less than unity where the sum is over the stars with lower  $\sigma_{\varpi, {\rm TGAS}}$. This suggests that the \citeauthor{Goea16} prescription gives uncertainties that are still overestimated for stars with small $\sigma_{\varpi, {\rm TGAS}}$ and underestimated for those with large $\sigma_{\varpi, {\rm TGAS}}$. 

Figure~\ref{fig:deflate} also shows two alternative scenarios. We show $\chi^2_{\rm red}$ corresponding to the best values of $\redTGAS$ and $\zTGAS$ (assuming that the RAVE uncertainties are correct), which are  $\redTGAS=0.95$ and $\zTGAS = 0.20$, which corresponds to $b=0$. Even when we do this (i.e, set the minimum uncertainty from TGAS to zero), the combined uncertainty for the stars with the lowest TGAS uncertainties is clearly too large. We therefore also consider the effect of arbitrarily reducing the RAVE uncertainties according to the formula
\[ \label{eq:deflateRAVE}
\sigma_{\varpi,{\rm RAVE, sc}} =  \redRAVE\; \sigma_{\varpi, {\rm RAVE}},
\]
and find that a value of  $\redRAVE=0.86$ (while keeping the quoted TGAS uncertainties) produces results that are roughly as good as the results we find when deflating the TGAS uncertainties. It is worth noting that, like the TGAS uncertainties, the RAVE stellar parameter uncertainties were designed to be conservative \citep{RAVEDR5}.

It is possible that the RAVE uncertainties tend to be overstated, particularly if the quoted external uncertainty estimates are overstated for most stars. While it certainly would not produce a systematic overestimate that was well described by eq~\ref{eq:deflateRAVE}, it could affect our estimates in a more complicated and subtle way. We are therefore not in a position to determine for sure whether it is the RAVE uncertainties or TGAS uncertainties (or both) that are overestimated. We would note that a comparison of DR5's parallax estimates to those from \Hipparcos\ did not suggest underestimated uncertainties in either instance \citep[][fig 25]{RAVEDR5}. We add that the dispersion in $\Delta$ is smaller than unity for both giants and dwarfs, considered independently. We conclude that our results are consistent with the TGAS uncertainties being underestimated, though probably not in quite the same way as the prescription of \cite{Goea16}. We will not attempt to correct for any overestimates of uncertainty when calculating the combined RAVE+TGAS estimates below.

\section{Combined distance estimates} \label{sec:Combined}

A fundamental element of Bayesian analysis is the updating of the probability of a hypothesis (for example, the hypothesised distance to a star) as more evidence becomes available. TGAS parallaxes provide new evidence regarding these distances, so we are required to take it into account when determining the distances. We can think of this as either an additional piece of input data, or as a prior on parallax for each star (in addition to the prior on distance implied by Equation~\ref{eq:priorofx}) -- the two statements are equivalent.

In previous sections we have investigated the properties of the RAVE distance pipeline in the absence of TGAS parallaxes, and developed an understanding of the problems with each dataset. We now incorporate the new evidence from these parallax measurements to obtain more accurate distance estimates than either can provide in isolation. We do this by including them in the set of inputs $(O_i,\sigma_i)$ in Equation~\ref{eq:maths}. It can be expected that the impact of the TGAS parallaxes will be greatest at $\teff$ values below the turnoff where there is serious ambiguity whether a star is on the main sequence or ascending the giant branch -- an uncertainty which is reflected in the bimodal pdfs which we are forced to represent using multi-Gaussian fits (Eq.~\ref{eq:defsfk}).

We have seen that the parallax estimates (from RAVE alone) for stars with $\log g$ values less than $\sim$$2.0$ appear to be particularly biassed in the sense that they are systematically lower than those found by TGAS. It is very likely that this is due to the RAVE $\log g$ values in this range being systematically underestimated, as is also suggested by a comparison to the $\log g$ values found by GALAH or APOGEE surveys for the same stars. We noted in Section~\ref{sec:Giants} that the TGAS parallax uncertainties for these stars are significantly larger than those found from the RAVE distance pipeline. Therefore we can not expect that our distance estimates for these stars are significantly de-biassed by including the TGAS parallax in the estimate.

In Figure~\ref{fig:FinalDelta} we show the average value of $\Delta$ (again as a function of $\teffIRFM$ for dwarfs and $\log g_{\rm DR5}$ for giants) for the combined parallax estimates, with the RAVE-only distance estimates (also using the IRFM temperatures) shown for comparison. One must be very careful not to  over-interpret these plots for several reasons (e.g., the RAVE+TGAS parallaxes are obviously not independent of the TGAS ones; the uncertainties for RAVE+TGAS, which enter the calculation of $\Delta$, are generally much smaller than those of RAVE alone), but they clearly indicate that the difference at low $\log g$ values is not removed when we include the TGAS parallax information. 

\subsection{Improvement} \label{sec:Improve}

\begin{figure}
   \centerline{
    \resizebox{\hsize}{!}{\includegraphics{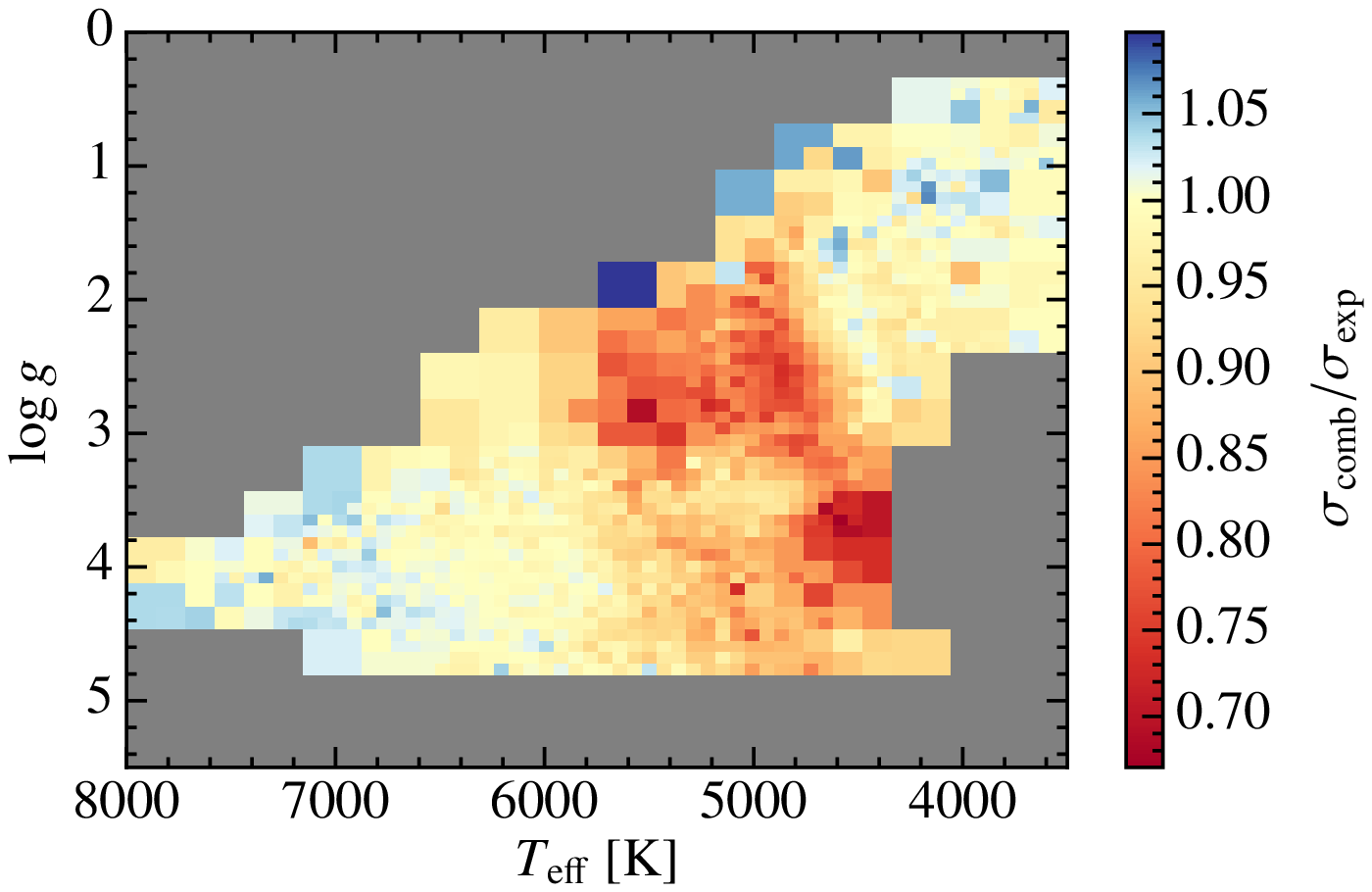}}}
    \centerline{
     \resizebox{\hsize}{!}{\includegraphics{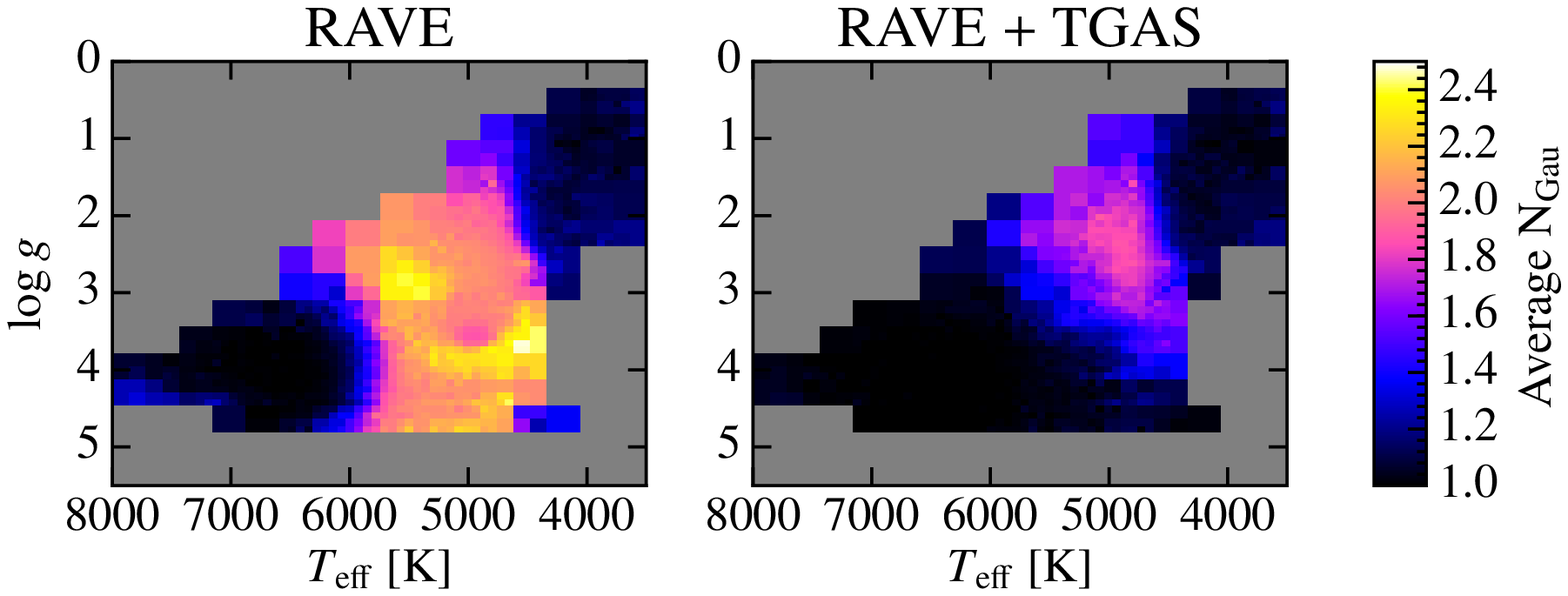}}}
  \caption{
  	The top panel shows the variation over the HR diagram of the ratio of the actually quoted uncertainty on the parallax when combining TGAS and RAVE data and the expected parallax uncertainty (Eq.\ref{eq:naive}) assuming Gaussian uncertainties. The $\teff$ values come from the IRFM.  In the region between the dwarf and giant branches and in the red clump the improvement on naive expectations is particularly clear. The lower panels provide an explanation: they show the number of Gaussian components required to represent the pdf in distance modulus (Eq.~\ref{eq:defsfk}) without TGAS parallaxes (left) and with them (right). Without the TGAS parallaxes we require a multi-Gaussian representation in $\sim$$45$ percent of cases, whereas with TGAS we only need it in $\sim$$23$ percent of cases.
	  	   \label{fig:Improvement}
}
\end{figure}

\begin{figure}
   \centerline{
    \resizebox{\hsize}{!}{\includegraphics{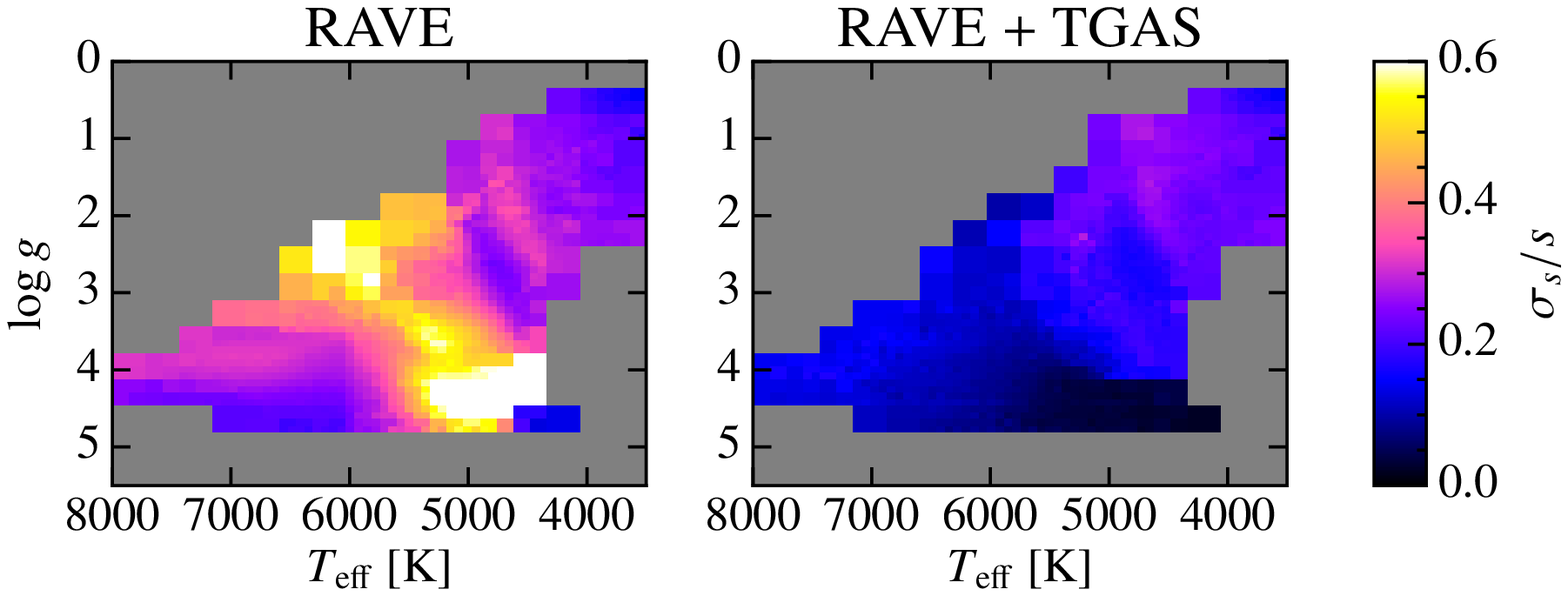}}}
  \caption{
  	Average fractional distance uncertainties across the HR diagram when we ignore TGAS (left) and when we use the TGAS parallax information (right). The improvement is particularly dramatic for cooler dwarfs and stars with $\teff\sim6000\,\mathrm{K}$, $\log g\sim2.5$. The $\teff$ values come from the IRFM. For low $\log g$ giants, the inclusion of TGAS parallaxes has little effect.
  \label{fig:ComboUncertHR}
}
\end{figure}

\begin{figure}
  \centerline{
    \resizebox{\hsize}{!}{\includegraphics{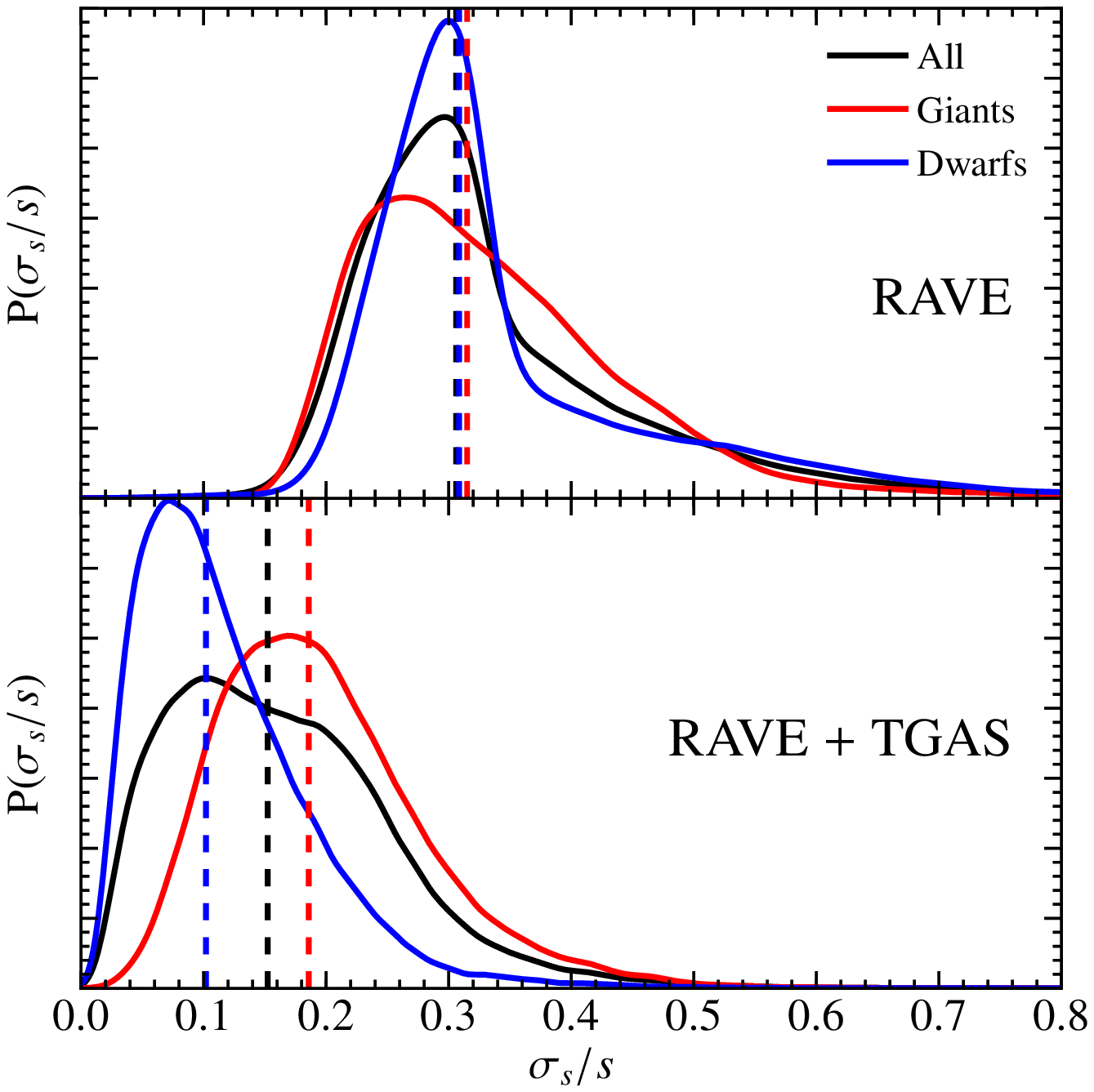}}}
  \caption{
  Fractional distance uncertainties for sources when we ignore TGAS parallaxes (upper panel) and when we use TGAS parallaxes (lower panel). In each case we show the pdfs for all sources (black), and separate ones for giants ($2.0<\log g<3.5$, red) and dwarfs ($\log g\ge3.5$, blue). The dashed lines show the median values in each case, (0.33 and 0.16 without TGAS and with TGAS, respectively) for all stars (i.e. 51 percent smaller with TGAS), 0.36 and 0.20 for giants (44 percent smaller) and 0.31 and 0.10 for dwarfs (66 percent smaller).
      	  	   \label{fig:ComboUncertDist}
}
\end{figure}

\begin{figure}
  \centerline{
    \resizebox{\hsize}{!}{\includegraphics{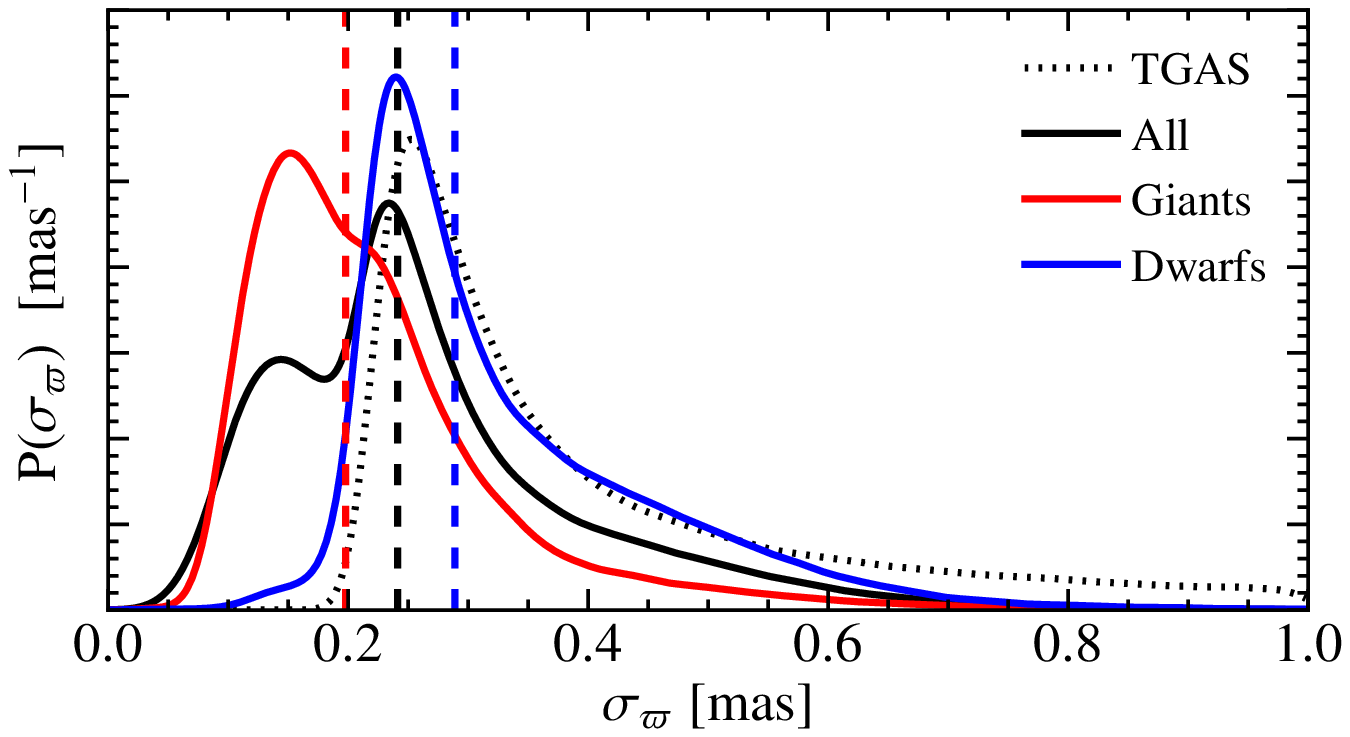}}}
  \caption{
  Parallax uncertainties when using the RAVE pipeline with TGAS parallaxes. The dotted curve is the pdf for all stars using just TGAS. The solid lines show the pdfs for all sources (black), and separate ones for giants ($2.0<\log g<3.5$, red) and dwarfs ($\log g\ge3.5$, blue). The dashed lines show the median values in each case which can be compared to the median TGAS uncertainty for these stars, which is 0.32$\mas$ (essentially independently of whether stars are dwarfs or giants). This median is 0.25$\mas$ for all stars (24 percent smaller than TGAS), 0.15$\mas$ for giants (54 percent smaller) and 0.29$\mas$ for dwarfs (9 percent smaller).
    \label{fig:ComboUncertPar}
}
\end{figure}

Including the TGAS parallaxes in our distance estimation inevitably leads to an improvement in the formal uncertainties. From the discussion of the previous sections, we can claim with some confidence that, outside a few regions of parameter space (e.g., low $\log g$, the stripe near the ecliptic pole), the combination does not introduce significant systematic errors into one dataset or the other.

We can make a na\"ive estimate of how the uncertainties will decrease when we combine the two datasets by approximating that the uncertainties from the RAVE-only distance pipeline ($\sigma_{\varpi,{\rm Sp}}$) are Gaussian, in which case we have a new expected uncertainty in parallax $\sigma_{\varpi,{\rm exp}}$ given by
\[ \label{eq:naive}
1/{\sigma^2_{\varpi,{\rm exp}}} = 1/{\sigma^2_{\varpi,{\rm Sp}}} + 1/{\sigma^2_{\varpi,{\rm TGAS}}}.
\]
Because the RAVE uncertainties are significantly non-Gaussian, we do significantly better than this is some regions of the HR diagram. This can be seen in Figure~\ref{fig:Improvement}, which shows the parallax uncertainty we find divided by that which we would naively expect. This is also reflected in the reduced number of stars for which the multi-Gaussian representations are required to describe the distance pdf (lower panel of Figure~\ref{fig:Improvement}). 

 In Figure~\ref{fig:ComboUncertHR} we show how the fractional distance uncertainty varies over the HR diagram, both with and without TGAS parallaxes. It is clear that the main improvement is for dwarfs, and for stars in the regions of the HR diagram where parallax information can break uncertainties regarding whether a star is a giant or a dwarf.

When we include TGAS parallaxes, the median fractional distance uncertainty (excluding stars with $\log g_{\rm DR5}<2.0$) falls to 15 percent, from 31 percent using spectrophotometric information alone. For dwarfs the median uncertainty is just 10 percent, while for giants it is 19 percent. The full pdfs of fractional distance uncertainty are shown in Figure~\ref{fig:ComboUncertDist}.

The improvement over TGAS alone is shown in terms of parallax uncertainty in  Figure~\ref{fig:ComboUncertPar}. In this case it is the giants for which the greatest improvement is found (again excluding stars with $\log g_{\rm DR5}<2.0$). The median TGAS uncertainty is $0.32\mas$ for either giants or dwarfs, while the median uncertainty for RAVE+TGAS is $0.20\mas$ for giants, and $0.24\mas$ for dwarfs.

\begin{figure}
  \centerline{
    \resizebox{\hsize}{!}{\includegraphics{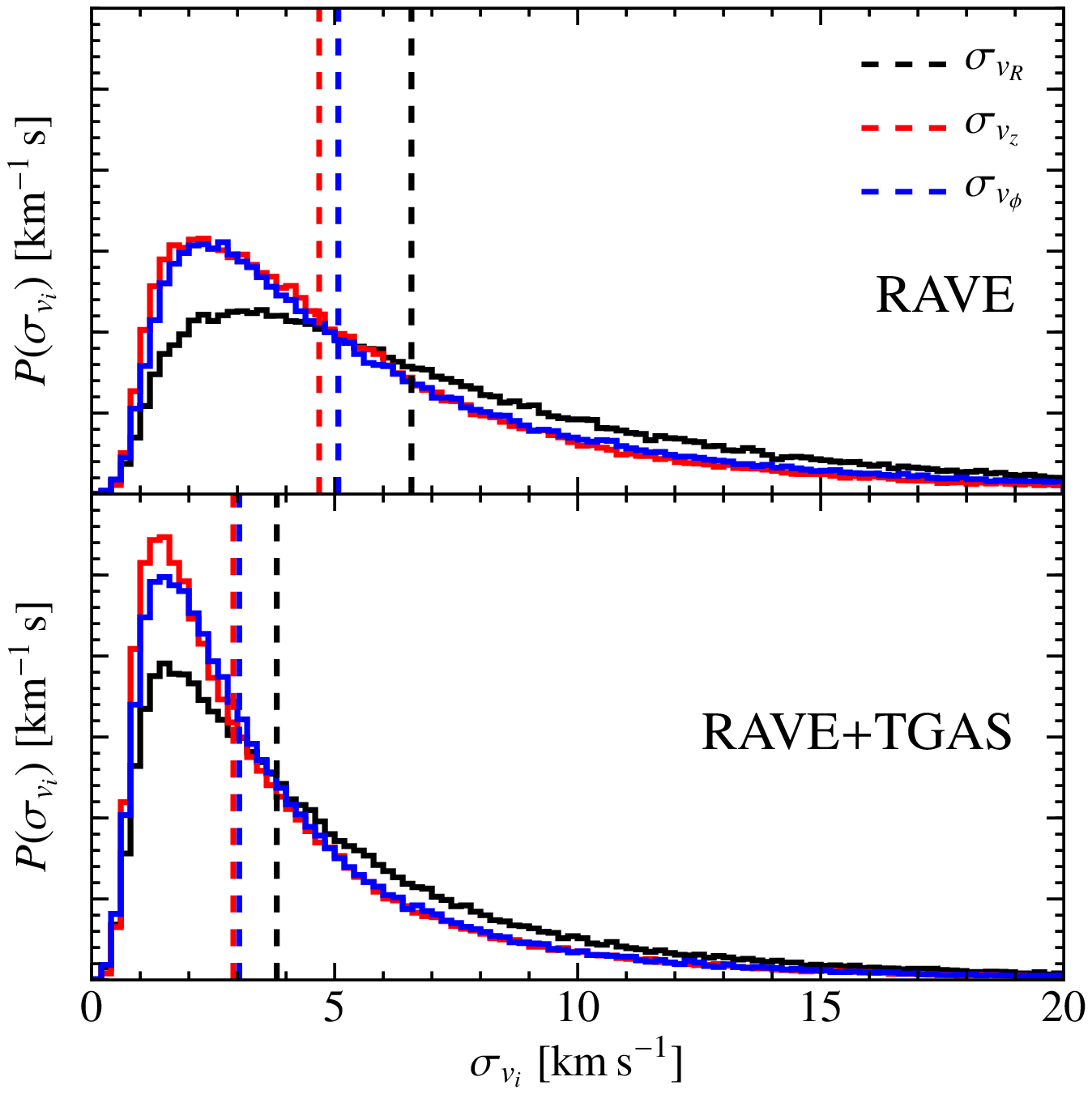}}}
  \caption{
Velocity uncertainties for sources when we ignore TGAS parallaxes (upper panel) and when we use TGAS parallaxes (lower panel). In each case we show the pdfs in $v_R$ (black), $v_z$ (red) and $v_\phi$ (blue). The dashed lines show the median values in each case, which are $6.6$ and $3.8\kms$ (without TGAS and with TGAS, respectively) for $v_R$, $4.7$ and $2.9\kms$ for $v_z$ and $5.1$ and $3.0\kms$ for $v_\phi$, i.e. the velocity uncertainty in each direction is reduced by $\sim 40$ percent. 
     	  	   \label{fig:VelocityUncert}	
}
\end{figure}

Using our combined estimates and the TGAS proper motions, we can convert this distance uncertainty into a velocity uncertainty. We take a simple Monte-Carlo approach to do this -- for each star we sample from the multi-Gaussian pdf in distance modulus, and from Gaussians in proper motion and radial velocity with the quoted uncertainties. We again assume that the Sun is $8.33\kpc$ from the Galactic centre and $15\pc$ from the Galactic plane.
If we characterise the resulting pdf in terms of a median value and a standard deviation (i.e. uncertainty) in each Galactocentric velocity component, we get the distribution of uncertainties shown in Figure~\ref{fig:VelocityUncert}. The introduction of TGAS parallaxes to our distance estimates improves the velocity accuracy by, on average, $\sim40$ percent in each direction.

{\breferee Finally we would like to estimate how we could correct our distance estimates to be unbiassed. Since we don't know the true values we will do this under the assumption that the TGAS values are unbiassed. We make the further approximation that -- at a given $\teff$ value for dwarfs or $\log g$ value for giants -- we can simply multiply all our RAVE+TGAS parallaxes by a correction factor ${\rm corr}_\varpi$ such that they are unbiassed. For values of  ${\rm corr}_\varpi\approx1$ it follows that the equivalent factor for distances is ${\rm corr}_s\approx2-{\rm corr}_\varpi$. We find the value of ${\rm corr}_\varpi$ by requiring that our statistic $\langle\Delta\rangle$ is zero if we compare ${\rm corr}_\varpi\varpi_{\rm RAVE+TGAS}$ and $\varpi_{\rm TGAS}$}

\begin{figure}
    \centerline{
    \resizebox{\hsize}{!}{\includegraphics{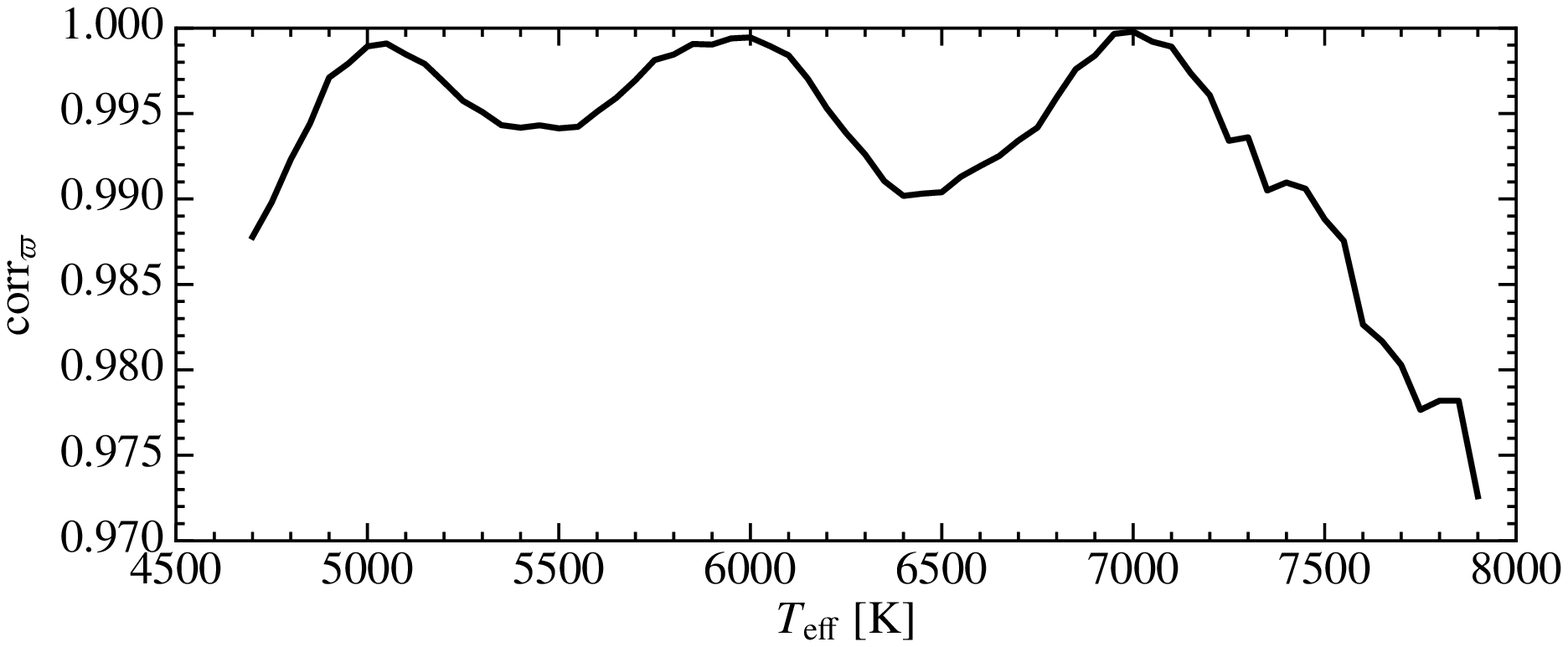}}}
  \centerline{
    \resizebox{\hsize}{!}{\includegraphics{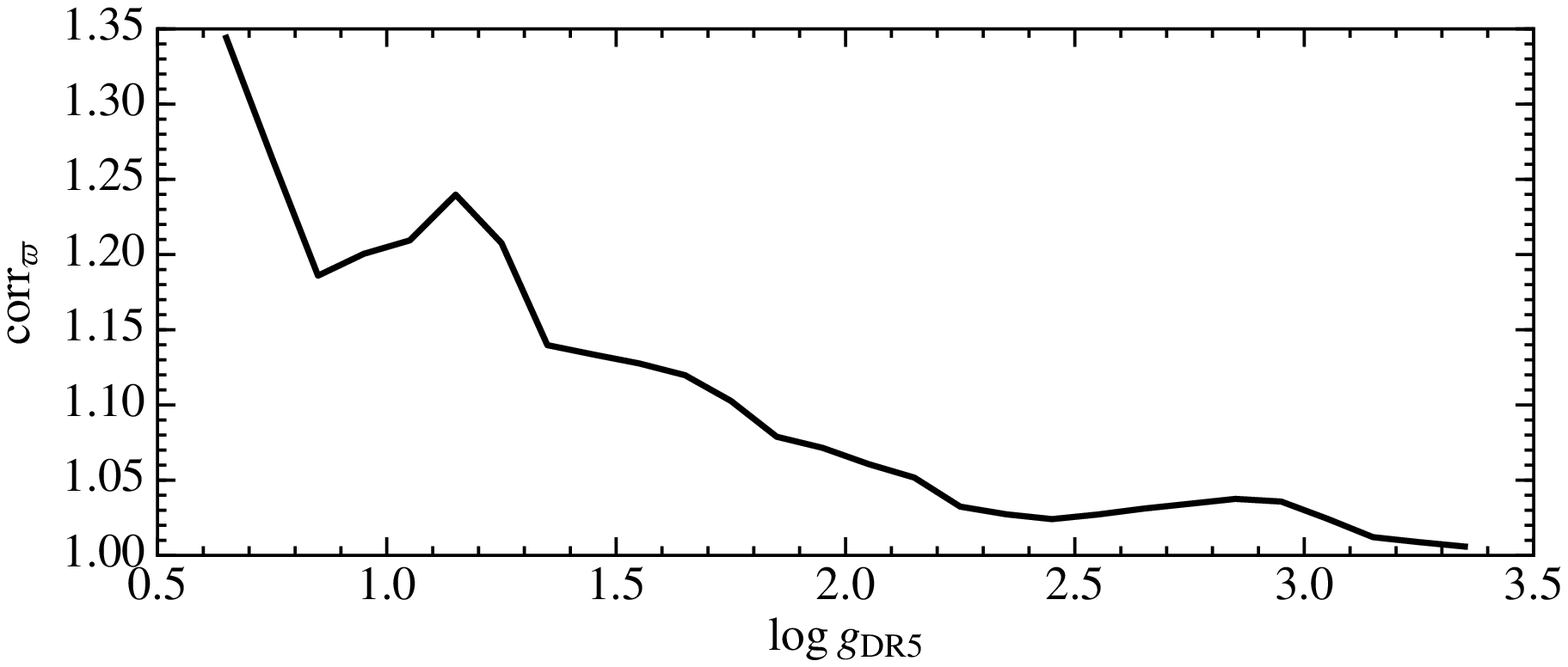}}} 
  \caption{
{\breferee Estimated parallax correction factors (${\rm corr}_\varpi$) for the RAVE+TGAS combined parallax estimates as a function of $\teff$ for dwarfs ($\log g\geq3.5$, upper) and $\log g$ for giants ($\log g<3.5$, lower).  Values are calculated in as a running average over a window of width $200\,{\rm K}$ or $0.3\dex$. If we multiply all the RAVE+TGAS $\varpi$ values in this window by ${\rm corr}_\varpi$, then $\Delta$ is, on average zero.}
     	  	   \label{fig:Corr}	
}
\end{figure}

{\breferee Figure~\ref{fig:Corr} shows the value of ${\rm corr}_\varpi$ we find as a function of $\teff$ value for dwarfs and $\log g$ for giants. The dwarfs require systematic changes of less than 1 percent in parallax (or distance) for all but the hottest stars. The giants seem to require systematic changes of more than 10 percent in parallax at $\log g<2.0$, up to around 35 percent at the lowest $\log g$ values. For these low $\log g$ stars, the approximation ${\rm corr}_s\approx2-{\rm corr}_\varpi$ becomes poor. }

\section{Age estimates} \label{sec:ages}

The classical method for determining the age of a star is by comparing the luminosity of an F or G star to that expected for stars of its colour on the main-sequence or turning off it. This is only possible if an independent estimate of its distance (e.g., a parallax) is available. By including TGAS parallaxes in the RAVE pipeline we are making precisely this comparison, with additional information and a sophisticated statistical treatment. We can therefore expect that the ages we derive are as reliable as any currently available for main-sequence stars.

While the original aim of this pipeline was to determine the distances to stars, an inevitable byproduct is that we also constrain the other `model' properties described in Section~\ref{sec:Bayes}, i.e., initial mass
${\cal M}$, age $\tau$, metallicity $\mh$ and line-of-sight extinction $A_V$. We can also produce new estimates of the other properties of the stars, such as $\teff$ and $\log g$, which we discuss below. 

Here we look at the improved estimates of $\tau$ that are made possible by including TGAS parallaxes. Age estimates from this pipeline were included in RAVE DR4 (in terms of $\log \tau$), but came with the strong caveat that the prior used (the Standard prior -- see our Section~\ref{sec:prior}) included a fixed relationship between metallicity and age (metal-poor stars are assumed to be old, metal-rich stars younger). In our case, we have now seen that we can use a prior without any explicit age-metallicity relationship and still produce reasonable results (at least in terms of parallaxes -- Figure 
\ref{fig:ComparePriors}). This gives us some confidence that we will not go too badly wrong using this prior when deriving ages.

\begin{figure}
   \centerline{
    \resizebox{\hsize}{!}{\includegraphics{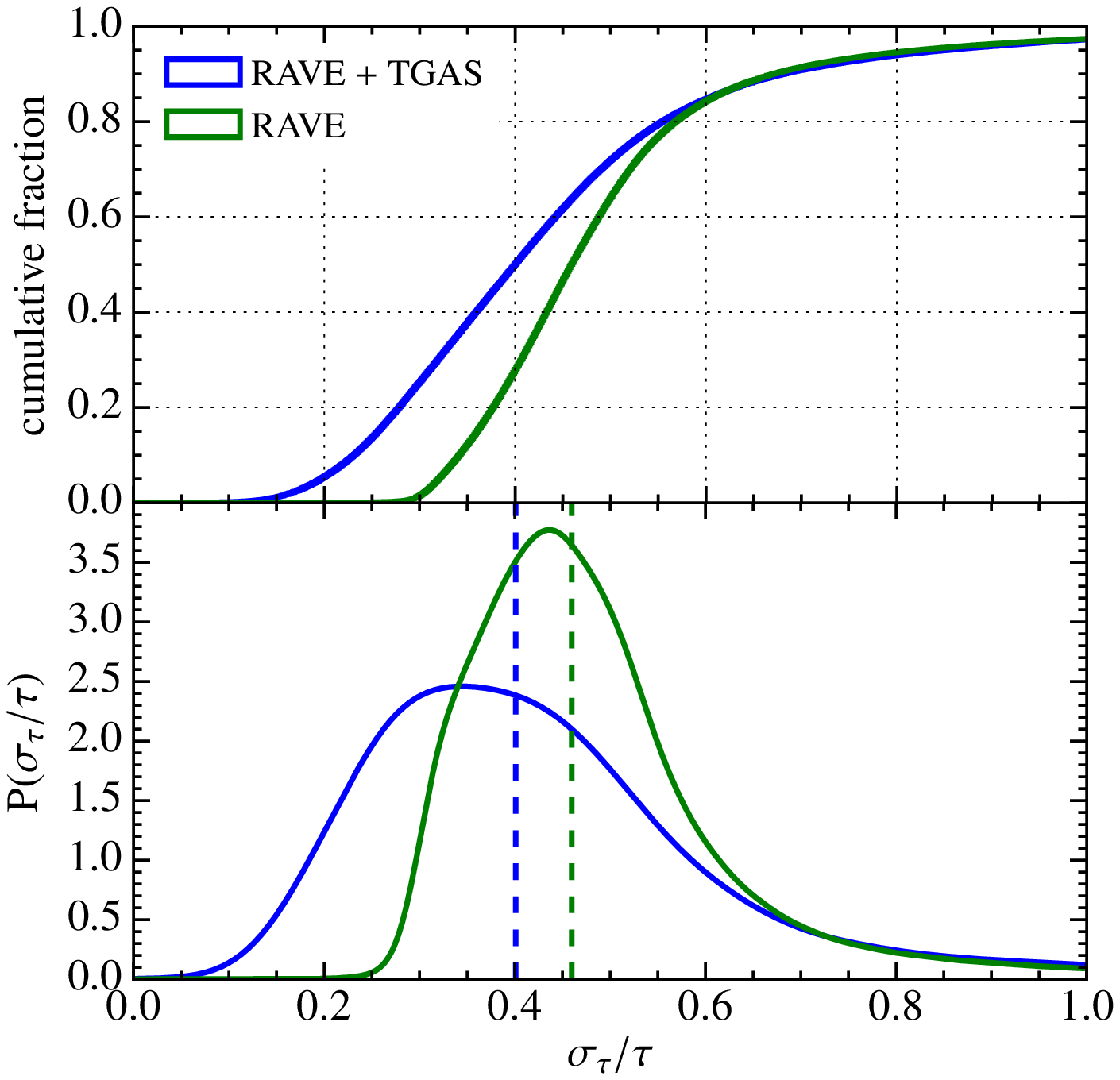}}}
  \caption{
  	The fraction of sources with a given fractional age uncertainty ($\sigma_\tau/\tau$) displayed as a pdf found using a kernel density estimation (lower panel), and as a cumulative distribution (upper panel). The median values are plotted as dashed lines. The plot shows the distribution of age uncertainties with and without TGAS parallaxes (blue and green curves, respectively). It is particularly clear that the inclusion of TGAS parallaxes allows us to derive age uncertainties of less than $30$ percent for a significant fraction of sources. 
	  	   \label{fig:AgeUncertDist}
}
\end{figure}

\begin{figure*}
  \centerline{
    \resizebox{0.33\hsize}{!}{\includegraphics{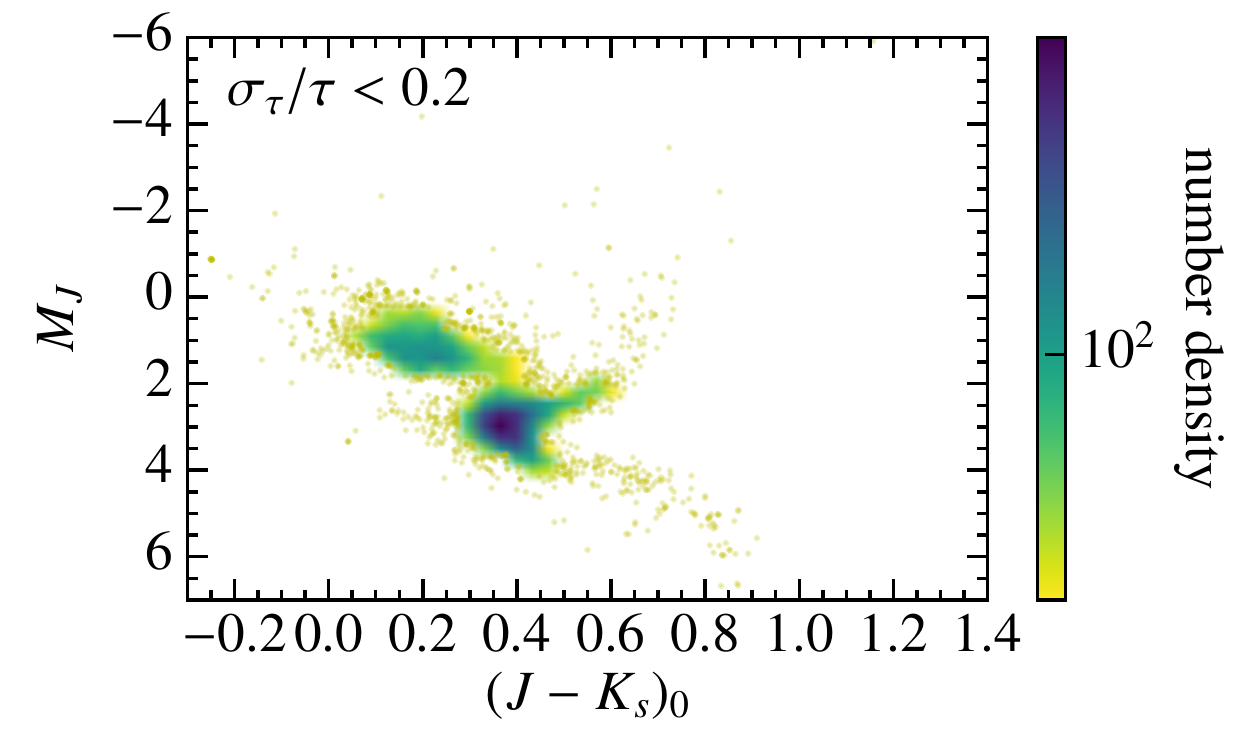}}
    \resizebox{0.33\hsize}{!}{\includegraphics{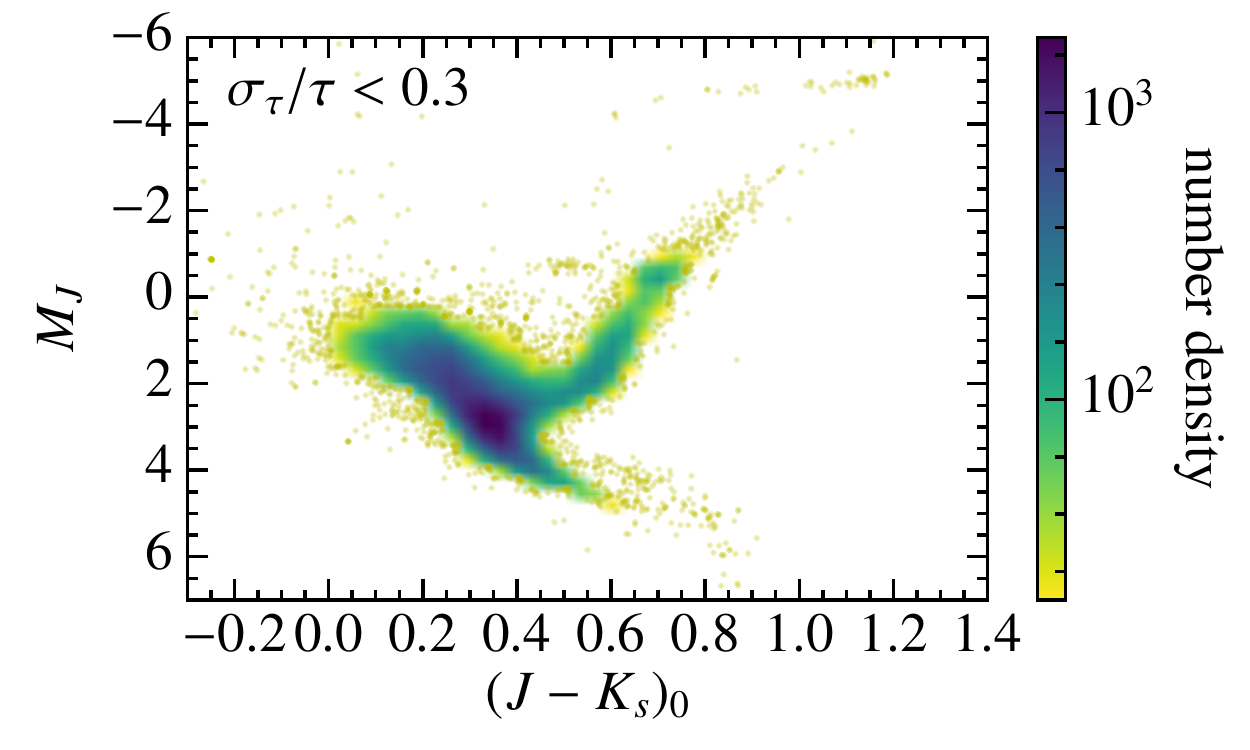}}
    \resizebox{0.33\hsize}{!}{\includegraphics{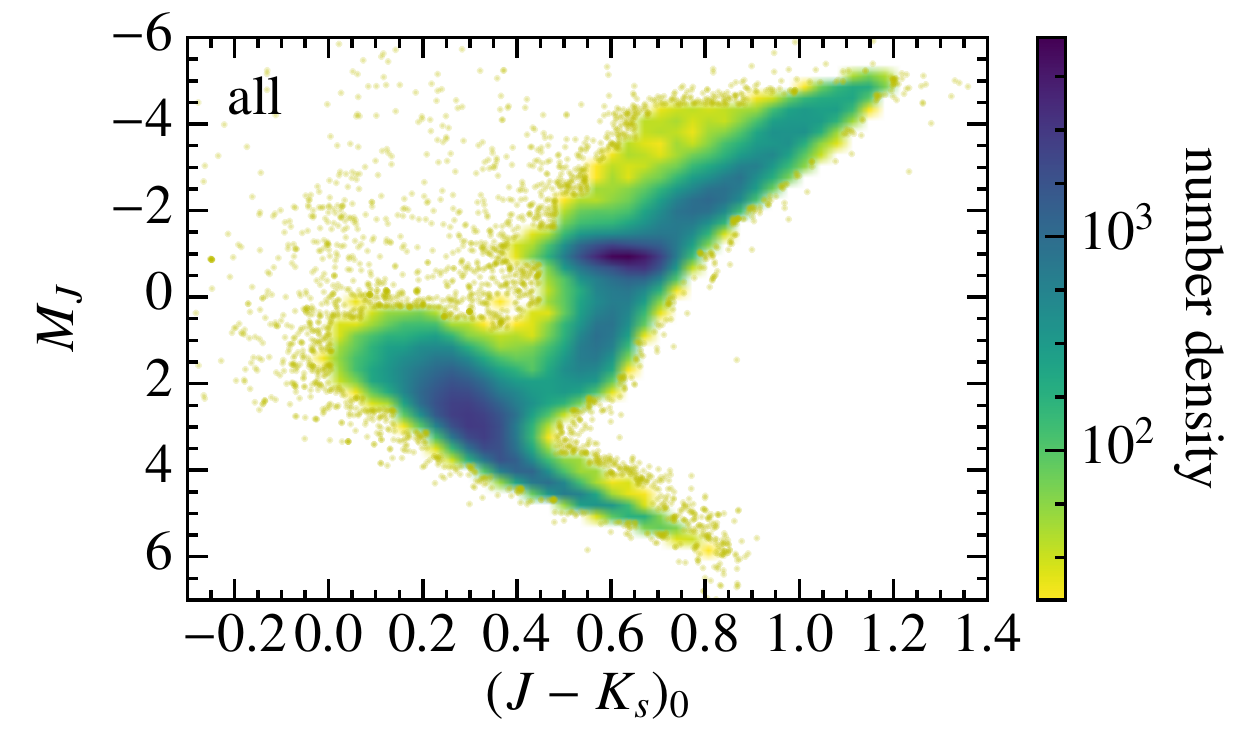}}}
  \caption{
  	The location of stars with small fractional age uncertainties in the HR diagram (with colour and absolute magnitude on the two axes in this case). Both the $(J-K_s)$ colour and the absolute magnitude in the $J$ band, $M_J$, have been corrected for extinction using the most likely $\log A_V$ value found by the distance pipeline.   The left figure shows those with age uncertainties less than 20 percent, the central figure those with age uncertainties less than 30 percent, and the right figure shows all stars (for comparison). The number density indicated by the colour bar corresponds to the numbers of stars in a pixel of height 0.1 magnitudes in $M_J$ and width 0.01 magnitudes in $(J-K_s)_0$. Unsurprisingly, the smallest fractional age uncertainties are for stars near the main-sequence turnoff.
	  	   \label{fig:AgeUncertHR}
}
\end{figure*}

We would expect that the addition of the TGAS parallax measurements provides us with substantial leverage when determining the ages of stars, and in Figure~\ref{fig:AgeUncertDist} we quantify this. It is clear that, particularly at the low-uncertainty end, we do have a substantial improvement in precision. Without TGAS only 1.5 percent of stars have fractional age uncertainties lower than 0.3, while with TGAS this increases to over 25 percent. In Figure~\ref{fig:AgeUncertHR} we show where in the HR diagram the stars with the smallest age uncertainties are found. As one would expect, they are primarily found near the main-sequence turnoff --  it is in this region that stars evolve quickly with age, and it is therefore possible to get an age estimate with small uncertainties even with imperfect observations.

{\breferee It is clear from Figure~\ref{fig:FinalDelta} that there are still some biasses in the distance estimates for dwarfs with $6000\lesssim\teff\lesssim7000\Kelvin$ (i.e.~in the main-sequence turnoff region), though Figure~\ref{fig:Corr} suggests that these are only at the 1 percent level. It is reasonable to ask whether this implies a bias in the age estimates. We can not know for sure, because we do not know what causes the bias. We have investigated the possible biasses by running the pipeline having either artificially decreased the input $\log g$ values by $0.4\dex$ or artificially decreased the input parallaxes by $50\muas$. Either change results in parallax estimates that are biassed in the opposite sense to that seen with the real data for these stars. In both cases the changes in stellar ages are small compared to the uncertainties. The change in input $\log g$ produces a typical change of $\sim0.5\Gyr$ (or 5 to 10 percent) but with no trend to higher or lower ages (i.e. no clear bias). The change in input $\varpi$ produces a smaller typical change of $\sim0.2\Gyr$ (or $\sim$4 percent) with a bias in the sense that cooler stars (with $\teff\sim6000$) have slightly lower ages than those originally quoted, by  $\sim0.1\Gyr$ (or $\sim$2 percent). These are negligible for most purposes, but it is entirely possible that other biasses in the analysis (for example in the metallicities or the stellar isochrones) have larger impacts on the age estimates. The study of the complex interplay of these different factors is beyond the scope of this paper.
}

We must caution that these age estimates are extremely hard to verify from external sources. A relatively small number of sources have age estimates from asteroseimology studies or because they are part of clusters with known ages,  and these are sources for which we have large age uncertainties. We can gain confidence from the facts that 1) the method we are using to determine distances and ages has been carefully tested with pseudo-data for accuracy by, amongst others, \cite{BuJJB10}; and 2) the application of this method to these data to find distances has been rigorously tested against TGAS parallaxes in this study, and we have found that it is generally successful (except for $\log g<2.0$ stars, where we believe that the problem lies in the quoted $\log g$ values).

In a forthcoming paper (Wojno et al., MNRAS submitted) we will use these age estimates to isolate young and old populations in the RAVE catalogue and study their properties. This will demonstrate the improvement over past studies \citep{Woea16} in understanding the relationship between age, metallicity and velocity of stars in the Solar neighbourhood which is made possible by the TGAS parallaxes.

\section{Stellar parameters} \label{sec:reverse}

\begin{figure}
  \centerline{
   \resizebox{\hsize}{!}{\includegraphics{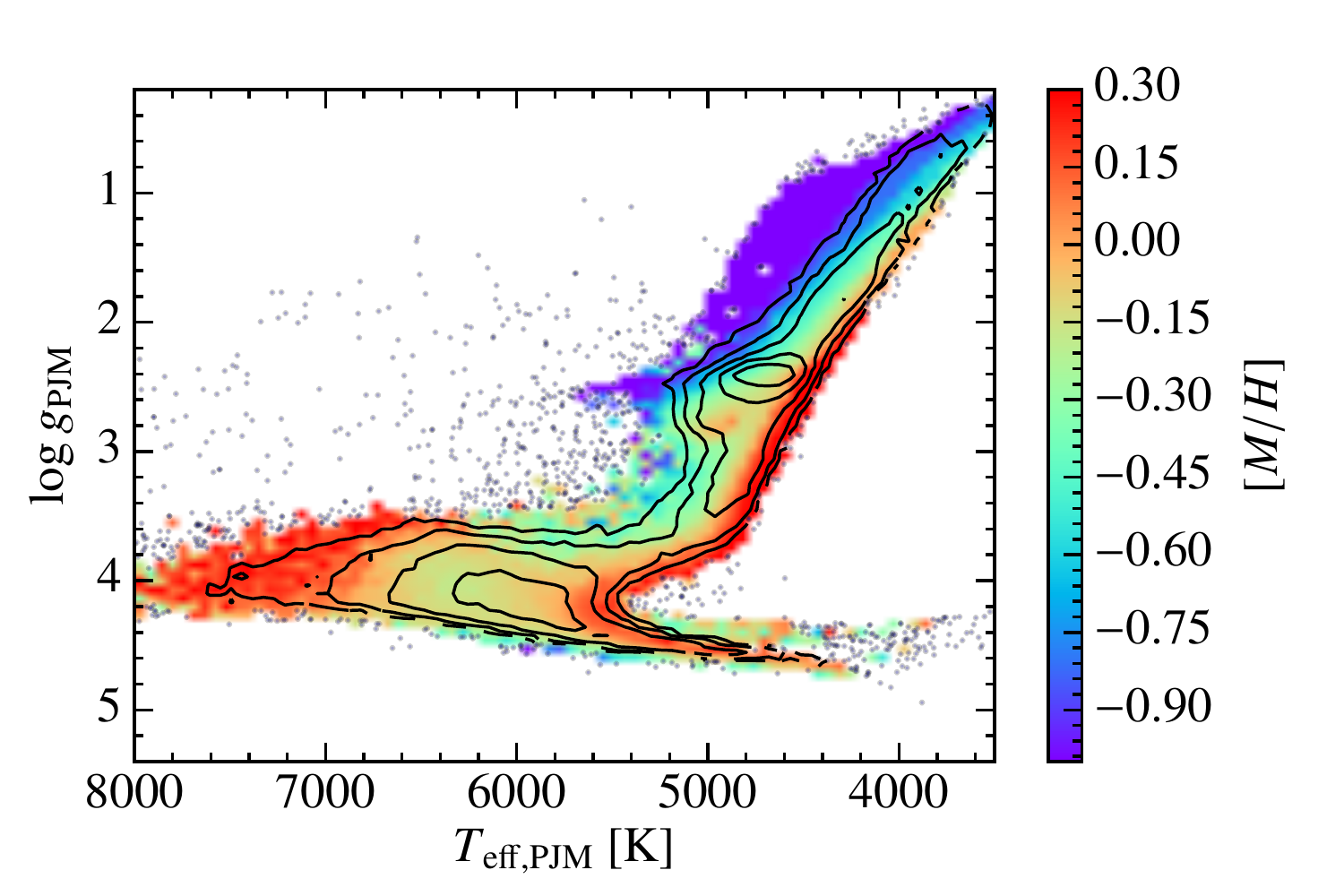}}}
     \caption{
  	Output from the `reverse pipeline', which finds the $\teff$ and $\log g$ values of the stars using the Bayesian method described in this paper. Pixels are coloured by median metallicity, and overlaid contours show the density (with a logarithmic scaling in density between contours).
	  	   \label{fig:HRReverse}
}
\end{figure}

The Bayesian pipeline takes $\log g$ and $\teff$ as inputs to the likelihood calculation, taken either directly from the spectroscopic pipeline or from the IRFM. It also, inevitably (if usually implicitly) determines a posterior probability distribution for these parameters. {\referee The increased information that we now have about the stars (primarily from TGAS) means that these posterior probability distributions are significantly better estimates of the stellar parameters than those we input. In future these may be used to provide estimates for use in the pipeline that determines the chemical abundances, and may be used without giving any input from the spectroscopic pipeline other than metallicity. Because the intention is to provide estimates of these stellar parameters, rather than take them as input, we refer to this as the {\it reverse pipeline}, though it is fundamentally the same machinery}.
The use of parallaxes to improve estimates of $\log g$ is far from new
\cite*[e.g.][]{BeFeLu03}, and here we simply extend it in much the same way as is being planned within the \Gaia\ consortium \citep[e.g.][]{CBJea13}.

Figure~\ref{fig:HRReverse} shows the HR diagram using the best estimates of $\teff$ and $\log g$ from the Bayesian pipeline (referred to as $T_\mathrm{eff, PJM}$, $\log g_{\rm PJM}$). We show the density of stars in this plane using a contour (showing a strong red clump) {\referee and colour regions of the diagram by the median metallicity of stars in that region. This can cause artifacts in regions with few stars, such as above the main sequence}. It is worth noting that the sources with $\log g_{DR5}<2$ do not have their $\log g$ values significantly shifted. This is because the TGAS parallaxes are too uncertain to have much of an effect (see Figure~\ref{fig:Respective}). Future \Gaia\ data releases will have smaller parallax uncertainties, so this approach is a viable one to improve the $\log g$ values for these stars after \Gaia\ DR2, on 25th April 2018. 

We caution that the stars found in regions of the HR diagram away from typical isochrones (e.g. above the main sequence) are likely to have rather untrustworthy parameters from our pipeline. This because our framework is not designed to deal with unusual objects such as binaries (which naturally lie above the main sequence in the colour-magnitude version of the HR diagram). The large majority of these stars are flagged in DR5 using the \cite{Maea12} approach {\creferee (for example, $\sim$900 of the $\sim$1000 found above the main sequence with $T_\mathrm{eff, PJM}<5500\Kelvin$ are flagged)}, and therefore flagged in our catalogue.

{\referee The clean appearance of this HR diagram is to a large extent by construction, because stellar models are used to determine a star's place on the diagram.  For a true test of the reliability of this method we must look at comparisons to external catalogues.}

\begin{figure}
  \centerline{
    \resizebox{\hsize}{!}{\includegraphics{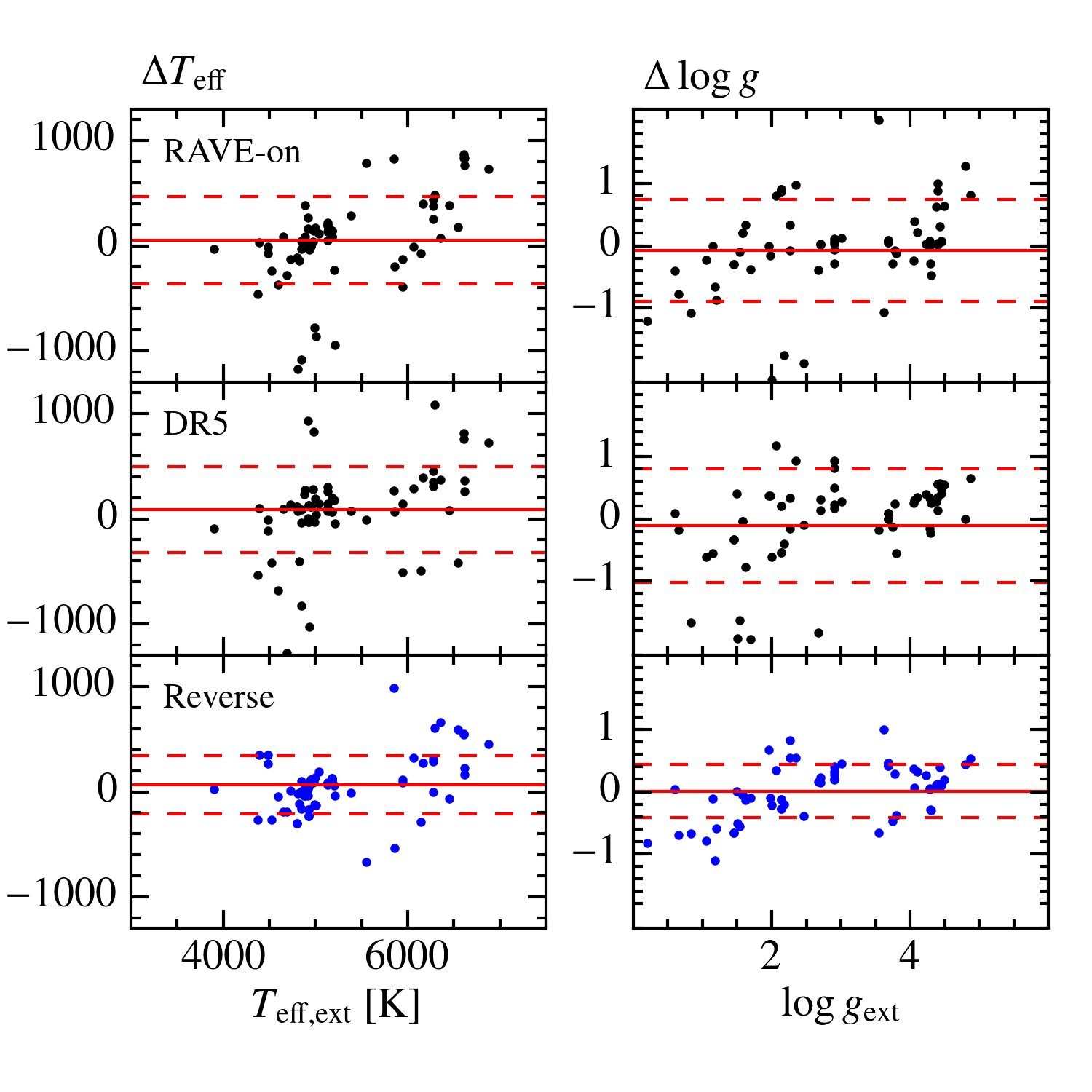}}}
  \caption{
  	{\referee Comparison of parameters from RAVE-on, DR5, and the reverse pipeline to those from high-resolution field stars studies (as described in the text). Note that the y-axis labels are placed at the top of the figure (i.e.,~$\Delta T_{\rm eff}$, $\Delta \log g$). The differences are given in the sense, e.g., $\log g_{\rm DR5}-\log g_{\rm ext}$. The solid red lines indicate the mean values, the dashed red lines are placed one standard deviation either side. In each case the reverse pipeline parameters show less bias and less spread. }
	  	   \label{fig:external1}
}
\end{figure}

\begin{figure}
  \centerline{
    \resizebox{\hsize}{!}{\includegraphics{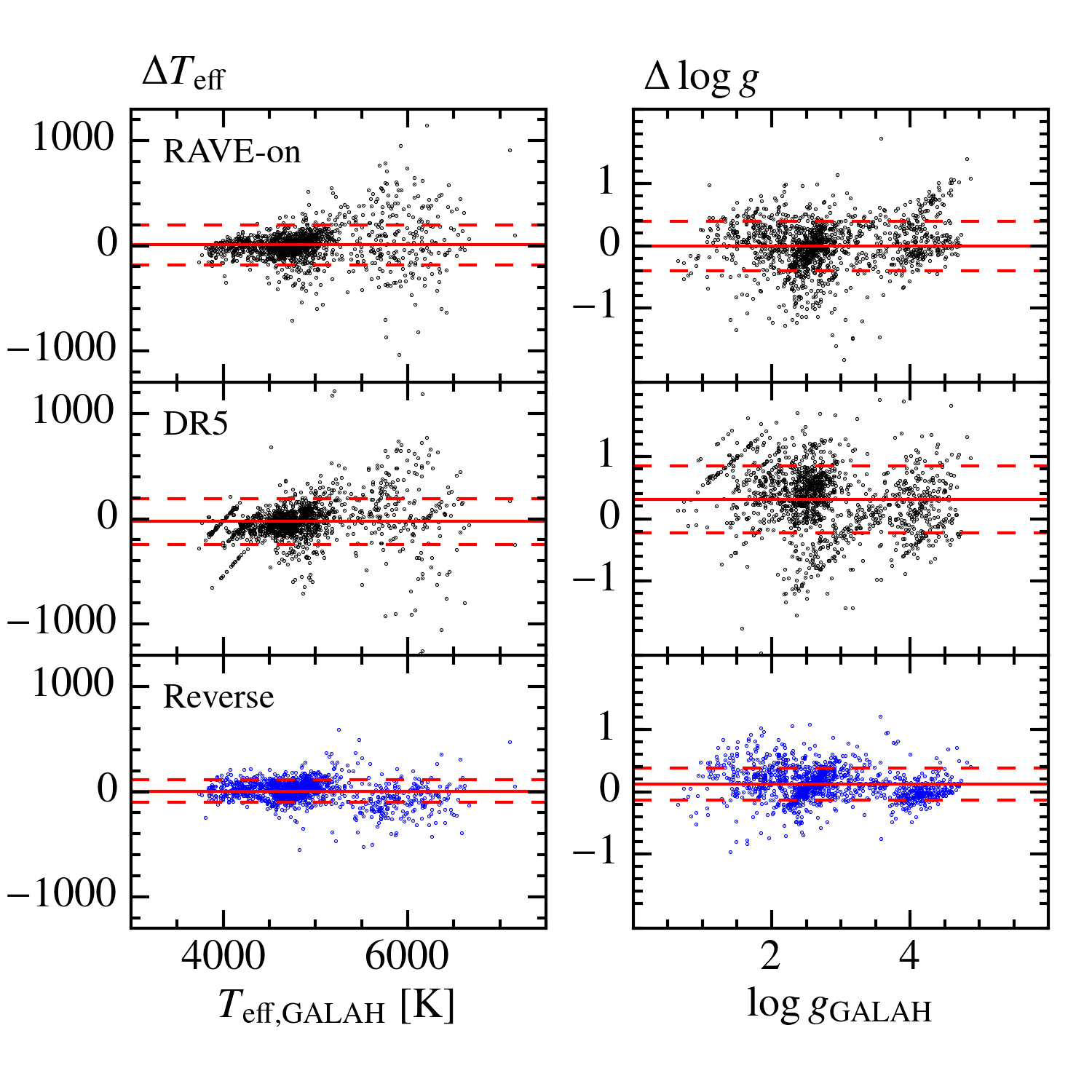}}}
  \caption{
  	{\referee As Figure~\ref{fig:external1}, except it uses parameters from GALAH for the comparison to parameters from RAVE-on, DR5, and the reverse pipeline. Again the spread of values from the reverse pipeline is far smaller than in either other case.}
	  	   \label{fig:external2}
}
\end{figure}

\begin{figure}
  \centerline{
    \resizebox{\hsize}{!}{\includegraphics{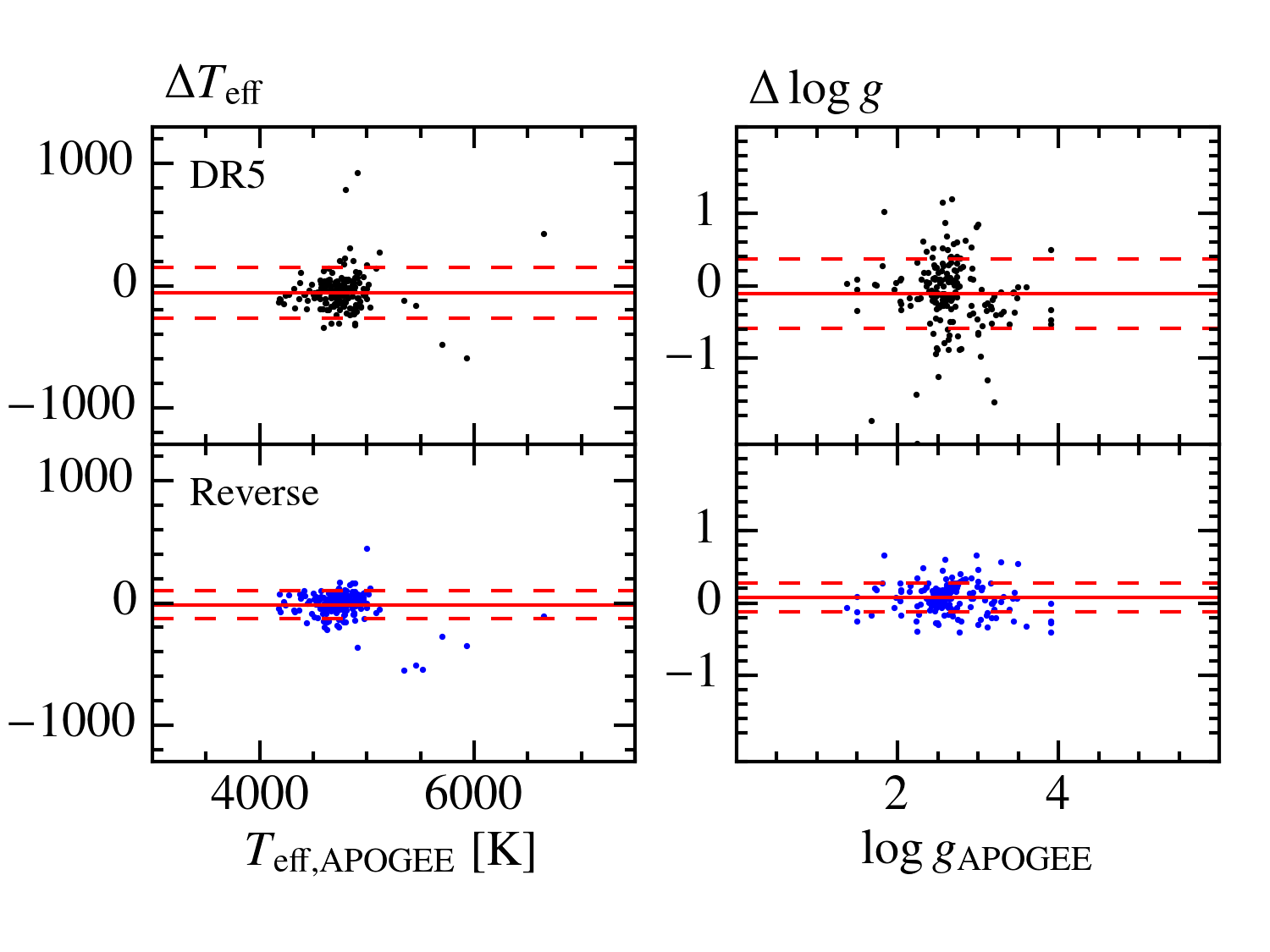}}}
  \caption{
  	{\referee As Figures~\ref{fig:external1}, and \ref{fig:external2} except it uses parameters from APOGEE for the comparison to parameters from DR5, and the reverse pipeline. We do not compare RAVE-on with APOGEE, as RAVE-on uses the RAVE-APOGEE overlap stars as part of their training sample. Again, we find that the scatter and bias are substantially reduced.}
	  	   \label{fig:external3}
}
\end{figure}

{\referee \cite{KuNice17} were the first to highlight the decrease in scatter in the $\teff$ and $\log g$ values from the reverse pipeline as compared to the RAVE pipeline alone.  Figures~\ref{fig:external1}, \ref{fig:external2} \& \ref{fig:external3} show a more detailed comparison in the $\teff$ and $\log g$ parameters from RAVE DR5, RAVE-on \citep[found from the RAVE spectra using a data driven approach by][]{RAVEon} and those presented here compared to high-resolution spectroscopy parameters.  First, Figure~\ref{fig:external1} shows the 67 stars presented here that could be matched with high-resolution field star studies from \cite{2014ApJ...797...13S, 2013ApJ...771...67I,2014AJ....147..136R}, from open and globular clusters Blanco 1 \citep{2005MNRAS.364..272F}, 47 Tuc \citep{2014ApJ...780...94C, 2008AJ....135.1551K, 2009A&A...505..139C}, Pleiades \citep{2009PASJ...61..931F}, NGC 2632 \citep{2015AJ....150..158Y} and IC 4651 \citep{2004A&A...422..951C, 2004A&A...424..951P}, as well as from the Gaia-ESO survey \citep{GaiaESO}.}

{\referee The difference in $\log g$ between these studies and those presented here is $0.0 \pm 0.42\dex$, as compared to $-0.08 \pm 0.83\dex$ and  $-0.11 \pm 0.92\dex$ from RAVE-on and RAVE DR5, respectively.  Upon selecting the 53 stars with SNR $> 40$, the difference in $\log g$ is reduced to $0.03 \pm 0.38\dex$, as compared to $-0.06\pm0.72\dex$ and  $0.00\pm0.83\dex$  from RAVE-on and RAVE DR5, respectively. }

{\referee The scatter in $\teff$ is also improved when adopting the temperatures presented here.  The difference in $\teff$ between the high-resolution studies and those presented here is $75 \pm 282\,\mathrm{K}$ as compared to $51\pm420\,\mathrm{K}$ and $86\pm410\,\mathrm{K}$ from RAVE-on and RAVE DR5, respectively.  Using the stars with SNR $> 40$, the difference in $\teff$ is $81\pm262\,\mathrm{K}$ as compared to $45\pm372\,\mathrm{K}$ and $87\pm390\,\mathrm{K}$ from RAVE-on and RAVE DR5, respectively.  }

{\referee Figure~\ref{fig:external2} shows how the $\teff$ and $\log g$ parameters from RAVE DR5, RAVE-on and those presented here compare to those from Galah DR1, a high-resolution (R$\sim$$28\,000$) spectroscopic survey. }

{\referee From 1379 overlap stars, the difference between Galah $\log g$ and that presented here is $0.12\pm0.26\dex$, compared to $0.0\pm0.40\dex$ from RAVE-on and $0.3\pm0.54\dex$ from RAVE DR5.  The 753 overlap stars with the best RAVE stellar parameters (i.e., AlgoConv $= 0$ and SNR $> 40$) the difference between Galah $\log g$ and that presented here is $0.11 \pm0.22\dex$, compared to $-0.03\pm0.33\dex$ from RAVE-on and $0.37\pm0.42\dex$ from RAVE DR5. }

{\referee Lastly, we present a comparison of the stellar parameters presented here to APOGEE (R$\sim$$22\,500$, Figure~\ref{fig:external3}).  Note that we do not compare RAVE-on with APOGEE, as RAVE-on uses the RAVE-APOGEE overlap stars as part of their training sample.  From 183 overlap stars, we find the difference between APOGEE $\log g$ and that presented here is $0.07\pm0.20\dex$, compared to $-0.11\pm$ 0.49$\dex$ from RAVE DR5.  The difference between APOGEE $\teff$ and that presented here is $-24 \pm124\,\mathrm{K}$, compared to $-58 \pm 210\,\mathrm{K}$ from RAVE DR5.  The 146 overlap stars with AlgoConv $= 0$ and SNR $> 40$ have a difference in $\log g$ between APOGEE and that presented here of $0.08 \pm 0.19\dex$, compared to $-0.06\pm 0.39\dex$ from RAVE DR5.  The difference between APOGEE $\teff$ and that presented here is $-23\pm101\,\mathrm{K}$, compared to $-69\pm 116\,\mathrm{K}$ from RAVE DR5.  }

{\referee Therefore, from a variety of different high-resolution studies, we conclude that the scatter in $\log g$ is a factor of 2 smaller when using surface gravities from the reverse pipeline as compared to both RAVE-on and RAVE DR5 parameters.  Also for stars with low SNR ($< 40$), the gravities and temperatures from the reverse pipeline are reliable, equal to or even better than the gravities and temperatures determined from the high SNR stars in RAVE DR5 and RAVE-on. }

It is our plan to use this method in an iterative fashion with the RAVE spectroscopic pipeline to improve the accuracy of our stellar parameters and therefore the RAVE abundance estimates. We anticipate that the results will be released as part of RAVE DR6.

\section{Conclusions} \label{sec:conclusions}

We have produced new distance, age and stellar parameter estimates for stars common to RAVE and TGAS which reflect new measurements of parallax and $\teff$ (from TGAS and the infra-red flux method, respectively). This allows us to produce distance estimates that are better than those that either RAVE or TGAS can achieve in isolation. {\referee It also allows us to make age estimates which have better than 30 percent precision for 25 percent of the stars in our sample, and estimates of the stellar parameters which are roughly twice as accurate as from RAVE spectra alone (when compared to external catalogues).}

RAVE is the spectroscopic survey with the largest number of sources in common with TGAS, and therefore this dataset has the largest number of sources with both radial velocities from spectroscopy and proper motions from space astrometry. The improvement in distance uncertainty due to this study provides a substantial decrease in the uncertainty on the 3D velocities of these stars. When combined with our age estimates, this gives new insight into the history of our Galaxy.

We have carefully tested the RAVE distance pipeline and the TGAS parallaxes against one another. From this comparison we can draw several conclusions:
\begin{enumerate}
\item The RAVE DR5 parallaxes were overestimated for dwarfs with $\teff\gtrsim5500\,\mathrm{K}$ and underestimated for giants with $\log g\lesssim2.0\dex$. This corresponds to a $\teff$ underestimate in the former case, and a $\log g$ underestimate in the latter. We can (mostly) correct for the former by using the Infrared Flux Method (IRFM) temperatures provided with DR5, but correcting for the latter is beyond the scope of this study.
\item When we use the IRFM $\teff$ values to find spectrophotometric parallaxes, the two parallax estimates agree well in the vast majority of cases, with systematic differences that are substantially smaller than the statistical ones.
\item A comparison as a function of position on the sky indicates that the TGAS parallaxes appear to be overestimated by $\sim0.3\mas$ in a region of the sky near Galactic coordinates $(l,b)=(100^\circ,0^\circ$ which is also near the ecliptic pole \citep[see also][]{GaiaDR1:Validation,ScAu17}.
\item The small random differences between the RAVE-only parallax estimates and the TGAS parallaxes, and the fact that this is found for many stellar types, suggests that the TGAS random uncertainties are overestimated by $\sim0.2\mas$.
\end{enumerate}

We provide flags with our distance estimates, as indicated in Table~\ref{tab:flags}. To use a `clean'  set of stars we recommend that users take only stars with the flag `flag\_all=0'. This yields a set of $137\,699$ stars.

As with previous distance estimates from RAVE, we characterise the output pdf from our distance pipeline by the expectation value and uncertainty in distance, distance modulus, and parallax, and by a multi-Gaussian pdf in distance modulus. This last option provides the most complete description of what the distance pipeline finds, though it is clearly less important here than it was before TGAS parallaxes became available (Fig~\ref{fig:Improvement}).

The apparatus we have used for this study is applicable to data from any spectroscopic survey. It is our intention to apply it to data from the APOGEE survey in the near future.

We will also produce distance estimates for RAVE stars that do not have TGAS parallaxes, using the AllWISE photometry and IRFM temperatures. These will have smaller systematic errors than the DR5 distances, particularly for hot dwarfs, because of the use of IRFM $\teff$ values. All of these distance estimates will be made available through the RAVE website (\url{http://dx.doi.org/10.17876/rave/dr.5/033} and \url{http://dx.doi.org/10.17876/rave/dr.5/034} for the distance estimates for sources with and without TGAS parallaxes, respectively). For TGAS sources they constitute a substantial improvement in distance and, therefore, velocity uncertainty over previous data releases. It is our hope that the new, more precise age and distance estimates are of great value in characterising the dynamics and history of our Galaxy.

\begin{table*}
\caption{Data flags specific to this study. In all cases 1 indicates a potential problem with the distance estimate.\label{tab:flags}}

\begin{center}
\begin{tabular}{ll}
\hline
Name & Explanation \\ \hline
flag\_low\_logg & $\log g_{\rm DR5}<2.0$ (see Section~\ref{sec:Giants}) \\
flag\_outlier & TGAS \& RAVE-only parallaxes differ by more than $4\sigma$ -- RAVE not used for final distances (see Section~\ref{sec:Outliers}) \\
flag\_N & Spectrum flagged as not normal by \cite{Maea12} \\
flag\_pole &  Source lies in the problematic region near the ecliptic pole ($165^\circ<\lambda<195^\circ,\beta<-30^\circ$, see Section~\ref{sec:TGAS}). \\
flag\_dup & A spectrum of the same star with a higher SNR is in RAVE \\
flag\_any & True if any of the above are true, otherwise false \\ \hline
\end{tabular}
\end{center}
\end{table*}


\section*{Acknowledgements}
The authors are grateful to the referee for numerous suggestions which improved the paper.
Paul McMillan is grateful to Lennart Lindegren for suggesting looking at the variation on-sky, and to Louise Howes for a careful reading of the draft.
Funding for the research in this study came from the Swedish
National Space Board, the Royal Physiographic Society in Lund, and some of the computations were performed on resources provided by the Swedish National Infrastructure for Computing (SNIC) at Lunarc under project SNIC 2016/4-17. Funding for RAVE has been provided by: the Australian Astronomical Observatory; the Leibniz-Institut fuer Astrophysik Potsdam (AIP); the Australian National University; the Australian Research Council; the French National Research Agency; the German Research Foundation (SPP 1177 and SFB 881); the European Research Council (ERC-StG 240271 Galactica); the Istituto Nazionale di Astrofisica at Padova; The Johns Hopkins University; the National Science Foundation of the USA (AST-0908326); the W. M. Keck foundation; the Macquarie University; the Netherlands Research School for Astronomy; the Natural Sciences and Engineering Research Council of Canada; the Slovenian Research Agency (research core funding No. P1-0188); the Swiss National Science Foundation; the Science \& Technology Facilities Council of the UK; Opticon; Strasbourg Observatory; and the Universities of Groningen, Heidelberg and Sydney.
The RAVE web site is at https://www.rave-survey.org. 
This work has made use of data from the European Space Agency (ESA) mission {\it Gaia} (\url{https://www.cosmos.esa.int/gaia}), processed by the {\it Gaia} Data Processing and Analysis Consortium (DPAC, \url{https://www.cosmos.esa.int/web/gaia/dpac/consortium}). Funding for the DPAC has been provided by national institutions, in particular the institutions participating in the {\it Gaia} Multilateral Agreement.

\bibliographystyle{mnras} \bibliography{new_refs}

\appendix

\setcounter{table}{3}
\begin{table*}
\begin{center}\caption{Catalogue description. Entries labelled~\mine\ are derived in this study. Entries labelled~\used\ are used to derive the values found in this study. Entries labelled~\DR5 were derived for RAVE DR5, and further explanation can be found in \protect\cite{RAVEDR5} \label{tab:Datatable}} 
\begin{tabular}{lcclll}

\hline 
Col    &   Units       &    Name &           Explanations \\           \hline
1         & -               & RAVE\_OBS\_ID                 & Target designation                \DR5                                     \\
2         & -               & RAVEID                      & RAVE target designation                  \DR5                               \\
3            & deg             & RAdeg                       & Right ascension                       \used      \DR5                            \\
4              & deg             & DEdeg                       & Declination                            \used     \DR5                            \\
5              & deg             & Glon                        & Galactic longitude                              \DR5                        \\
6              & deg             & Glat                        & Galactic latitude                                 \DR5                      \\
7               & $\kms$            & HRV                         & Heliocentric radial velocity         \DR5                                   \\
8               & $\kms$            & eHRV                        & HRV error                                      \DR5                         \\                                     
 9 & $\masyr$ & pmRA\_TGAS & Proper motion in RA from TGAS ($\dot{\alpha}\cos\delta$)
\\
10 & $\masyr$ & pmRA\_error\_TGAS & Standard uncertainty in proper motion in RA from TGAS
\\
11 & $\masyr$ & pmDE\_TGAS & Proper motion in DE from TGAS ($\dot{\delta}$)
\\
12 & $\masyr$ & pmDE\_error\_TGAS & Standard uncertainty in proper motion in DE from TGAS
\\
13 & $\pc$ & distance & Distance estimate \mine
\\
14 & $\pc$ & distance\_err & Distance uncertainty \mine
\\
15 & ${\rm yr}$ & age & Stellar age  estimate \mine
\\
16 & ${\rm yr}$  & age\_err & Stellar age uncertainty \mine
\\
17 & $\msun$ & mass & Stellar mass estimate \mine
\\
18 &  $\msun$ & mass\_err & Stellar mass uncertainty \mine
\\
19 & -- & log\_A\_V & $\log_{10} A_V$ extinction estimate \mine
\\
20 & -- & log\_A\_V\_err & $\log_{10} A_V$ extinction uncertainty \mine
\\
21 & $\mas$ & parallax & Parallax estimate \mine
\\
22 &$\mas$  & parallax\_err & Parallax uncertainty estimate \mine
\\
23 & -- & dist\_mod & Distance modulus estimate \mine
\\
24 & -- & dist\_mod\_err & Distance modulus uncertainty \mine
\\
25 & -- & Teff\_PJM & $\teff$  estimate \mine
\\
26 & -- & Teff\_PJM\_err & $\teff$ uncertainty \mine
\\
27 & -- & logg\_PJM & $\log g$ estimate \mine
\\
28 & -- & logg\_PJM\_err & $\log g$ uncertainty \mine
\\
29 & -- & number\_of\_Gaussians\_fit & Number of components for multi-Gaussian fit ($N_{\rm Gau}$ in Eq.~\ref{eq:defsfk}) \mine
\\
30 & -- & mean\_1 & Parameter for multi-Gaussian fit ($\widehat{\mu_1}$ in Eq.~\ref{eq:defsfk}) \mine
\\
31 & -- & sig\_1 & Parameter for multi-Gaussian fit (${\sigma_1}$ in Eq.~\ref{eq:defsfk}) \mine
\\
32 & -- & frac\_1 &  Parameter for multi-Gaussian fit (${f_1}$ in Eq.~\ref{eq:defsfk}) \mine
\\
33 &  -- & mean\_2 & Parameter for multi-Gaussian fit ($\widehat{\mu_2}$ in Eq.~\ref{eq:defsfk}) \mine
\\
34 & --  & sig\_2 & Parameter for multi-Gaussian fit (${\sigma_2}$ in Eq.~\ref{eq:defsfk}) \mine
\\
35 & --  & frac\_2 & Parameter for multi-Gaussian fit (${f_2}$ in Eq.~\ref{eq:defsfk}) \mine
\\
36 & --  & mean\_3 & Parameter for multi-Gaussian fit ($\widehat{\mu_3}$ in Eq.~\ref{eq:defsfk}) \mine
\\
37 &  -- & sig\_3 & Parameter for multi-Gaussian fit (${\sigma_3}$ in Eq.~\ref{eq:defsfk}) \mine
\\
38 &  -- & frac\_3 & Parameter for multi-Gaussian fit (${f_3}$ in Eq.~\ref{eq:defsfk}) \mine
\\
39 & & FitQuality\_Gauss & Quality of multi-Gaussian fit, as described by Eq.~\ref{eq:defsF} \mine
\\
40 & & Fit\_Flag\_Gauss & Quality flag for multi-Gaussian fit, as discussed in Section~\ref{sec:Bayes} \mine
\\
41 &  -- & AV\_Schlegel & Extinction used in prior ($A_V^{\rm pr}(\infty,l,b)$ in Eq.~\ref{eq:priorExtinction}) \used
\\
42 &  -- & logg\_N\_K & RAVE DR5 Calibrated $\log$ gravity \used \DR5
\\
43 &  -- & elogg\_K & RAVE DR5 Internal uncertainty $\log$ gravity \used \DR5
\\
44 & K & Teff\_IR & Temperature from infrared flux method \used \DR5
\\
45 & K & eTeff\_IR & Internal uncertainty on temperature from infrared flux method \used \DR5
\\
46 &$\dex$ & Met\_N\_K & Calibrated metallicity $\mh$ \used \DR5
\\
47 &$\dex$ & eMet\_K & Internal uncertainty on calibrated metallicity $\mh$ \used \DR5
\\
48 & $\mas$ & parallax\_TGAS & Parallax from TGAS \used
\\
49 &$\mas$  & parallax\_error\_TGAS & Quoted uncertainty on parallax from TGAS \used
\\
50 & -- & Jmag\_2MASS & $J$-magnitude from 2MASS \used
\\
51 & --  & eJmag\_2MASS & Uncertainty on $J$-magnitude from 2MASS \used
\\
52 & --  & Hmag\_2MASS &  $H$-magnitude from 2MASS \used
\\
53 &  -- & eHmag\_2MASS & Uncertainty on $H$-magnitude from 2MASS \used
\\
54 &  -- & Kmag\_2MASS &  $K$-magnitude from 2MASS \used
\\
55 & --  & eKmag\_2MASS & Uncertainty on $K$-magnitude from 2MASS \used
\\
56 &  -- & W1mag\_ALLWISE &  $W1$-magnitude from AllWISE \used
\\
57 &  -- & eW1mag\_ALLWISE & Uncertainty on $W1$-magnitude from AllWISE \used
\\
58 &  -- & W2mag\_ALLWISE & $W2$-magnitude from AllWISE \used
\\
59 &  -- & eW2mag\_ALLWISE & Uncertainty on $W2$-magnitude from AllWISE \used
\\
60 &$\dex$ & Mg & Abundance of Mg $[{\rm Mg}/{\rm H}]$ \DR5
\\
61 & & Mg\_N & Number of spectral lines used for calculation of abundance \DR5
\\
62 &$\dex$ & Al & Abundance of Al $[{\rm Al}/{\rm H}]$ \DR5
\\
63 & & Al\_N & Number of spectral lines used for calculation of abundance \DR5
\\
\end{tabular}
\end{center}
\end{table*}

\setcounter{table}{3}
\begin{table*}
\caption{Catalogue description (continued)} 
\begin{center}
\begin{tabular}{lcclll}
\hline 
Col    &   Units       &    Name &           Explanations \\           \hline

64 &$\dex$ & Si &  Abundance of Si $[{\rm Si}/{\rm H}]$ \DR5
\\
65 & & Si\_N & Number of spectral lines used for calculation of abundance \DR5
\\
66 &$\dex$ & Ti &  Abundance of Ti $[{\rm Ti}/{\rm H}]$ \DR5
\\
67 & & Ti\_N & Number of spectral lines used for calculation of abundance \DR5
\\
68 &$\dex$ & Fe &  Abundance of Fe $[{\rm Fe}/{\rm H}]$ \DR5
\\
69 & & Fe\_N & Number of spectral lines used for calculation of abundance \DR5
\\
70 &$\dex$ & Ni &  Abundance of Ni $[{\rm Ni}/{\rm H}]$ \DR5
\\
71 & & Ni\_N & Number of spectral lines used for calculation of abundance \DR5
\\
72 & & c1 &  Spectral flag following \cite{Maea12} \DR5                                        
\\
73 & & c2 & Spectral flag following \cite{Maea12} \DR5                 
\\
74 & & c3 & Spectral flag following \cite{Maea12} \DR5                 
\\
75 & & c4 & Spectral flag following \cite{Maea12} \DR5                 
\\
76 & & c5 & Spectral flag following \cite{Maea12} \DR5                 
\\
77 & & c6 & Spectral flag following \cite{Maea12} \DR5                 
\\
78 & & c7 & Spectral flag following \cite{Maea12} \DR5                 
\\
79 & & c8 & Spectral flag following \cite{Maea12} \DR5                 
\\
80 & & c9 & Spectral flag following \cite{Maea12} \DR5                 
\\
81 & & c10 & Spectral flag following \cite{Maea12} \DR5                 
\\
82 & & c11 & Spectral flag following \cite{Maea12} \DR5                 
\\
83 & & c12 & Spectral flag following \cite{Maea12} \DR5                 
\\
84 & & c13 & Spectral flag following \cite{Maea12} \DR5                 
\\
85 & & c14 & Spectral flag following \cite{Maea12} \DR5                 
\\
86 & & c15 & Spectral flag following \cite{Maea12} \DR5                 
\\
87 & & c16 & Spectral flag following \cite{Maea12} \DR5                 
\\
88 & & c17 & Spectral flag following \cite{Maea12} \DR5                 
\\
89 & & c18 & Spectral flag following \cite{Maea12} \DR5                 
\\
90 & & c19 & Spectral flag following \cite{Maea12} \DR5                 
\\
91 & & c20 & Spectral flag following \cite{Maea12} \DR5                 
\\
92 & & SNR &  Signal to Noise value \DR5
\\
93 & & Algo\_Conv\_K & Quality Flag for Stellar Parameter pipeline [0..4] \DR5
\\
94 & & flag\_lowlogg & Quality flag (see Table~\ref{tab:flags}) \mine
\\
95 & & flag\_outlier &  Quality flag (see Table~\ref{tab:flags}) \mine
\\
96 & & flag\_N &  Quality flag (see Table~\ref{tab:flags}) \mine
\\
97 & & flag\_pole &  Quality flag (see Table~\ref{tab:flags}) \mine
\\
98 & & flag\_dup &  Quality flag (see Table~\ref{tab:flags}) \mine
\\
99 & & flag\_any &  Quality flag (see Table~\ref{tab:flags}) \mine
\\

\end{tabular}
\end{center}
\end{table*}

\bsp	
\label{lastpage}
\end{document}